\documentclass[12pt,a4]{article}
\usepackage[utf8]{inputenc}
\usepackage[english]{babel}
\usepackage{amsthm,amsmath,latexsym,amssymb,amsfonts,amscd}
\usepackage{graphics,lscape,fancyhdr,array,stmaryrd,euscript}
\usepackage{ifthen,empheq,slashed}
\usepackage{ifpdf}
\usepackage{verbatim,slashed}
\numberwithin{equation}{section}

\newcommand{\bep}{\begin{picture}}
\newcommand{\eep}{\end{picture}}

\newcounter{YoungHeight}\newcounter{YoungWidth}

\newcounter{Mul1}\newcounter{Mul2}\newcounter{Mul3}\newcounter{Mul4}
\newcounter{A0}\newcounter{A1}\newcounter{A2}
\newcounter{B3}
\newcounter{C3}\newcounter{C4}
\newcounter{D1}\newcounter{D2}\newcounter{D3}
\newcounter{T0}\newcounter{T1}

\newlength{\txtHShift}

\newlength{\txtWidth}

\newcommand{\HalfLength}[2]{\setcounter{Mul1}{#1}\setcounter{Mul2}{#1}\addtocounter{Mul1}{\value{Mul2}}\addtocounter{Mul1}{\value{Mul2}}%
\addtocounter{Mul1}{\value{Mul2}}\addtocounter{Mul1}{\value{Mul2}}\setcounter{#2}{\value{Mul1}}}

\newcommand{\Add}[3]{\setcounter{#1}{#2}\addtocounter{#1}{#3}}

\newcommand{\Length}[1]{#10}
\newcommand{\YoungScale}{}

\newcommand{\shiftedText}[2]{{\hspace{#1}#2}}
\newcommand{\calcHShift}[1]{\settowidth{\txtWidth}{#1}\setlength{\txtHShift}{-0.5\txtWidth}}

\newcommand{\TextCenter}[3]{{\HalfLength{#2}{T0}%
\HalfLength{#3}{T1}\addtocounter{T1}{-3}\calcHShift{#1}%
\put(\value{T0},\value{T1}){\shiftedText{\txtHShift}{#1}}}}

\newcommand{\TextCenterB}[3]{{\calcHShift{#1}\HalfLength{#2}{T0}\Add{T1}{\Length{#3}}{-7}\put(\value{T0},\value{T1}){\shiftedText{\txtHShift}{#1}}}}

\newcommand{\TextTop}[3]{{\calcHShift{#1}\HalfLength{#2}{T0}\Add{T1}{\Length{#3}}{-9}\put(\value{T0},\value{T1}){\shiftedText{\txtHShift}{#1}}}}

\newcommand{\BlockA}[2]{{\YoungScale\bep(\Length{#1},\Length{#2}){\Add{A1}{#1}{1}\Add{A2}{#2}{1}}%
\multiput(0,0)(10,0){\value{A1}}{\line(0,1){\Length{#2}}}\multiput(0,0)(0,10){\value{A2}}{\line(1,0){\Length{#1}}}%
\setcounter{YoungHeight}{\Length{#2}}\setcounter{YoungWidth}{\Length{#1}}\eep}}

\newcommand{\RectT}[3]{\bep(\Length{#1},\Length{#2})\put(0,0){\line(1,0){\Length{#1}}}\put(0,0){\line(0,1){\Length{#2}}}%
\put(\Length{#1},\Length{#2}){\line(-1,0){\Length{#1}}}\put(\Length{#1},\Length{#2}){\line(0,-1){\Length{#2}}}#3{#1}{#2}\eep}

\newcommand{\RectARow}[2]{{\bep(\Length{#1},10)\put(0,0){\RectT{#1}{1}{\TextTop{#2}}}\eep}}

\newcommand{\RectARowUp}[2]{{\bep(\Length{#1},10)\put(0,0){\RectT{#1}{1}{\TextCenterB{#2}}}\eep}}

\newcommand{\RectBRowUp}[4]{{\bep(\Length{#1},20)\put(0,0){\RectT{#2}{1}{\TextCenterB{#4}}}%
\put(0,10){\RectT{#1}{1}{\TextCenterB{#3}}}\eep}}

\newcommand{\RectCRowUp}[6]{{\bep(\Length{#1},30)\put(0,0){\RectT{#3}{1}{\TextCenterB{#6}}}%
\put(0,10){\RectT{#2}{1}{\TextCenterB{#5}}}\put(0,20){\RectT{#1}{1}{\TextCenterB{#4}}}\eep}}

\newcommand{\YoungA}{\BlockA{1}{1}}

\newcommand{\BlockApar}[2]{\parbox{\Length{#1}pt}{\YoungScale\bep(\Length{#1},\Length{#2}){\Add{A1}{#1}{1}\Add{A2}{#2}{1}}%
\multiput(0,0)(10,0){\value{A1}}{\line(0,1){\Length{#2}}}\multiput(0,0)(0,10){\value{A2}}{\line(1,0){\Length{#1}}}%
\setcounter{YoungHeight}{\Length{#2}}\setcounter{YoungWidth}{\Length{#1}}\eep}}

\newcommand{\BlockBpar}[4]{\parbox{\Length{#1}pt}{\YoungScale\Add{B3}{\Length{#2}}{\Length{#4}}%
\bep(\Length{#1},\value{B3})\put(0,\Length{#4}){\BlockA{#1}{#2}}%
\put(0,0){\BlockA{#3}{#4}}\setcounter{YoungHeight}{\value{B3}}\setcounter{YoungWidth}{\Length{#1}}\eep}}

\newcommand{\YoungpA}{\BlockApar{1}{1}}
\newcommand{\YoungpB}{\BlockApar{2}{1}}

\newcommand{\YoungpAA}{\BlockApar{1}{2}}
\newcommand{\YoungpBA}{\BlockBpar{2}{1}{1}{1}}
\newcommand{\YoungpBB}{\BlockApar{2}{2}}

\usepackage{hyperref,setspace}

\usepackage[numbers,sort&compress]{natbib}
\usepackage[nottoc]{tocbibind}

\newcommand{\pl}{\partial}
\newcommand{\be}{\begin{equation}}
\newcommand{\ee}{\end{equation}}

\newcommand{\ualpha}{{\ensuremath{\underline{\alpha}}}}

\newcommand{\ua}{{\ensuremath{\underline{a}}}}
\newcommand{\ub}{{\ensuremath{\underline{b}}}}
\newcommand{\uc}{{\ensuremath{\underline{c}}}}
\newcommand{\mm}{{\ensuremath{\underline{m}}}}
\newcommand{\nn}{{\ensuremath{\underline{n}}}}

\newcommand{\la}{{\ensuremath{\mathsf{a}}}}
\newcommand{\lb}{{\ensuremath{\mathsf{b}}}}
\newcommand{\lc}{{\ensuremath{\mathsf{c}}}}

\newcommand{\fud}[2]{{}^{#1}{}_{#2}\,}

\newcommand{\besubeqs}{\begin{subequations}}
\newcommand{\esubeqs}{\end{subequations}}

\newcommand{\Verma}[3]{\ensuremath{\mathcal{D}\left(#1;#2,#3\right)}}
\newcommand{\Rac}{\ensuremath{\mathrm{Rac}}}
\newcommand{\Di}{\ensuremath{\mathrm{Di}}}
\newcommand{\Wi}{\ensuremath{\mathrm{Wi}}}
\newcommand{\Wib}{\ensuremath{\bar{\mathrm{Wi}}}}

\newcommand{\sWb}{\bar{\mathrm{W}}}
\newcommand{\sW}{{\mathrm{W}}}
\newcommand{\sD}{{\mathrm{D}}}
\newcommand{\Yy}[1]{\mathbb{Y}\left(#1\right)}

\newcommand{\commute}[2]{\left[ #1 \, , \, #2 \right]}
\newcommand{\anticommute}[2]{\left\{ #1 \, , \, #2 \right\}}
\newcommand{\Romans}{{|\Omega_0\rangle}}

\newcommand{\Qb}{\bar{Q}}
\newcommand{\Bb}{\bar{B}}

\begin{document}
\hfill
\begin{flushright}
    {LMU-ASC 41/16}
\end{flushright}
\vskip 0.015\textheight
\begin{center}

{\Large\bfseries Exceptional F(4) Higher-Spin Theory in AdS${}_6$ at One-Loop \\
\vspace{0.4cm}
and other Tests of Duality} \\

\vskip 0.03\textheight

Murat \textsc{G\"{u}naydin},${}^{1}$ Evgeny \textsc{Skvortsov},${}^{2,4}$ Tung \textsc{Tran},${}^{3}$

\vskip 0.03\textheight

{\em ${}^{1}$Institute for Gravitation and the Cosmos \\
 Physics Department ,
Pennsylvania State University\\
University Park, PA 16802, USA}\\
\vspace*{5pt}
{\em ${}^{2}$ Arnold Sommerfeld Center for Theoretical Physics\\
Ludwig-Maximilians University Munich\\
Theresienstr. 37, D-80333 Munich, Germany}\\
\vspace*{5pt}
{\em ${}^{3}$ Department of Physics, Brown University \\
Providence, Rhode Island 02912, USA} \\
\vspace*{5pt}
{\em ${}^{4}$ Lebedev Institute of Physics, \\
Leninsky ave. 53, 119991 Moscow, Russia}

\vskip 0.02\textheight

{\bf Abstract }

\end{center}
\begin{quotation}
\noindent
\indent We study  the higher-spin gauge theory in six-dimensional anti-de Sitter space $AdS_6$ that is based on the exceptional  Lie superalgebra $F(4)$. The relevant higher-spin algebra was  constructed in \href{https://arxiv.org/abs/1409.2185}{arXiv:1409.2185 [hep-th]}.
We determine the spectrum of the theory and show that it contains the physical fields  of the Romans $F(4)$ gauged supergravity. The full spectrum consists of  an infinite tower of unitary supermultiplets of $F(4)$ which extend the Romans multiplet to higher spins plus a single short supermultiplet.   \newline
\indent Motivated by applications to this novel supersymmetric higher-spin theory as well as to other theories, we extend the known one-loop tests of $AdS/CFT$ duality in various directions. The spectral zeta-function is derived for the most general case of fermionic and mixed-symmetry fields, which allows one  to test the Type-A and B theories and supersymmetric extensions thereof in any dimension. We also study  higher-spin doubletons and partially-massless fields. While most of the tests are successfully passed, the Type-B theory in all even dimensional anti-de Sitter spacetimes presents an interesting puzzle: the free energy as computed from the bulk is not equal to that of the free fermion on the CFT side, though  there is some systematics to the discrepancy.
\end{quotation}

\newpage

\tableofcontents

\section{Introduction}
\label{sec:Intro}
$AdS/CFT$ duality implies the equivalence of  M/Superstring theory formulated on the product of anti-de Sitter spacetimes $AdS_{d+1}$ with some compact manifold
and certain superconformal field theories on $d$-dimensional Minkowskian spacetimes which correspond to the boundaries of $AdS_{d+1}$ \cite{Maldacena:1997re,Gubser:1998bc,Witten:1998qj}.
The spectrum of M/Superstring theory consists of  a finite number of massless states and an infinite set of massive states.

Higher-spin theories differ from M/Superstring theory in one fundamental way, namely they involve massless fields of arbitrarily high spins and furthermore they favour $AdS$ backgrounds for their consistent formulations \cite{Fradkin:1986qy}.  Higher-spin theories are models of AdS/CFT correspondence that should be considerably simpler than the full-fledged strings on $AdS_5\times S^5$ vs. maximally super-symmetric gauge theory in four dimensions while sharing some of the main features with string theory --- dynamical graviton and fields of arbitrarily high spin.

The basic properties of higher-spin (HS) AdS/CFT dualities  include: (i) higher-spin theories are in most cases duals of CFT's with matter in fundamental representation rather than in adjoint \cite{Klebanov:2002ja}, which simplifies the spectrum of single-trace operators and reduces the field content of HS theories as compared to string theory;
(ii) unbroken higher-spin theories are expected to be dual to free CFT's  \cite{Sundborg:2000wp,Sezgin:2002rt,Klebanov:2002ja,Fernando:2015tiu};
(iii) models of HS AdS/CFT dualities exist in any spacetime dimension \cite{Fernando:2015tiu};
(iv) interacting CFT's, like the Wilson-Fisher $O(N)$ model, can be duals of the same higher-spin theories for a different choice of boundary conditions \cite{Klebanov:2002ja,Sezgin:2003pt,Leigh:2003gk}; (v) the duals of CFT's with matter in the adjoint representation, e.g. $\mathcal{N}=4$ SYM at zero coupling, should also be certain HS theories coupled to matter,  \cite{Sundborg:2000wp,Skvortsov:2015pea, Bae:2016rgm,Bae:2016hfy}.

Singletons and doubletons and their supermultiplets play a fundamental role in the construction of the Kaluza-Klein spectra of $11d$ supergravity \cite{Gunaydin:1984wc,Gunaydin:1985tc} and type IIB supergravity \cite{Gunaydin:1984fk} and
 in the formulation of  higher-spin theories  \cite{Flato:1978qz,Fronsdal:1981gq,Fradkin:1986qy,Gunaydin:1989um,Sezgin:2001yf,Sezgin:1998eh,Govil:2014uwa,Govil:2013uta,Fernando:2015tiu}. They are massless conformal fields: scalar, fermion, spin-one in $4d$ etc. As was shown for $AdS_{4,5,7}$ in \cite{Flato:1978qz, Fronsdal:1981gq,Gunaydin:1984fk,Gunaydin:1984wc}, massless representations of $AdS$ groups and their supersymmetric extensions can all be obtained by tensoring (supermultiplets of) singleton or doubleton representations. The Poincare limits of singletons and doubletons are singular and their field theories  live on the boundaries of $AdS$ spacetimes as conformally invariant theories \cite{Flato:1978qz, Fronsdal:1981gq,Gunaydin:1984fk,Gunaydin:1984wc,Gunaydin:1985tc}. Since HS theories involve massless fields of all spins in $AdS$ spacetimes they naturally fit into the framework of $AdS/CFT$ dualities. Higher tensor products of singletons and doubletons generate the massive KK spectra of various compactifications \cite{Gunaydin:1984fk,Gunaydin:1985tc,Gunaydin:1984wc}.

The simplest free conformal fields provide the basic examples of HS AdS/CFT dualities: free scalar field is dual to Type-A HS theory with spectrum made of totally-symmetric HS fields and free fermion is dual to Type-B whose spectrum contains specific mixed-symmetry fields that include totally-symmetric HS fields too. The results of \cite{Fernando:2009fq,Fernando:2010dp,Fernando:2014pya,Fernando:2015tiu} establishing a one-to-one correspondence between higher-spin theories and supersymmetric extensions thereof and massless conformal fields and conformal supermultiplets imply that this duality extends to all unbroken higher-spin theories and their supersymmetric extensions.

Symmetries of gauged supergravities in $AdS_{3,4,5,\slashed{6},7}$ are well covered by the classical Lie superalgebras of type $OSp(M|N)$ or $SU(N|M)$ \cite{Kac:1977em,Nahm:1977tg,Salam:1989fm}. 
The gap in $AdS_6$ gauged supergravities was bridged by Romans in \cite{Romans:1985tw} where the relevant superalgebra turned out to be the exceptional superalgebra $F(4)$.\footnote{We should note that the simple exceptional Lie algebra of rank 4 is denoted as $F_4$ and does not contain the even subalgebra $SO(7) \oplus SU(2)$ of the exceptional Lie superalgebra $F(4)$.}  Later it was shown that Romans gauged supergravity arises in a warped $S^4$ compactification of the massive $IIA$ supergravity \cite{Cvetic:1999un} as well as from type $IIB$ supergravity \cite{Jeong:2013jfc}. In general much less is known about $AdS_6$ in the context of AdS/CFT dualities\footnote{ See e.g. \cite{Ferrara:1998gv,Brandhuber:1999np,Nishimura:2000wj,Alday:2014bta,Karndumri:2014lba} and references therein.} and HS theories than in other dimensions.

One of the original motivations \cite{Fradkin:1986qy}  for higher-spin theories had been to overcome the $\mathcal{N}\leq8$ restriction on the number of super-symmetries in $d=4$ supergravities. In $AdS_{3,4,5,\slashed{6},7}$ one does find infinite families of anti-de Sitter  superalgebras with any number of supersymmetries. We find it remarkable that there exists \cite{Fernando:2014pya} an $AdS_6$ higher-spin algebra whose maximal finite-dimensional subalgebra is the exceptional Lie superalgebra $F(4)$, which is a unique supersymmetric extension of the $5d$ conformal algebra $SO(5,2)$. Even subalgebra of $F(4)$  is $SO(5,2)\oplus SU(2)$ and the corresponding  HS algebra  can be realized as the universal enveloping algebra of the minimal unitary supermultiplet of $F(4)$ (super-singleton)  obtained via the quasiconformal approach \cite{Gunaydin:2000xr,Gunaydin:2006vz,Gunaydin:2005zz} that consists of two complex scalars in a doublet of R-symmetry group $SU(2)_R$  and a symplectic Majorana spinor field. The HS dual  contains a tower of totally-symmetric bosonic and fermionic HS fields that are dual to HS conserved currents and super-currents as well as a tower of mixed-symmetry fields. The lowest $F(4)$ supermultiplet in this infinite tower, as we will show, is exactly the Romans' supergravity multiplet.

Relying upon AdS/CFT, higher-spin theories should be consistent quantum field theories, which requires a proof. At present, the full action of any of the higher-spin theories is not yet known.\footnote{ The cubic action in de-Donder gauge was recently reconstructed \cite{Sleight:2016dba} in any dimension for the Type-A theory by the AdS/CFT matching with some partial results in \cite{Bekaert:2014cea,Skvortsov:2015pea}, see also \cite{Kessel:2015kna} for $3d$. A part of the on-shell quartic action is known in $d=4$ thanks to \cite{Bekaert:2015tva}. There are also alternative approaches to the action problem: a generalized Hamiltonian sigma-model action \cite{Boulanger:2011dd}, where the Fronsdal kinetic terms are absent, but the theory still can be quantized. See also \cite{Leigh:2014qca} and \cite{Koch:2014aqa}. The spectrum of HS theories is determined by HS algebras and for that reason is consistent with linearized Vasiliev equations whenever they are available \cite{Vasiliev:1990en,Vasiliev:2003ev}. }  The part of the action that is known does not allow to compute the full one-loop self-energy\footnote{See \cite{Ponomarev:2016jqk} for the promising partial results that indicate that the quartic vertex has good chances to cancel all the infinities coming from the bubble made of two cubic vertices.} or beta-function that would provide an access to the quantum properties of HS theories and also a link to the anomalous dimensions of the higher-spin currents in the Wilson-Fisher vector-model. Fortunately, knowledge of kinetic terms is sufficient to perform many nontrivial consistency checks on both sides of the duality by matching various quantities that can be extracted from one-loop partition functions. Another important ingredient of one-loop computations is the knowledge of the spectrum, which can be inferred from the list of higher-spin algebras \cite{Konshtein:1988yg,Vasiliev:2004cm,Govil:2014uwa,Govil:2013uta,Fernando:2014pya,Fernando:2015tiu}. A simpler way to calculate the spectrum is to enumerate single-trace operators in various free CFT's, which increases considerably the number of examples.

Many one-loop tests have already been performed in a series of papers \cite{Giombi:2013fka,Giombi:2014iua,Giombi:2014yra,Beccaria:2014xda,Beccaria:2014jxa,Beccaria:2014zma,Beccaria:2014qea,Beccaria:2015vaa,Beccaria:2016tqy,Bae:2016rgm,Bae:2016hfy}, see also \cite{Gupta:2012he,Gaberdiel:2011zw} for the $3d$ case. The main lessons are as follows. Each of the fields in the spectrum of HS theories contributes a certain amount to one of the computable quantities: sphere free energy, Casimir Energy, $a$- and $c$-anomaly coefficients. The sum over all spins is formally divergent and requires a regularization. Refined in this way the sum over spins becomes finite and matches the corresponding quantity on the CFT side, which in many cases leads to nontrivial tests rather than $0=0$  equalities.

Motivated by our study of the exceptional $F(4)$ higher-spin theory in $AdS_6$ we extend the one-loop tests to a number of cases: (i) we derive the spectral zeta-function for arbitrary mixed-symmetry bosonic and fermionic fields; (ii) we compute one-loop determinants for Type-A and Type-B theories; (iii) we study the contributions of fermionic HS fields in diverse dimensions, which is crucial for the consistency of SUSY HS theories; (iv) in $AdS_5$ we study Type-D,E,... HS theories that are supposed to be dual to higher-spin doubletons with spin greater than one and find that they do not pass the one-loop test; (v) partially-massless fields are also briefly discussed; (vi) a simple expression for the $a$-anomaly of an arbitrary-spin free field is found; (vii) with the help of the heat kernel technique it is argued that a part of the tadpole diagram of the Type-A theory should vanish; (viii) the spectrum of the $F(4)$ HS theory is worked out and is shown to contain the Romans supermultiplet.

Extending the findings of \cite{Giombi:2013fka} we discover that the Type-B theories in all even dimensions lead to puzzling results that call for a better understanding of the duality, the bulk result, however,  still can be represented as a change of the $F$-energy.

The outline is as follows. In Section \ref{sec:HSTheoriesatOneLoop} we review the basic facts about Higher-Spin AdS/CFT correspondence and recall what can be extracted at one-loop order given the fact that the full action is not known. In Section \ref{sec:tests} one-loop tests are performed for a number of cases: fermionic HS fields that are necessary present in SUSY HS theories and for mixed-symmetry fields that are omnipresent in $AdS_5$ and higher. In Section \ref{sec:ffour} the properties of the exceptional $F(4)$ Higher-Spin Theory are studied.  Technicalities are collected in numerous Appendices. The summary of the results and discussion can be found in Section \ref{sec:conclusions}.

\section{Higher-Spin Theories at One-Loop}
\label{sec:HSTheoriesatOneLoop}
As discussed below, one-loop computations in higher-spin (HS) theories require one simple ingredient as an input data: a CFT with infinitely many conserved higher-rank tensors --- higher-spin currents. Such CFT's are very special --- they are free or $N\rightarrow\infty$ limits of certain interacting ones, which again behave like free theories in the strict $N=\infty$ limit. The algebra of HS currents determines the field content of the dual HS theory and allows one to perform many one-loop tests. We briefly review basic facts about higher-spin theories and the scheme of one-loop tests.

\subsection{Higher-Spin Theories}
\label{sec:HSTheories}
The intrinsic definition of higher-spin (HS) theories is that they are field theories with infinitely many massless higher-spin fields. A systematic approach is via the Noether procedure, i.e. one starts with the free fields and then tries to add interaction vertices and deform gauge transformations as to maintain gauge invariance of the action.

The AdS/CFT correspondence provides an easier approach to HS theories --- HS theories can be thought of as duals to free CFT's \cite{Sundborg:2000wp,Sezgin:2002rt,Klebanov:2002ja,Fernando:2015tiu}. Indeed, HS gauge fields are dual to conserved tensors of rank greater than two, i.e HS conserved tensors\footnote{We use $a,b,c,...=0,...,d-1$ do denote $CFT^d$ Lorentz indices and $\ua,\ub,\uc,...=0,...,d$ for $AdS_{d+1}$ bulk Lorentz indices. }
\begin{align}
    \pl^mJ_{mabc...}&=0 &&\Longleftrightarrow && \delta \Phi_{\mm\ua\ub\uc...}=\nabla_\mm\xi_{\ua\ub\uc...}+...
\end{align}
The presence of an at least one conserved HS tensor in addition to  the stress-tensor in a $CFT^d$ in $d\geq3$ makes this CFT a free one in disguise \cite{Maldacena:2011jn,Alba:2013yda,Boulanger:2013zza,Stanev:2013qra,Alba:2015upa,Fernando:2015tiu}. In particular, it implies that conserved tensors of arbitrarily high rank are present. Conserved tensors generate charges and for that reason such CFT's have infinite-dimensional algebras of symmetries, higher-spin algebras, see \cite{Fradkin:1986ka} for the first occurrence of the HS algebra concept in the literature. On the CFT side HS algebra is the algebra of global symmetries and contains the conformal algebra as a subalgebra.

HS currents together with the stress-tensor and few other matter fields that are produced by acting with HS charges form a representation of the HS algebra they generate. On the AdS side the global symmetry should become a gauge symmetry, i.e. a given HS algebra needs to be gauged. The spectrum of the dual HS theory is induced by the single-trace CFT operators: HS currents and certain other operators as will become clear below.

Let us consider two simplest examples of free CFT's: free scalar and free fermion, whose duals are usually called Type-A and Type-B, respectively, and then extend them to more general cases including the super-symmetric ones.

\paragraph{Type-A.} A free scalar field $\square \phi=0$ as a representation of the conformal algebra is usually called $\Rac$. With one complex scalar one can construct conserved higher-spin currents, which are totally-symmetric tensors:
\begin{align}
    J_{s}&= \bar{\phi} \pl^s \phi +...\,, && \Delta=d+s-2\,,\\
    J_0&= \bar{\phi}\phi\,, && \Delta=d-2\,,
\end{align}
where we also add the 'spin-zero current' $\bar{\phi}\phi $. If the scalar is real then the currents of odd ranks vanish. For a free theory doing operator product expansion (OPE)  is practically equivalent to computing the tensor product of the conformal algebra representations, which in the case of $\bar{\phi}\phi$ OPE leads to \cite{Flato:1978qz,Vasiliev:2004cm,Dolan:2005wy}:
\begin{align}
    \Rac\otimes \Rac&= \sum_s J_s\,.\label{typeaope}
\end{align}
More generally, one can take the scalar field with values in some representation $V$ of some Lie group $G$ and impose the singlet constraint, i.e. project onto $G$-invariants. In the representation theory language the fundamental fields belong to $S=\Rac\otimes V$ and the spectrum of bilinear operators corresponds to the $G$-invariant part of the tensor product $S\otimes S$. Technically what matters is the symmetry of $J_s$ with respect to exchange of two fields and the symmetry of the $G$-invariant tensors. For example, if $\phi^i$ are $SO(N)$-vectors and $N$ is large, then the relevant invariant tensor is $\delta_{ij}$, which is symmetric. Noting that $P(J_s)=(-)^s J_s$, where $P$ exchanges the two scalar fields, we observe that all HS currents with odd spins are projected out and the $SO(N)$-invariant single-trace operators belong to $(\Rac\otimes\Rac)_S$, i.e. have even spins. Therefore, the $SO(N)$-singlet constraint distinguishes between (anti)-symmetric parts of $\Rac\otimes \Rac$,  \cite{Flato:1978qz,Vasiliev:2004cm,Dolan:2005wy}:
\begin{align}
    (\Rac\otimes \Rac)_S&= \sum_k J_{2k}\,, &
    (\Rac\otimes \Rac)_A&= \sum_k J_{2k+1}\,.
\end{align}

In accordance with \eqref{typeaope} the spectrum of the Type-A theory is made of bosonic totally-symmetric HS fields that are duals of $J_s$, known as Fronsdal fields \cite{Fronsdal:1978rb}, and an additional scalar field $\Phi_0$ that is dual to $\phi^2$. At the free level Fronsdal fields $s=0,1,2,3,...$ obey\footnote{Abbreviation $a(s)$ or $\ua(s)$ is for the group of $s$ symmetric/to be symmetrized indices $a_1...a_s$. For simplicity we impose the transverse traceless (TT) gauge: the field is a traceless tensor and is $\nabla$-transverse. To be consistent with the TT-gauge the gauge parameter is also TT.}
\begin{align}
    \left(-\nabla^2 +M^2_{s}\right)\left(\Phi^{\ua(s)}+\nabla^\ua\xi^{\ua(s-1)}\right)&=0\,,&
    M^2_{s}&=(d+s-2)(s-2)-s\,,
\end{align}
where $\xi^{\ua(s-1)}$ represents gauge modes. The value of the mass-like term follows directly from the conformal weight of the conserved HS current it is dual to, as usual.

HS theory of totally-symmetric HS fields, $s=0,1,2,3,4,...$ is called the non-minimal Type-A, which is the $U(N)$-singlet projection, and the one with even spins only, $s=0,2,4,...$ is the minimal Type-A, which is the $O(N)$-singlet projection. One can also define the $usp(N)$-singlet theory whose spectrum is made of three copies of odd spins and one copy of even spins \cite{Giombi:2013fka}.

\paragraph{Type-B.} Analogously, one can take a free fermion $\slashed {\pl}\psi=0$, which is called $\Di$. The spectrum of single-trace operators is more complicated \cite{Flato:1978qz,Vasiliev:2004cm,Dolan:2005wy,Alkalaev:2012rg,Alkalaev:2012ic}. They have the symmetry of all hook Young diagrams $\mathbb{Y}(s,1^p)$:\footnote{Throughout the paper we will often use the language and pictures of Young diagrams to refer to $so(d)$ representations. If it has  weight $(s_1,...,s_n)$ for $d=2n$ or $d=2n+1$, then we will denote it by the Young diagram with rows of lengths $s_i$ (the rows of zero length omitted). Notation $1^p$ means $p$ rows of length one.  }
\begin{align}
    J_{s,p}&= \bar{\psi} \gamma...\gamma \pl^{s-1}\psi+...\,.
\end{align}
In more detail, the mixed-symmetry currents are irreducible tensors $J_{a(s),m[p]}$ that are symmetric in $a_1...a_s$ and anti-symmetric in $m_1...m_p$, obey the Young condition, have vanishing traces and are conserved:\footnote{Note that the conservation is not simply $\pl\cdot J=0$ due to the Young symmetry. One has to project onto the right irreducible component, otherwise there are no solutions or unitarity is lost. The projection is done by anti-symmetrizing over all $m$ indices in the second line.}
\begin{align}
J_{a_1...a_s,m_1...m_p}&= \bar{\psi} \gamma_{a_s m_1...m_p} \pl_{a_1...a_{s-1}} \psi+...\,,\label{typebcurrent}\\ \notag
    \begin{aligned}
    &\text{conservation:} & \pl^n J_{a(s-2)mn,m[p]}&=0\,,  \\
    &\text{Young:}& J_{a(s),am[p-1]}&=0\,,\\
    &\text{tracelessness:}& J\fud{b}{a(s-2)b,m[k]}&=0\,.
    \end{aligned} &&
    \parbox{60pt}{{\bep(60,50)\unitlength=0.38mm%
    \put(0,40){\RectARowUp{6}{$s$}}%
    \put(0,0){\RectT{1}{4}{\TextCenter{$p$}}}\eep}}
\end{align}
Conserved currents correspond to $s\geq2,\forall p$ and $s=1,p=0$, the latter is a usual conserved current $\bar{\psi}\gamma_a\psi$.
In particular, the totally-symmetric HS currents, including the stress-tensor are still there. Also, there are anomalous, i.e. not obeying any conservation law, anti-symmetric tensors and an additional scalar operator $\bar{\psi}\psi$:
\begin{align}\label{masssivepforms}
    J_{m[p]}&= \bar{\psi} \gamma_{m_1}...\gamma_{m_p} \psi\,, && p=0,2,3,4,...\,,
\end{align}
which are degenerate cases of the same expression \eqref{typebcurrent}. The spectrum of single-trace operators can equivalently be computed as $\Di\otimes \Di$ \cite{Flato:1978qz,Vasiliev:2004cm,Dolan:2005wy}:
\begin{align}
    \Di\otimes \Di &= \sum_{s,p} J_{s,p}\,.
\end{align}
The height, $(p+1)$, of the hook Young diagrams cannot exceed dimension $d$. Moreover, with the help of the $\epsilon$-tensor the hooks with $p+1>d/2$ can be dualized back to $p+1\leq d/2$. In the case of $d$ even one may also decompose the hooks with $p+1=d/2$ into two irreducible components. We will give a more detailed description of the Type-B spectrum in Section \ref{sec:tests}.

Accordingly, the spectrum of the Type-B theory is made of bosonic mixed-symmetry gauge fields with spin $\mathbb{Y}(s,1^p)$, $s>1,\forall p$ or $s=1,p=0$:\footnote{Young symmetry requires to add $\nabla \xi$-terms with different permutations, which are hidden in $...$.}
\begin{align}
    &\left(-\nabla^2 +M^2_{s,1^p}\right)\left(\Phi^{\ua(s),\mm[p]}+\nabla^\ua \xi^{\ua(s-1),\mm[p]}+...\right)=0\,,\\
    &M^2_{s,1^p}=(d+s-2)(s-2)-s-p\,.
\end{align}
We will refer to such fields simply as {\it hooks}, having in mind the shape of Young diagrams $\mathbb{Y}(s,1^p)$. The general formula for the mass-like term was found in \cite{Metsaev:1995re, Metsaev:1997nj}. The anti-symmetric tensors \eqref{masssivepforms} are dual to massive\footnote{There are different definitions of masslessness in anti-de Sitter space. As far as $s>\tfrac12$ fields are concerned, the most natural definition seems to be the one where massless fields are those that have gauge symmetries which reduce the number of physical degrees of freedom. The same fields can also be found in the tensor product of two singletons/doubletons, which is the definition of masslessness adopted in \cite{Gunaydin:1984fk,Gunaydin:1984wc,Gunaydin:1985tc}. As for matter fields with $s=0,\tfrac12$ one can either adopt the latter definition or refer to conformally coupled fields as massless instead. Massive $h$-forms of Type-B theories do not have any gauge symmetries.}  anti-symmetric fields, including the scalar $\Phi$:
\begin{align}
    \left(-\nabla^2 +M^2_{1^h}\right)\Phi^{\mm[h]}&=0\,, &
    M^2_{1^h}&=-(d-1)-h\,, && h=0,2,3,4,...\,.
\end{align}
This is the spectrum of the non-minimal Type-B and one can extend the discussion to the duals of (symplectic)(Majorana)-Weyl fermions. It is worth stressing that Type-A theory is in no sense a sub-theory of Type-B. In particular, the cubic couplings are different \cite{Bekaert:2014cea,Skvortsov:2015pea}, the only exception being the $d=3$ case where there are no mixed-symmetry fields and the HS algebras generated by free boson and free fermion are the same.

\paragraph{SUSY HS Theories.} The simplest super-symmetric HS theories result from CFT's made of a number of free scalars and fermions. The single-trace operators contain those of Type-A and Type-B combined. Also, there are super-currents:\footnote{As primaries the currents must be traceless in $a(s)$ and $\gamma$-traceless in $a(s);\alpha$, the former being a consequence of the latter.}
\begin{align}
    J_{s=m+\tfrac12}&= \phi \pl^m\psi+... && \Longleftrightarrow&&J_{a(m);\alpha}=\phi \pl_{a_1}...\pl_{a_m}\psi_\alpha+...\,.
\end{align}
The super-currents, as representations of the conformal algebra, belong to $\Di\otimes \Rac$  \cite{Flato:1978qz,Vasiliev:2004cm,Dolan:2005wy}:
\begin{align}
   \Di\otimes \Rac&=\sum_{m=0} J_{s=m+\tfrac12}\,.
\end{align}
The super-currents are dual to totally-symmetric fermionic HS fields, Fang-Fronsdal fields \cite{Fang:1978wz,Fang:1979hq}:
\begin{align}
    (\slashed{\nabla}+m)\left(\Phi^{\ua(s);\ualpha}+
    \nabla^{\ua}\xi^{\ua(s-1);\ualpha}\right)&=0\,, & m^2&=-\left(s+\tfrac{d-4}{2}\right)^2\,.
\end{align}
For the purpose of computing the determinants we need to know the square of the HS Dirac operators
\begin{align}
    (-\slashed{\nabla}+m)(+\slashed{\nabla}+m)&=\left(-\nabla^2 +M^2_{s}\right)\,,&    M^2_{s}&=m^2+s+\frac{d(d+1)}4\,,
\end{align}
where the mass-like terms were found in \cite{Metsaev:1998xg} for fermionic fields of any symmetry type.

Therefore, in the simplest super-symmetric HS theory the spectrum is Type-A plus Type-B plus fermionic HS fields, which  can be packed symbolically into super-matrices of the form
\begin{align}
    \begin{pmatrix}
    \text{Type-A}=\Rac\otimes \Rac & \Rac\times \Di\\
    \Di\times \Rac & \text{Type-B}=\Di\otimes \Di
    \end{pmatrix}=
    \sum
    \begin{pmatrix}
    \Phi^{\ua(s)} & \Psi^{\ua(s-\tfrac12);\ualpha}\\
    \Psi^{\ua(s-\tfrac12);\ualpha} & \Phi^{\ua(s),\mm[p]}
    \end{pmatrix}
\end{align}
Again, one can take a number of $\phi$'s and $\psi$'s and impose the singlet constraint with respect to some global symmetry group. Note that the $d=3$ case is special in that there are no mixed-symmetry fields, i.e. $p=0$: both Type-A and Type-B have totally-symmetric HS fields only, but $\phi^2$ has weight $\Delta=1$ while $\bar{\psi}\psi$ has $\Delta=2$, which corresponds to the same mass-like term $M^2=-2$.

\paragraph{More general HS theories. } The general scheme is the following. Given some $d$ there is a list $L$ of free conformal fields that can exist in $CFT^d$. Generically, $L$ always contains free scalar and free fermion. Also, one can add free conformal fields\footnote{For a comprehensive list of conformally-invariant equations we refer to \cite{Shaynkman:2004vu}.} $\phi^{\mathbb{S}}$ with any spin $\mathbb{S}$ obeying $\square^k \phi^{\mathbb{S}}+...=0$, $k=1,2,...$ equations of motion. However, these are usually non-unitary, which may not be an obstruction to HS AdS/CFT. In even dimension $d=2n$ doubletons $S_j$ with spin-$j$ are also available\footnote{Formally, the $so(d=2n)$-spin of  doubletons is $\mathbb{Y}(j,...,j)=\mathbb{Y}(j^n)$, but we abbreviate it simply as spin-$j$. } \cite{Gunaydin:1984wc,Gunaydin:1984fk,Metsaev:1995jp,Bekaert:2009fg,Fernando:2015tiu}, where $j=0,\tfrac12$ are the usual $\Rac$ and $\Di$. The $j=1$ case corresponds to $\tfrac{d}{2}$-forms, e.g. the Maxwell field-strength $F_{ab}$ in $d=4$. It can be further projected onto (anti)-selfdual components, $S^{\pm}_1$. Therefore, there is some list of free conformal fields of interest in dimension $d$:
\begin{align}
    L&=\{\Di,\Rac,....\}
\end{align}
In order to build a more general free CFT one can select a number of distinct free fields $L_i$. For every field one can pick some group $H_i$ and let it take values in some representation of $H_i$. Also, one should choose some group $F\in H_i$ that will be used to impose the singlet constraint, i.e. by projecting onto the invariants of $F$. The higher-spin symmetry or the spectrum of the $AdS$-dual theory is then generated by all bilinear quasi-primary operators that are $F$-singlets. In the case of the adjoint duality one has to consider long trace operators that are dual to certain matter-like massive (HS) fields that couple to the massless sector. Therefore, the duals of free CFT's with matter in adjoint representations look like the duals of CFT's with fundamental matter coupled to certain matter multiplets \cite{Sundborg:2000wp,Skvortsov:2015pea, Bae:2016rgm,Bae:2016hfy}.

For example, one can take $F=u(N)$, $L_1=\Rac$, $H_1=u(n)\times u(N)$, $L_2=\Di$, $H_2=u(m)\times u(N)$ and fields $L_{1,2}$ to take values in the $Nn$ and $Nm$ dimensional representations. The resulting $F$-singlet spectrum has HS fields of $\Rac\otimes \Rac$ with values in $u(n)$, fields of $\Di\otimes \Di$ with values in $u(m)$ and $2nm$ fermionic HS fields, see \cite{Konstein:1989ij} for $d=3$.

As another example, one can define Type-C \cite{Beccaria:2014zma} as the dual of the spin-$j=1$ doubleton $S_1$ for $d=2n$, i.e. $AdS_5/CFT^4$, $AdS_7/CFT^6$, etc. The spectrum of Type-C contains more complicated mixed-symmetry fields. It is also possible to cook up extended multiplets $n_b\Rac+n_f\Di+n_vS_1$ \cite{Beccaria:2014xda}. In $AdS_7/CFT^6$ one can take \cite{Beccaria:2014qea} the $(2,0)$ tensor supermultiplet that contains $\Rac$, $\Di$ and a self-dual rank-three tensor $T=S_1$, which is spin-one doubleton \cite{Gunaydin:1984wc}.

The HS algebra based on $\sum_i n_i L_i$ contains diagonal elements $L_i\otimes L_i$ and off-diagonal ones $L_i\otimes L_j$. Fermionic HS fields, if any, are always placed into off-diagonal blocks since they result from $\text{fermion}\otimes\text{boson}$ products. There are some other fields that can only arise in off-diagonal blocks, e.g. partially-massless fields of even depths \cite{Alkalaev:2014nsa}.

In dimensions $d=3,4$ and $d=6$ there exist conformal superalgebras, namely $OSp(N|4,\mathbb{R})$, $SU(2,2|N)$ and $OSp(8^*|2N)$,  with arbitrary number of supersymmetry generators.  Since there is no constraint on the spins of the particles in higher-spin theories there is no constraint on the number of supersymmetry generators of HS superalgebras in these dimensions \cite{Konstein:1989ij,Gunaydin:1989um,Vasiliev:2004cm,Govil:2014uwa,Govil:2013uta}. More general HS superalgebras in higher than six dimensions can also be defined \cite{Konstein:1989ij,Vasiliev:2004cm}. However the supersymmetry in these theories do not obey the usual spin and statistics connection.  Only in  dimensions $d\leq 6$  the HS superalgebras contain the usual spacetime conformal superalgebras as finite-dimensional subalgebras.

Also, as was noted in \cite{Chang:2012kt} in the case of $AdS_4/CFT^3$ and in \cite{Gomez:2014dwa} for $AdS_3/CFT^2$ the AdS/CFT truncates the number of super-symmetries back to the usual one by the boundary conditions. The same is expected to be true in any other dimension where the super-symmetric CFT duals can exist.\footnote{E.S. is grateful to Kostya Alkalaev for the discussion on the truncations of SUSY HS theories by boundary conditions.} In other words, usual AdS/CFT restricts the number of super-symmetries in HS theories not to exceed that of supergravities. Dualities between HS theories with any number of SUSY's and free CFT's made of a number of scalars and fermions may still work in any $d$, though not having standard superalgebras behind.

More recent work has shown that $\Rac$'s, which are singletons of $SO(d,2)$ for odd $d$  and scalar doubletons of $SO(d,2)$ for even $d$, are simply the minimal unitary representations of $SO(d,2)$. For odd $d$ they admit a single deformation (spinor singleton), $\Di$, and for even $d$ they admit an infinite family of deformations (doubletons) \cite{Fernando:2009fq,Fernando:2010dp,Fernando:2014pya,Fernando:2015tiu}. Furthermore there exists a one-to-one correspondence between the minimal unitary representations of $SO(d,2)$ and their deformations and massless conformal fields in $d$ dimensional Minkowskian spacetimes \cite{Fernando:2015tiu}. These results were obtained by quantization of the geometric realization of $SO(d,2)$ as quasiconformal groups \cite{Gunaydin:2000xr,Gunaydin:2006vz,Gunaydin:2005zz}. The geometric realizations of noncompact groups as quasiconformal groups that leave invariant a quartic light-cone was discovered in \cite{Gunaydin:2000xr}. The quantization of the geometric quasiconformal realization of a noncompact group
leads directly to its minimal unitary representation \cite{Gunaydin:2001bt,Gunaydin:2005zz}.

\paragraph{Quadratic action.} Combining the ingredients together the quadratic gauge fixed action of the simplest SUSY HS theory that is cooked up from $\Rac$'s and $\Di$'s should have the form
{\allowdisplaybreaks\begin{align}
    S_0&=\frac{1}{G}\int\left[N_AS_A+N_BS_B+N_FS_{F}\right]\,,\\
    S_{A}&=\frac12 \sum_s\int \Phi_{\ua(s)}\left(-\nabla^2 +M^2_{s}\right)\Phi^{\ua(s)}\,,\\
    S_{B}&=\frac12 \sum_{s,p}\int \Phi_{\ua(s),\mm[p]}\left(-\nabla^2 +M^2_{s,1^p}\right)\Phi^{\ua(s),\mm[p]}\,,\\
    S_{F}&= \sum_{s}\int \bar{\Psi}_{\ua(s-\tfrac12)}\left(\slashed{\nabla}+m_s\right)\Psi^{\ua(s-\tfrac12)}\,,
\end{align}}\noindent
where the multiplicities $N_A$, $N_B$, $N_F$ depend on the multiplet chosen and also, for specific multiplets, can depend on whether spin is even or odd, but HS fermions enter all together $s=\tfrac12, \tfrac32, \tfrac52,...$. On general grounds the bulk coupling constant $G$ should be related to the fraction $N$ of the fields removed by the singlet constraint as $G^{-1}\sim N$, at least in the large $N$ limit. It was observed \cite{Giombi:2013fka,Giombi:2014iua} that in some of the cases this relation should be $G^{-1}= a(N+\text{integer})$. It can happen that there are higher-order $N^{-1}$-corrections as well.

\subsection{One-Loop Tests}
\label{sec:onelooptests}
The idea of the one-loop tests of HS AdS/CFT was explained  in \cite{Giombi:2013fka,Giombi:2014iua}. The AdS partition function
\begin{align}
    Z_{AdS}&=\int \prod_k D\Phi_k\, e^{S[\Phi_s]}\,,
\end{align}
as a function of the bulk coupling $G$ should lead to the following expansion of the free energy $F_{AdS}$:
\begin{align}
    -\ln Z_{AdS}&=F_{AdS}=\frac{1}{G} F^0_{AdS} +F^1_{AdS} +G F^2_{AdS}+...\,,
\end{align}
where the first term is the classical action evaluated at an extremum. $F^1$ stands for one-loop corrections, etc. The large-$N$ counting suggests that $G^{-1}\sim N$. Moreover, $N$ is expected to be quantized \cite{Maldacena:2011jn}, which is not yet seen in the bulk.\footnote{It is an interesting question whether  the quantization of the bulk HS coupling can be understood as a consequence of invariance under large higher-spin transformations as in Chern-Simons theory.} On the dual CFT side there should be a similar expansion for the CFT free energy $F_{CFT}$:
\begin{align}
    -\ln Z_{CFT}&=F_{CFT}=N F^0_{CFT}+F^{1}_{CFT}+\frac1{N}F^2_{CFT}+...\,.
\end{align}
A nice property of free CFT's is that all but the first term are zero, which should match $F^0_{AdS}$. However, since the classical action is not known, one cannot compute $F^0_{AdS}$ and compare it to $F^0_{CFT}$. Still, one can check that the second term, $F^1_{AdS}$ vanishes identically or produces a contribution proportional to $F^0_{CFT}$, which can be compensated by modifying the simplest relation $G^{-1}=N$ to  $G^{-1}=a(N+\text{integer})$ \cite{Giombi:2013fka,Giombi:2014iua}.

This basic idea allows to perform several non-trivial tests thanks to the fact $F$ can be computed on different backgrounds. The simplest ones include ${S}^d$, $\mathbb{R}\times {S}^{d-1}$ and ${S}^1\times {S}^{d-1}$ that are the boundaries of Euclidean $AdS_{d+1}=\mathbb{H}^{d+1}$, global $AdS_{d+1}$ and thermal $AdS_{d+1}$, respectively.\footnote{Note that on more complicated backgrounds one encounters the problem of light states \cite{Gaberdiel:2012uj,Banerjee:2012gh}.} In addition, due to the appearance of $\log$-divergences on both sides of AdS/CFT more numbers should agree.

\paragraph{CFT Side.} The free energy computed on $d$-sphere ${S}^d$ of radius $R$ is a well-defined number in odd $d$ provided the power divergences are regularized away and is $a_d\log R$ in even $d$, where $a$ is the Weyl anomaly coefficient, see e.g. \cite{Casini:2010kt} for conformal scalar.

The free energy on ${S}_\beta^1\times {S}^{d-1}$ with the radius of the circle playing the role of inverse temperature $\beta$ should have the form
\begin{align}
    F&= a_d \log l\Lambda +\beta E_c +F_\beta\,,
\end{align}
where $a_d$ is the anomaly and it vanishes for odd $d$ and also for $\Rac$ and $\Di$ on $\mathbb{R}\times {S}^{d-1}$ and ${S}^1\times {S}^{d-1}$.
The last term $F_\beta$ goes to zero when $\beta\rightarrow\infty$, i.e. for $\mathbb{R}\times {S}^{d-1}$, and can be easily computed in a free CFT:
\begin{align}
    F_\beta&=\mathrm{tr}\, \log[1\mp e^{-H \beta}]^{\mp1}=\mp\sum_m \frac{(\pm)^m}{m}Z_0(m\beta)\,.
\end{align}
Here $Z_0(\beta)$ is one-particle partition function
\begin{align}
    Z_0&= \mathrm{tr}\, e^{-\beta H}=\sum_n d_n e^{-\beta \omega_n}\,,
\end{align}
where $d_n$ and $\omega_n$ are degeneracies and eigen values of the free CFT Hamiltonian. The second term, which is proportional to $\beta$, is the Casimir Energy. It is given by a formally divergent sum
\begin{align}
    E_c&=(-)^F\frac12 \sum_n d_n \omega_n =(-)^F \frac12 \zeta_0(-1)\,,  & \zeta_0(z)&=\sum_n \frac{d_n}{\omega_n^z}\,,
\end{align}
which is usually regularized via $\zeta$-function. For free fields it vanishes for odd $d$. The Mellin transform maps $Z_0$ into $\zeta_0$. See Appendix \ref{app:MrCasimir} for many explicit values.

It is crucial to impose the singlet constraint on the CFT side. In a free CFT, e.g. $\Rac$, $F_\beta$ is constructed from the character $Z_0$ of $\Rac$. After the singlet constraint is imposed, one finds, see e.g. \cite{Giombi:2014yra}, that
$F_\beta$ is built from the character $Z$ of the singlet sector instead of the $\Rac$-character $Z_0$, i.e. from the character of $\Rac\otimes \Rac$ if the CFT is just $\Rac$.
Also, the Casimir Energy is $E^{\text{sing}}_c=NN_f \beta E_c$,
where $N_fN$ is the total number of free fields with the factor of $N$ removed by the singlet constraint.

\paragraph{AdS Side.} The one-loop free energy for a number of (massless) fields in $AdS_{d+1}$ is given by determinant of the bulk kinetic terms
\begin{align}
    (-)^FF^1_{AdS}&=\frac12\sum_s \mathrm{tr}\,\log|-\nabla^2+M^2_\Phi|-\frac12\sum_s\mathrm{tr}\,\log|-\nabla^2+M^2_\xi|\,,
\end{align}
where the sum is over all fields $\Phi_s$ with the ghost contribution\footnote{See \cite{Buchbinder:1995ez} for an earlier discussion of quantization of higher-spin fields in $AdS_4$.} subtracted by the second term if $\Phi_s$ is a gauge field. There is an additional minus for fermions. It can be computed by the standard zeta-function regularization \cite{Dowker:1975tf,Hawking:1976ja} of one-loop determinants and leads to
\begin{align}
    (-)^FF_{AdS}^1&=-\frac12 \zeta'(0)- \zeta(0)\log l\Lambda\,,
\end{align}
where $l$ is the $AdS$ radius, $\Lambda$ is a UV cutoff.

In Euclidean $AdS_{d+1}$ the $\zeta$-function is proportional to the regularized volume of $AdS_{d+1}$ space, which is a well-defined number for $AdS_{d=2n+2}$ and contains $\log R$ for $AdS_{d=2n+1}$. Another $\log$-term, which is $\log l\Lambda$ is present in $AdS_{d=2n+2}$ and is related to the conformal anomaly. The one-loop free energy on the thermal $AdS_{d+1}$ with boundary $S^1_\beta\times S^{d-1}$ is expected to be
\begin{align}
    F&=\beta[a_{d+1}\log l\Lambda +E_c]+F_\beta\,,
\end{align}
where $F_\beta$ vanishes in the high temperature $\beta\rightarrow0$ limit. In thermal $AdS_{2n+1}$ the $a_{d+1}$-anomaly is zero, while in $AdS_{2n+2}$ it should be the same as in Euclidean $AdS$ \cite{Giombi:2014yra}. Therefore, it can be computed from the free energy in Euclidean $AdS_{d+1}$ with boundary $S^d$, i.e. $\mathbb{H}^{d+1}$. In the latter case only the total anomaly coefficient can vanish, as was shown in \cite{Giombi:2013fka,Giombi:2014iua}. Therefore, once $a_{d+1}=0$ one can scrutinize the rest of the one-loop contribution, which is now well-defined.

The $N^0$ part of the free energy, $F_\beta$, counts the spectrum of states and should be automatically the same on both sides of the duality. Indeed, the spectrum of HS theories is determined by the representation theory of HS algebra. In its turn the HS algebras are constructed from free fields. The spectrum of single-trace operators is the same as the spectrum of HS fields and is given by the tensor product of appropriate (multiplets of) singletons/doubletons. Therefore, the $F_\beta$ part can be ignored on both sides for a moment: it can be attributed to generalized Flato-Fronsdal theorems, see e.g. \cite{Giombi:2014yra} for some checks. While the representation theory guarantees that the spectra should match, a direct path-integral proof is needed.

It is important that the leftover order-$N^0$ correction, i.e. Casimir Energy $E_c$, does not vanish sometimes (for the minimal theories or for the Type-C \cite{Beccaria:2014zma}), which requires to modify $G^{-1}=N$.

What needs to be checked depends heavily on whether $d$ is even or odd.
\paragraph{Tests in $\boldsymbol{AdS_{2n+2}/CFT^{2n+1}}$.} The CFT partition function on a sphere is a number, while $F^1_{AdS}$ in Euclidean $AdS$ contains $\log l\Lambda$-divergences for individual fields, which have to cancel for the right multiplet, otherwise the finite part is ill-defined. Then the finite part, $-\tfrac12 \zeta'(0)$, should be compared to $F_{CFT}^1$, which is zero in free CFT's. If $F^1_{AdS}$ is found to be non-zero, then one can try to adjust the relation between $N$ and bulk coupling $G$ as to make the two sides agree, assuming that $F^0_{AdS}=F^0_{CFT}$ and $F^1_{AdS}= \text{integer multiple of }F^0_{CFT}$, the latter requirement is due to the quantization of the bulk coupling. It was found \cite{Giombi:2013fka} that this is the case for the minimal models with even spins and $F^1_{AdS}$ is equal to $F^0_{CFT}$ for a free field that is behind the duality \cite{Klebanov:2011gs}.

Another test is for Casimir Energy $E_c$. It vanishes on the CFT side, while every field contributes a finite amount on the $AdS$-side. Therefore, only appropriately regularized sum over spins can vanish.

\paragraph{Tests in $\boldsymbol{AdS_{2n+1}/CFT^{2n}}$.} The regularized volume of $AdS$-space contains $\log R$, while the sphere free energy $F_{CFT}=a_d\log R$ is given by the $a$-coefficient of the Weyl anomaly. Here there is no $\log l\Lambda$-term since it vanishes for every field individually. Again, $F^1_{AdS}$ either vanishes or should be equal to an integer multiple of the $a$-anomaly of the dual free CFT, $F^0_{CFT}$, and can be compensated by modifying $G^{-1}=N$. The same computation then gives the anomaly for the conformal HS fields --- Fradkin-Tseytlin fields, $-2a_{HS}=a_{CHS}$, \cite{Fradkin:1985am,Tseytlin:2013jya,Giombi:2013yva,Giombi:2014iua}.

The Casimir Energy test is more non-trivial since it does not have to vanish on the CFT side either. $F^1_{AdS}$ corresponds to the order-$N^0$ corrections in CFT, which are absent for free CFT's.

It is also important to note that all the tests must be mutually consistent. In particular, if a modification of $G^{-1}=N$ is needed, it must be the same for all the tests in a given theory.

\section{One-Loop Tests}
\label{sec:tests}
In this section we perform the one-loop tests reviewed in Section \ref{sec:HSTheoriesatOneLoop}. The main emphasis is on the cases that have not yet been widely studied: even dimensions, spectral zeta-function for fermionic and mixed-symmetry HS fields. Less conventional cases of partially-massless fields and higher-spin doubletons are discussed in Appendix \ref{app:strangeHS}.

The spectrum of SUSY HS theories is made of bosonic and fermionic HS fields. In the simplest case one takes free CFT made of $n$ scalars and $m$ fermions, $S=n\Rac\oplus m\Di$. By imposing different singlet constraints the spectrum of bosonic HS fields can be truncated, for example, to even spins only, resulting in minimal theories. The spin of fermionic HS fields, if any, runs over all half-integer values $s=\tfrac{1}{2},\tfrac{3}{2},\tfrac{5}{2},...$. In the minimal theories the order $N^0$ one-loop corrections usually do not vanish and it is important for the consistency of SUSY HS theories that the modifications of $G^{-1}=N$ required
for consistency of Type-A and Type-B are the same, which was observed for $a$, $c$, $E_c$ in $AdS_{5,7}$  \cite{Beccaria:2014qea,Beccaria:2014xda} and for $E_c$ in all $AdS_{2n+1}$ \cite{Giombi:2014yra}.

\subsection{Casimir Energy Test}
\label{sec:CasimirTest}
The Casimir Energy tests are the simplest since the computation of $E_c$ is not difficult and we refer to Appendix \ref{app:MrCasimir} for technicalities. Each field contributes some finite amount to the Casimir Energy. It is important to use the same regularization that has been already applied for Type-A and Type-B models.

We will discuss HS fermions only, since pure Type-A and Type-B have been already checked. Vanishing of the Casimir energy can be seen after summation over spins with the exponential regulator $\exp[-\epsilon(s+(d-3)/2)]$. For example, in $AdS_6$ the summation of $E_c$ over all totally-symmetric HS fermionic fields leads to
\begin{align}
     -\sum_{m=0}^{}\textstyle \frac{(m+1) (m+2) \left(1344 m^6+12096 m^5+39760 m^4+57120 m^3+31388 m^2+420 m-2449\right)}{967680}e^{-\epsilon(m+(d-2)/2)}\Big|_{\text{fin.}}&=0\,,
\end{align}
where $|_{\text{fin.}}$ means to take the finite $\epsilon$-part of the sum evaluated with the exponential regulator. The same can be seen directly from the character of $\Di\otimes\Rac$ in any dimension:
\begin{align}
    \chi(\Di)\chi(\Rac)&=\cosh \left(\frac{\beta }{2}\right) \sinh ^{2-2 d}\left(\frac{\beta }{2}\right) 2^{\left[\frac{d}{2}\right]-2 d+3}\,,
\end{align}
which is manifestly even in $\beta$ and therefore the Casimir Energy vanishes. For completeness let us recall \cite{Giombi:2014yra} that for the same reason the Casimir energy vanishes for non-minimal Type-A,B and is equal to that of $\Rac$ and $\Di$ for the minimal ones. The Casimir Energy for the fermionic subsector is bounded to always vanish, which is what we observed.

The Casimir Energy tests for more complicated mixed-symmetry fields and partially-massless fields are also discussed in Appendix \ref{app:strangeHS}. Let us note that the computation of the Casimir energy for individual fields can be considerably simplified thanks to several observations.

First, it is sufficient to know the Casimir Energy of a single weight-$\Delta$ conformal scalar operator $O_\Delta$, the character being $q^\Delta(1-q)^{-d}$. Indeed, for generic $\Delta$ the number of physical degrees of freedom factorizes out in the character. For critical $\Delta$ that corresponds to appearance of singular sub-modules (equations of motion) the Casimir Energy can be obtained by following the exact sequence of $so(d,2)$-modules that determines the irreducible conformal representation.

Second, the Casimir Energy and its first derivative can be shown to vanish for $\Delta=d/2$ for $d$ even/odd:
\begin{align}
    E_c(\Delta=\tfrac{d}2)&=0\,, && d=2k\,,\\
    \frac{\pl}{\pl \Delta}E_c(\Delta=\tfrac{d}2)&=0\,, && d=2k+1\,.
\end{align}
Moreover, the second derivative of $E_c$ with respect to the conformal weight has a very simple form:
\begin{align}
    \frac{\pl^2}{\pl \Delta^2} E_c (\Delta)&=\frac{(-)^d \Gamma (\Delta )}{2 \Gamma (d) \Gamma (\Delta-d +1)}\,.
\end{align}

\subsection{Laplace Equation and Zeta Function}
The eigenvalue problem of Laplace operator is closely related to construction of zeta-function. We first discuss how to compute the eigenvalues and degeneracies for the Laplace operator on a sphere and then proceed to zeta-function on Euclidean $AdS_{d+1}$, i.e. on hyperbolic space $\mathbb{H}^{d+1}$, which can be obtained from that on a sphere up to few important details.

\subsubsection{Laplace Eigenvalue Problem}
We are interested in the spectrum of Laplacian on $S^N=SO(N+1)/SO(N)$:
\begin{align}
    (-\nabla^2+M^2) \Phi^{\mathbb{S}}_{n}=\lambda_n^{\mathbb{S}} \Phi^{\mathbb{S}}_n\,,
\end{align}
where $M^2$ is the mass-like term and $\Phi^{\mathbb{S}}$ is a transverse, traceless field with Lorentz spin $\mathbb{S}$, where $\mathbb{S}$ can be any representation which we label by a Young diagram, $\mathbb{S}=\mathbb{Y}(s_1,...,s_k)$. As is well-known, the eigenvalues $\lambda_n$ are given by the difference of two Casimir operators with a trivial shift by $M^2$:
\begin{align}
    -\lambda_n&=C^{so(N+1)}_2(\mathbb{S}_n)-C^{so(N)}_2(\mathbb{S})+M^2\,,\\
    d_n&=\mathrm{\dim}\, \mathbb{S}_n\,,
\end{align}
 Here the Young diagrams $\mathbb{S}_n$ of representations that contribute are obtained from $\mathbb{S}$ by adding a row of extra length $n$ as the first row:\footnote{In general, there are many more representations that contain $\mathbb{S}$ upon reduction to $so(N)$. The restriction to transverse and traceless fields reduces this freedom to one number, which is $n$. Transverse and traceless fields result from imposing gauges on the off-shell fields.}
\begin{align}
\mathbb{S}&=\parbox{100pt}{{\bep(80,40)\unitlength=0.38mm%
\put(0,0){\RectBRowUp{5}{4}{$...$}{$s_n$}}%
\put(0,20){\RectBRowUp{8}{7}{$s_1$}{$s_2$}}\eep}}
&
\mathbb{S}_n&=\parbox{85pt}{{\bep(80,50)\unitlength=0.38mm%
\put(0,0){\RectBRowUp{5}{4}{$...$}{$s_n$}}%
\put(0,20){\RectCRowUp{12}{8}{7}{$s_1+n$}{$s_1$}{$s_2$}}\eep}}
\end{align}
The degeneracy $d_n$ is just the dimension of $\mathbb{S}_n$. For example, for the scalar Laplacian with $M^2=0$ we have
\begin{align}
    \lambda_n&=n(N+n-1)\,, & d_n&=\mathrm{dim}^{so(N+1)}\, \mathbb{Y}(n)\,,
\end{align}
where $d_n$ is the number of components of the totally-symmetric
rank-$n$ tensor of $so(N+1)$. Analogously, for totally-symmetric rank-$s$ tensor fields we find
\begin{align}
    \lambda_n&=M^2+E(E-N+1)-s\,, \qquad \qquad E=N+s+n-1\,, \\
    d_n&=\mathrm{dim}^{so(N+1)}\, \mathbb{Y}(s+n,s)\,.
\end{align}

\subsubsection{Spectral Zeta-function}
Knowing eigen values $\lambda_n$ and degeneracy $d_n$ one can compute the spectral $\zeta$-function on $S^{d+1}$:
\begin{align}
    \zeta(z)&= \mathrm{vol}\, S^{d+1} \times \sum_n \frac{d_n}{(\lambda_n)^z}\,.
\end{align}
Extension to hyperbolic space $\mathbb{H}^{d+1}$ requires some work, see e.g. \cite{Camporesi:1992wn,Camporesi:1992tm,Camporesi:1993mz,CAMPORESI199457,Camporesi:1994ga,Camporesi:1995fb,Camporesi:1990wm,Lal:2012ax,Gupta:2012he,Gopakumar:2011qs}. The cases of $\mathbb{H}^{2n+1}$ and $\mathbb{H}^{2n}$ are very different.  Here $\zeta(z)$ is the spectral $\zeta$-function, which is the Mellin transform of the traced heat kernel at coincident points:
\begin{align}
    \zeta(z)&=\frac{1}{\Gamma[z]}\int_0^{\infty} dt \, t^{z-1}\, K(x,x;t)\,.
\end{align}
In homogeneous spaces the heat kernel at coincident points $K(x,x;t)$ does not depend on coordinates and the volume of the space factorizes out. The volume factor is a source of additional divergences.

The eigenvalues can be computed in a rather simple way for any irreducible representation of weight $\Delta$. The rule established on many examples, see e.g. \cite{Camporesi:1993mz,CAMPORESI199457} is to replace $s_1+n$, which is the length of the first row, by $i\lambda-\tfrac{d}2$ where $\lambda$ is non-negative and real:
\begin{align}
    -\lambda_n&=C^{d+2}(i\lambda-\tfrac{d}2,s_1,s_2,...)-C^{d+1}(s_1,s_2,...)+M^2=\frac{1}{4} (d-2 \Delta )^2+\lambda ^2+m^2\,,\\
    M^2&=m^2+\Delta(\Delta-d)-s_1-s_2-...\,,
\end{align}
where we took the standard normalization of the mass-like term, see e.g. \cite{Metsaev:1995re}: for $\Delta$ corresponding to gauge fields, both unitary \cite{Metsaev:1995re} and non-unitary \cite{Metsaev:1995re,Skvortsov:2009zu}, we have $m^2=0$.

The heat kernel contains only a contribution of the principal series in the odd dimensional case $\mathbb{H}^{2k+1}$. In the even dimensional case $\mathbb{H}^{2k}$ a discrete series can contribute \cite{CAMPORESI199457} too, depending on the type of representation. Effectively, the appearance of the discrete series contribution results in a shift by a constant ---  the formal degree of the discrete series. A contribution from discrete series arises for higher-spin doubletons --- fields in $AdS_{2k}$ that can be lifted \cite{Metsaev:1995jp} to representations of the conformal algebra $so(2k,2)$. For such fields the Young diagram $\mathbb{S}$ has $n$ rows of non-zero length. The case of $n$-forms was studied in \cite{CAMPORESI199457}. In what follows we will ignore the contribution of discrete series, but it would be interesting to understand if they play any role in HS AdS/CFT in $d>2$.

Zeta-function naturally has several different factors and the general expression is usually written in the following form:
\begin{align}
    \zeta&= \frac{\mathrm{vol}(\mathbb{H}^{d+1})}{\mathrm{vol}(S^d)} v_d g(s)\int_0^\infty d\lambda\, \frac{\mu(\lambda)}{\left[\frac{1}{4} (d-2 \Delta )^2+\lambda ^2\right]^z}\,,\label{zetageneral}
\end{align}
where $\mu(\lambda)$ is the spectral density that is normalized to its flat-space value:
\begin{align}
    \mu(\lambda)|_{\lambda\rightarrow\infty}&=w_d \lambda^{d}\,, &w_d&=\frac{\pi}{[2^{d-1}\Gamma(\tfrac{d+1}2)]^2}\,.
\end{align}
$g(s)$ is the number of components of the irreducible transverse traceless tensor that corresponds to the spin of the field. The volume factors are self-evident. There is an extra factor, which is a leftover:
\begin{align}
    v_d&= \frac{2^{d-1}}{\pi}\,, && u_d=v_dw_d=\frac{(\mathrm{vol}(S^d))^2}{(2 \pi )^{d+1}}\,.
\end{align}

\paragraph{Odd dimensions.} In the case of odd dimensions, $\mathbb{H}^{2k+1}$, $d=2k$, the $\zeta$-function is obtained by a simple replacement $s_1+n\rightarrow i \lambda-\tfrac{d}2$:
\begin{align}
    \boldsymbol{\mu}(\lambda)&=\frac{1}{\mathrm{vol} ({S}^{2k+1})}\mathrm{dim}^{so(d+1,1)}\,\left[i\lambda-\tfrac{d}2,\mathbb{S}\right]\,,
\end{align}
where the boldface $\boldsymbol{\mu}(\lambda)$ contains all the factors from \eqref{zetageneral} except for the ratio of volumes. We then extract $g(s)$, $v_d$ and $w_d$ factors. For example, for any even $d$ we find for totally-symmetric spin-$s$ bosonic fields, spin $s=m+\tfrac12$ fermionic fields and for bosonic fields with the shape of $\mathbb{Y}(s,1^p)$-hook:
\begin{align}
    \text{bosons}&: && \mu^B(\lambda)=w_d \left(\left(\frac{d-2}{2}+s\right)^2+\lambda ^2\right)\prod_{j=0}^{\frac{d-4}{2}} \left(j^2+\lambda ^2\right)\,,\\
    \text{fermions}&: && \mu^F(\lambda)=w_d \left(\left(\frac{d-1}{2}+m\right)^2+\lambda ^2\right)\prod _{j=0}^{\frac{d-4}{2}} \left(\left(j+\frac{1}{2}\right)^2+\lambda ^2\right)\,,\\
    \text{hooks}&: && \mu^H(\lambda)=w_d\frac{\left(\left(\frac{d-2}{2}+s\right)^2+\lambda ^2\right)}{\left(\lambda ^2+\left(\frac{d}2-p-1\right)^2\right)}\prod _{j=0}^{\frac{d-2}{2}} \left(j^2+\lambda ^2\right)\,,
\end{align}
where the spin factors are:
\begin{align}
    g^B(s)&=\frac{(d+2 s-2) \Gamma (d+s-2)}{\Gamma (d-1) \Gamma (s+1)}=\mathrm{dim}^{so(d)}\, \mathbb{Y}(s)\,,\\
    g^F(m)&=\frac{\Gamma (d+m-1) 2^{\left[\frac{d}{2}\right]}}{\Gamma (d-1) \Gamma (m+1)}=\mathrm{dim}^{so(d)}\, \mathbb{Y}_{\tfrac12}(m)\,, \\
    g^H(s,p)&=\frac{(d+2 s-2) \Gamma (d+s-1)}{(p+s) \Gamma (p+1) \Gamma (s) (d-p+s-2) \Gamma (d-p-1)}=\mathrm{dim}^{so(d)}\, \mathbb{Y}(s,1^p)\,.
\end{align}
The $s=1$ case of hooks corresponds to $(p+1)$-forms studied in \cite{CAMPORESI199457}; spin-$s$ bosons were investigated in \cite{Camporesi:1994ga}. The most general case in $AdS_5$ and $AdS_7$ was studied in \cite{Beccaria:2014xda,Beccaria:2014qea}.

\paragraph{Even dimensions.} In the case of even dimensions, $\mathbb{H}^{2k+2}$, $d=2k+1$, there are two complications: there can be additional discrete modes and the Plancherel measure is not a polynomial. In the cases we are interested in the discrete modes should not contribute and the spectral density is a product of a formally continued dimension $d_n$ and a hyperbolic function
\begin{align}
    \boldsymbol{\mu}(\lambda)&=\frac{i}{\mathrm{vol} ({S}^{2k+2})}\,\mathrm{dim}^{so(d+1,1)}\,\left[i\lambda-\tfrac{d}2,\mathbb{S}\right]h(\lambda)\,,\\
    h(\lambda)&=\begin{cases}
                    \tanh{\pi\lambda}\,, & \text{bosons}\,,\\
                    \coth{\pi \lambda}\,, & \text{fermions}\,.
                \end{cases}
\end{align}
For example, for any even $d$ we find for totally-symmetric spin-$s$ bosonic fields, spin $s=m+\tfrac12$ fermionic fields and for bosonic fields with the shape of $\mathbb{Y}(s,1^p)$-hook:
\begin{align}
    \text{bosons}&: && \mu^B(\lambda)=w_d\lambda  \tanh (\pi  \lambda ) \left(\left(\frac{d-2}{2}+s\right)^2+\lambda ^2\right)\prod_{j=1/2}^{\frac{d-4}{2}} \left(j^2+\lambda ^2\right)\,,\\
    \text{fermions}&: && \mu^F(\lambda)=w_d\lambda  \coth (\pi  \lambda ) \left(\left(\frac{d-1}{2}+m\right)^2+\lambda ^2\right)\prod _{j=1/2}^{\frac{d-4}{2}} \left(\left(j+\frac{1}{2}\right)^2+\lambda ^2\right)\,,\\
    \text{hooks}&: && \mu^H(\lambda)=w_d\lambda  \tanh (\pi  \lambda )\frac{\left(\left(\frac{d-2}{2}+s\right)^2+\lambda ^2\right)}{\left(\lambda ^2+\left(\frac{d}2-p-1\right)^2\right)}\prod _{j=1/2}^{\frac{d-2}{2}} \left(j^2+\lambda ^2\right)\,,
\end{align}
where the spin factors are the same. Degenerate hooks with $s=1$ again correspond to $(p+1)$-forms studied in \cite{CAMPORESI199457}. For symmetric bosonic fields we refer to \cite{Camporesi:1994ga}.

\paragraph{Mixed-Symmetry Fields.} As one more example of interest let us take a mixed-symmetry field of shape $\mathbb{Y}(s_1,s_2)$:
\begin{align}
   \mu^M(\lambda)&=w_d \left(\left(\frac{d-2}{2}+s_1\right)^2+\lambda ^2\right) \left(\left(\frac{d-4}{2}+s_2\right)^2+\lambda ^2\right)\times f_{E/O}\,,\\
   g^M(s_1,s_2)&=\mathrm{dim}^{so(d)}\, \mathbb{Y}(s_1,s_2)\,,\\
   f_O&=\prod _{j=0}^{\frac{d-6}{2}} \left(j^2+\lambda ^2\right)\,, \qquad\qquad \qquad \qquad \text{odd dimensions}\,,\\
   f_E&=\prod _{j=1/2}^{\frac{d-6}{2}} \left(j^2+\lambda ^2\right)\lambda  \tanh (\pi  \lambda )\,,\qquad\quad \text{even dimensions}\,.
\end{align}
The expression for the most general mixed-symmetry field with spin defined by $so(d)$ Young diagram $\mathbb{Y}(s_1,s_2,...,s_k)$ with $k$ rows follows the same pattern:
{\allowdisplaybreaks\begin{align}
   \mu^M(\lambda)&=w_d \prod_{i=1}^{i=k}\left(\left(\frac{d-2i}{2}+s_1\right)^2+\lambda ^2\right) \times f_{E/O}\,,\\
   g^M(s_1,s_2,...,s_k)&=\mathrm{dim}^{so(d)}\, \mathbb{Y}(s_1,s_2,...,s_k)\,,\\
   f_O&=\prod _{j=0}^{\frac{d-2k-2}{2}} \left(j^2+\lambda ^2\right)\,, \qquad\qquad \qquad \qquad \text{odd dimensions}\,,\\
   f_E&=\prod _{j=1/2}^{\frac{d-2k-2}{2}} \left(j^2+\lambda ^2\right)\lambda  \tanh (\pi  \lambda )\,,\qquad\quad \text{even dimensions}\,.
\end{align}}

For fermionic mixed-symmetry fields one has to correct $f_{E/O}$ factors only:
{\allowdisplaybreaks\begin{align}
   f_O&=\prod _{j=0}^{\frac{d-2k-2}{2}} \left((j+\tfrac12)^2+\lambda ^2\right)\,, \qquad\qquad \qquad \qquad \text{odd dimensions}\,,\\
   f_E&=\prod _{j=1/2}^{\frac{d-2k-2}{2}} \left((j+\tfrac12)^2+\lambda ^2\right)\lambda  \coth (\pi  \lambda )\,,\qquad\qquad \text{even dimensions}\,.
\end{align}}

Let us collect the relevant formulae with all factors now added to $\mu(\lambda)$, which we call $\tilde{\mu}(\lambda)$.
The complete spectral zeta-function is
\begin{align}
    \zeta(z)&=\int_{0}^{\infty}d\lambda\,\frac{\tilde{\mu}(\lambda)}{\left[\lambda^2+\left(\Delta-\tfrac{d}{2}\right)^2\right]^z}\,.
\end{align}
It is worth stressing that these are the zeta-functions for transverse, traceless tensors and the ghost contribution is not yet subtracted. Ghosts for massless fields always come with $\Delta+1,s-1$ as compared to $\Delta,s$ of the fields themselves.

\paragraph{Four Dimensions.} In four-dimensions there are no mixed-symmetry fields and bosons and fermions are described by almost the same formulae \cite{Camporesi:1993mz}
\begin{align}
    \text{bosons/fermions}&: &\tilde{\mu}(\lambda)&=\frac{\lambda  (2 s+1)  \left(\lambda ^2+\left(s+\frac{1}{2}\right)^2\right)}{6}\times \begin{cases}
                    \tanh{\pi\lambda}\,, & \text{bosons}\,,\\
                    \coth{\pi \lambda}\,, & \text{fermions}\,.
                \end{cases}
\end{align}

\paragraph{Five Dimensions.} The explicit formulae in five dimensions, i.e. $AdS_5$, are, see also \cite{Beccaria:2014xda}:
{\allowdisplaybreaks\begin{align*}
    \text{bosons}&: &\tilde{\mu}(\lambda)&=\log R \frac{\lambda ^2  (s+1)^2 \left(\lambda ^2+(s+1)^2\right)}{12 \pi }\,,\\
    \text{fermions}&: &\tilde{\mu}(\lambda)&=\log R\frac{\left(\lambda ^2+\frac{1}{4}\right)  (2 s+1) (2 s+3) \left(\lambda ^2+(s+1)^2\right)}{24 \pi }\,,\\
    \text{height-one hooks}&: &\tilde{\mu}(\lambda)&=\log R\frac{\left(\lambda ^2+1\right)  s (s+2) \left(\lambda ^2+(s+1)^2\right)}{6 \pi }\,,\\
    \text{two-row}&: &\tilde{\mu}(\lambda)&=\log R\frac{ \left(\lambda ^2+(s_1+1)^2\right) (s_1-s_2+1) (s_1+s_2+1) \left(\lambda ^2+s_2^2\right)}{6 \pi }\,.
\end{align*}}\noindent
\paragraph{Six Dimensions.} For application to HS theory based on  $F(4)$ we are also interested in six-dimensional anti-de Sitter space:{\footnotesize
{\allowdisplaybreaks\begin{align*}
    \text{bosons}&: &\tilde{\mu}(\lambda)&=-\frac{\lambda  \left(\lambda ^2+\frac{1}{4}\right) (s+1) (s+2) (2 s+3) \tanh (\pi  \lambda ) \left(\lambda ^2+\left(s+\frac{3}{2}\right)^2\right)}{720 }\,,\\
    \text{fermions}&: &\tilde{\mu}(\lambda)&=-\frac{\lambda  \left(\lambda ^2+1\right) \left(s+\frac{1}{2}\right) \left(s+\frac{3}{2}\right) \left(s+\frac{5}{2}\right) \coth (\pi  \lambda ) \left(\lambda ^2+\left(s+\frac{3}{2}\right)^2\right)}{180 }\,,\\
    \text{hooks}&: &\tilde{\mu}(\lambda)&=-\frac{\lambda  \left(\lambda ^2+\frac{9}{4}\right) s (s+3) (2 s+3) \tanh (\pi  \lambda ) \left(\lambda ^2+\left(s+\frac{3}{2}\right)^2\right)}{ 240}\,,\\
    \text{two-row}&: &\tilde{\mu}(\lambda)&=-\frac{\lambda  (2 s_1+3) (2 s_2+1) \tanh (\pi  \lambda ) (s_1-s_2+1) (s_1+s_2+2) \left(\lambda ^2+\left(s_1+\frac{3}{2}\right)^2\right) \left(\lambda ^2+\left(s_2+\frac{1}{2}\right)^2\right)}{720 }\,.
\end{align*}}}\noindent Note that for fermions we use spin $s$, rather than integer $m=s-\tfrac12$. The only hooks in $AdS_6$ are of shape $\mathbb{Y}(s,1)$. Also, the bosonic cases are all mutually consistent and follow from the two-row one.  Note that fermions cannot be obtained as $s\rightarrow s+1/2$ from bosons in this case, contrary to $d=3$.

\subsection{Zeta Function Tests: Odd Dimensions}
\label{sec:zetatestsodd}
Odd dimensions are easier since evaluation of $\zeta(0)$ and $\zeta'(0)$ is of no technical difficulty. In particular, $\zeta(0)=0$ for each field individually. The new results are on mixed-symmetry fields that belong to Type-B theories and fermionic HS fields, where all the tests are successfully passed. Also, we found a general formula for the $a$-anomaly. The zeta-function for the whole multiplet of some HS theory is denoted as $\zeta_{HS}$.

\subsubsection{Fermionic HS Fields}
\label{sec:fermionictestsO}

Firstly, $\zeta_s(0)=0$ for any $s$ and therefore the bulk result is well-defined. It is proportional to $\log R$ due to the regularized volume of $AdS_{2k+1}$. On the boundary it should be equal to the Weyl anomaly coefficient, $a\log R$, but this has been already accounted for by the contribution of bosonic HS fields. Therefore, we should check that $\zeta'_{HS}(0)=0$. To give few examples, in $AdS_5$, see also \cite{Beccaria:2014xda}, we find that
\begin{align*}
    \frac{\zeta'_s(0)}{\log R}&=\frac{ (2 s+1)^2 (2 s (s+1) (28 s (s+1)-31)-7)}{1440}\,, && s>\frac12\,,\\
        \frac{\zeta'_s(0)}{\log R}&=-\frac{11}{180}\,,  && s=\frac12\,.
\end{align*}
Using the same exponential cut-off $\exp[-\epsilon(s+\tfrac{d-3}2)]$ we find the total $a$-coefficient to vanish
\begin{align}
    \zeta_{HS}'(0)&=\sum_{s=\tfrac32,\tfrac52,...} \zeta'_s(0)+\zeta'_{\tfrac12}(0)=0\,.
\end{align}
In $AdS_7$ we have a more complicated formulae, but fortunately with the same result that $\zeta_{HS}'(0)=0$, see also \cite{Beccaria:2014qea}:
\begin{align*}
    \frac{\zeta'_s(0)}{\log R}&=\frac{ (2 s+1) (2 s+3)^2 (2 s+5) (2 s (s+3) (16 s (s+3) (11 s (s+3)-1)-981)-695)}{9676800}\,, && s>\frac12\,,\\
        \frac{\zeta'_s(0)}{\log R}&=-\frac{13}{280}\,,   && s=\frac12\,.
\end{align*}
In general dimension the computation can be simplified by introducing $P_d(\lambda)=P_d(-\lambda)$:
\begin{align}
    P_d(\lambda)&=\sum_k \alpha_k \lambda^k =\prod _{j=0}^{\frac{d-4}{2}} \left(\left(j+\frac{1}{2}\right)^2+\lambda ^2\right)\,.
\end{align}
Then, with the help of the simple integration formula
\begin{align}
    a(z)=\int_0^\infty d\lambda\, \frac{\lambda ^k}{\left(b^2+\lambda ^2\right)^z}&= \frac{\Gamma \left(\frac{k+1}{2}\right) b^{k-2 z+1} \Gamma \left(-\frac{k}{2}+z-\frac{1}{2}\right)}{2 \Gamma (z)}\,,
\end{align}
where $b=\Delta-d/2$, one finds that $\zeta(0)=0$ and $\zeta'(0)$ can be obtained from (only even $k$ matters)
\begin{align}
    \pl_z a(z)\Big|_{z=0}&=\frac{-i^k(\Delta-\tfrac{d}2)^{k+1}}{4(k+1)}\,.
\end{align}
Then, it can be effortlessly checked up to any given dimension that the total $\zeta'_{HS}(0)$ vanishes identically. In fact, it also vanishes when restricted to 'even half-integer' spins  $s=\tfrac12+2n$.

\subsubsection{Symmetric HS Fields}
\label{sec:bosonictestsO}
The case of Type-A was studied in \cite{Giombi:2014iua,Giombi:2013fka,Beccaria:2014qea,Beccaria:2014zma,Beccaria:2014xda}. Let us quote the results. As always in odd dimensions $\zeta_s(0)=0$, while  $\zeta'_s(0)$ can be computed the same way as we did for fermions. The final output is
\begin{align}
    \zeta'_{HS, \text{non-min.}}(0)&=0\,,\\
    \zeta'_{HS, \text{min.}}(0)&= -2 a_\phi\log R\,,
\end{align}
where $a^d_\phi$ is the Weyl-anomaly coefficient of the free scalar field in $CFT^d$, for which one finds, see e.g. \cite{Casini:2010kt},
\begin{align}
    a_\phi^{4}&=\frac{1}{90}\,, &
    a_\phi^{6}&=-\frac{1}{756}\,, &
    a_\phi^{8}&=\frac{23}{113400}\,, &
    a_\phi^{10}&=-\frac{263}{7484400}\,. &
\end{align}

\subsubsection{Mixed-Symmetry HS Fields}
\label{sec:mixedsymmetrytestsO}
We will discuss various versions of the Type-B theory that contains mixed-symmetry fields with Young diagrams of hook shape \eqref{typebcurrent}.
The contribution of certain mixed-symmetry fields has been already studied in lower-dimensional cases of $AdS_{5,7}$ in \cite{Beccaria:2014xda,Beccaria:2014qea,Beccaria:2014zma}. With the help of the general formula for the zeta-function we can extend these results for the Type-B theory to any dimension. Here we should find that $F^1_{AdS}$ is either zero or is a multiple of the free fermion Weyl anomaly $a^d_{\psi}$, see e.g. \cite{Aros:2011iz}:
\begin{align}
       a^4_\psi&=\frac{11}{180}\,, &a^6_\psi&=-\frac{191}{7560}\,, & a^8_\psi&=\frac{2497}{226800}\,,&
       a^{10}_\psi&=-\frac{14797}{2993760}\,.
\end{align}
First of all, the spectrum of the non-minimal theory is given by the tensor product of Dirac free fermion $\Di$ that decomposes into a direct sum $\Wi\oplus\Wib$ of two Weyl fermions. With the help of Appendix \ref{app:dimschars} one finds for $AdS_{2k+1}$:
\begin{align}
    \Di\otimes\Di&=\bigoplus_{n}\Yy{n,1^{k-1}}_+\oplus\bigoplus_{n}\Yy{n,1^{k-1}}_-\oplus2\bigoplus_{n=1,i=1} \Yy{n,1^{k-i-1}}\oplus 2\bullet\,,
\end{align}
where we indicate the spin of the fields only as the conformal weight/$AdS$ energy is obvious.

For example, in seven dimensions the contribution of the scalar field and the total contributions of hooks of height $p=0,1,2$ are:\footnote{The zeta-function for hooks with $p+1>d/2$ is the same as for the dual fields with $p+1<d/2$.}
\begin{align}
    \zeta'_0(0)&=\frac{8}{945}\,, & \zeta'_p(0)&=\left\{\frac{1}{756},-\frac{8}{945},-\frac{1}{378}\right\}\label{Adsseven}\,,
\end{align}
while in nine dimensions the contribution of the scalar field and the total contributions of hooks of height $p=0,1,2,3$ are:
\begin{align}
    \zeta'_0(0)&=\frac{9}{1400}\,, & \zeta'_p(0)&= \left\{\frac{13}{14175},-\frac{353}{56700},-\frac{13}{14175},-\frac{23}{56700}\right\}\,,\label{Adsnine}
\end{align}
the total sum being zero, as is expected.

As for the minimal theories, there are several surprises. First of all, one can take just $U(N)$-singlet sector of $\Wi$. With the help of Appendix \ref{app:dimschars} the spectrum is
\begin{align}
    so(d=4k)&: &&
    \left\{\begin{aligned}
    (\Wi\otimes \Wi)=&\bigoplus_n\Yy{n,1^{2k-1}}_+\oplus\bigoplus_{n,i} \Yy{n,1^{2k-4i-1}}\oplus\\
    &\bigoplus_{n,i} \Yy{n,1^{2k-4i-3}}\oplus \bullet
    \end{aligned}\right.  \\
    so(d=4k+2)&: &&
    \left\{\begin{aligned}
    (\Wi\otimes \Wi)=&\bigoplus_n\Yy{n,1^{2k}}_+\oplus\bigoplus_{n,i} \Yy{n,1^{2k-4i}}\oplus\\
    &\bigoplus_{n,i} \Yy{n,1^{2k-4i-2}}
    \end{aligned} \right.
\end{align}
We see that for $d=4k$, i.e. $AdS_{4k+1}$, the spectrum does not contain symmetric higher-spin fields at all. In particular, there is no graviton. Nevertheless, the total $\zeta'_{HS}(0)$ can be found to vanish. For example, consider $AdS_9$, for which the results on the row-by-row basis were quoted in \eqref{Adsnine}. The spectrum of $U(N)$ Weyl fermion $\Wi$ is
\begin{align}
    \Wi\otimes \Wi&=\bullet \oplus \bigoplus_{n} \Yy{n,1}\oplus \Yy{n,1,1,1}_+\,,
\end{align}
and we see that $9/1400-(353/56700)-(23/113400)=0$. The same is of course true for the $\Wi\otimes \Wib$ sub-sector: $13/14175-(13/14175)=0$. The latter sector contains symmetric HS fields, including the graviton:
\begin{align}
    \Wi\otimes \Wib&=\bigoplus_{n} \Yy{n}\oplus \Yy{n,1,1}\,.
\end{align}
For $d=4k+2$, i.e. $AdS_{4k+3}$, the $U(N)$ Weyl fermion does include totally-symmetric HS fields, so the theory looks healthy. The spectrum of the two parts is
\begin{align}
    \Wi\otimes \Wi&= \bigoplus_{n} \Yy{n}\oplus \Yy{n,1,1}_+\,,\\
    \Wi\otimes \Wib&=\bullet\oplus\bigoplus_{n} \Yy{n,1}\,.
\end{align}
Again, the two sub-sectors result in $\zeta'_{HS}(0)=0$ independently: $1/756-(1/756)=0$ and $8/945-(8/945)=0$.

As for the minimal Type-B theory there are several options. Firstly, one can take the anti-symmetric part of $\Di\otimes\Di$, which would be the minimal Type-B. Secondly, one can take the anti-symmetric part of only $\Wi\otimes\Wi$, which would be the minimalistic option. The spectrum of the minimalistic Type-B theory is even more peculiar. We refer to Appendix \ref{app:dimschars} for more detail, while giving two examples here-below. In $AdS_7$ we find, see also \cite{Beccaria:2014qea},
\begin{align}
    (\Wi\otimes \Wi)_{O(N)}&= \bigoplus_{n} \Yy{2n+1}\oplus \Yy{2n,1,1}_+\,.
\end{align}
The total $\zeta'_{HS}(0)$ is $-(1/378)+211/7560=191/7560$, which is in accordance with the $a$-anomaly of one Weyl fermion on ${S}^6$, see also Appendix \ref{app:MrCasimir}. In $AdS_9$ the spectrum of the minimalistic Type-B is
\begin{align}
    (\Wi\otimes \Wi)_{O(N)}&=\bigoplus_{n} \Yy{2n+1,1}\oplus \Yy{2n,1,1,1}_+\,,
\end{align}
and the contribution to $\zeta'_{HS}(0)$ is $23/5400-(3463/226800)=2497/226800$, which is again in accordance with the $a$-anomaly of the free fermion. The contribution of the symmetric part of the tensor product
\begin{align}
    (\Wi\otimes \Wi)_{S}&=\bullet\oplus\bigoplus_{n} \Yy{2n,1}\oplus \Yy{2n+1,1,1,1}_+\,,
\end{align}
which would be relevant for the $usp(N)$-singlet theory comes with the opposite sign, $-2497/226800$. The latter is obvious, of course, without any computation since the total anomaly was found to vanish.

The same pattern can be observed in other dimensions. According to the quite general law \cite{Gubser:2002vv, Diaz:2007an,Tseytlin:2013jya}, the $a$-anomaly of conformal HS fields on the boundary can be computed from the $AdS$ side according to $a_{CHS}=-2a_{HS}$, which is related to more general results on the ratio of determinants \cite{Barvinsky:2005ms}. Therefore, vanishing of total $a_{HS}$ for the mixed-symmetry fields of Type-B implies the one-loop consistency of the conformal higher-spin theory with spectrum of conformal HS fields given by the sources to the single-trace operators built out of free fermion. As in the case of Type-A conformal HS theory \cite{Fradkin:1985am,Segal:2002gd}, the action is given by the $\log \Lambda$-part of the generating function of correlators of mixed-symmetry currents $J_{s,p}$, \eqref{typebcurrent}:
\begin{align}
     S_{CHS}[\varphi_{s,p}]&= \log \Lambda\text{-part of} \log \int D\bar{\psi} D\psi\, e^{\int \bar{\psi}\slashed{\pl} \psi + \sum_{s,p} J_{s,p}\varphi_{s,p}}\,,
\end{align}
where $\varphi_{s,p}$ are the sources for $J_{s,p}$.

\subsubsection{Simplifying a-anomaly}
\label{sec:aanomaly}
The examples above reveal that $\zeta'(0)$, which is related to the boundary $a$-anomaly, $-2a \log R=\zeta'(0)$, is a quite complicated expression. However, it comes from a very simple formula. Following earlier results \cite{Giombi:2013yva,Giombi:2014iua,Beccaria:2014qea,Beccaria:2014xda}, consider the formula
\begin{align}
    a'(\Delta)=\frac{1}{\log R}\frac{1}{2\Delta-d}\frac{\pl} {\pl \Delta} \zeta'_\Delta(0)\,,
\end{align}
for any $\Delta$ and any irreducible representation $\mathbb{S}$ defined by some Young diagram $\Yy{s_1,...,s_n}$ with $n$ rows. Then we find that
\begin{align}
   a'(\Delta)&=(-)^{n+1}\mathrm{dim}\, \Yy{s_1,...,s_n}\frac{\Gamma[\Delta-n]\prod_{i=1}^{n}(\Delta+s_i-i)(d+s_i-\Delta-i)}{\Gamma[\Delta-d+n+1]\Gamma[d+1]}\,.
\end{align}
$a$ does not have a nice factorized form, but it is always proportional to $\Delta-d/2$, i.e. it vanishes at $\Delta=d/2$, which is a boundary condition for the integral that allows to reconstruct $a$ from $a'$:
\begin{align}
    a(\Delta)=\frac{1}{\log R}\zeta'_\Delta(0)&=\int_{d/2}^{\Delta}dx\, (2x-d) a'(x)\,.
\end{align}

\subsection{Zeta Function Tests: Even Dimensions}
\label{sec:zetatests}
Even dimensional $AdS_{2n+2}$ spaces are much harder due to the complexity of spectral density that is not a simple polynomial, but contains the functions $\tanh$ or $\coth$. Moreover, $\zeta(0)$ is generally non-zero for each field (which is due to the conformal anomaly for the case of conformally-invariant fields). Below we present the main results with the technicalities devoted to Appendices. The most interesting case is that of mixed-symmetry fields from the Type-B theory.

\subsubsection{Fermionic HS Fields}
\label{sec:fermionictestsE}
Let us start with few examples. Computation of $\zeta(0)$ is not too difficult thanks to a handful of papers \cite{Camporesi:1993mz,Camporesi:1991nw,Giombi:2013fka}. For example, in $AdS_4$ and $AdS_6$ the sum over all fermions is zero
\begin{align*}
    \sum_{m=0}&\frac{-1200 m^4-2400 m^3-1560 m^2-360 m-47}{2880}=0\,,\\
    -\sum_{m=0}&\frac{(m+1) (m+2) \left(2016 m^6+18144 m^5+60704 m^4+92064 m^3+56462 m^2+42 m-9061\right)}{483840}=0\,.
\end{align*}
The same can be checked for any dimension, see Appendices for the details. As different from odd dimensions, the sum over all 'even half-integer' spins does not vanish.

The computation of $\zeta'(0)$ is trickier, see Appendices, but it can be shown on a dimension by dimension basis that for $AdS_{4,6,8,...}$ one finds $\zeta_{\text{fermions}}'(0)=0$. Therefore, adding fermionic HS fields is consistent to a given order, which is a necessary condition for the existence of SUSY HS theories.

\subsubsection{Symmetric HS Fields}
\label{sec:bosonictestsE}
The case of symmetric HS fields was already studied in \cite{Giombi:2013fka,Giombi:2014iua}. The summary is that $\zeta_{HS}(0)=0$ both for minimal and non-minimal Type-A theories while $\zeta'(0)$ does not vanish for the minimal Type-A and is equal to the sphere free energy of one free scalar:
\begin{align}
    \zeta_{HS, non-min}(0)&=0\,, & \zeta_{HS, min.}(0)&=0\,,\\
    -\tfrac12\zeta'_{HS, non-min}(0)&=0\,, & -\tfrac12\zeta'_{HS, min.}(0)&=F^{\phi}_{d}\,.
\end{align}
As before, the minimal Type-A requires $G^{-1}=N-1$.

\subsubsection{Mixed-Symmetry HS Fields}
\label{sec:mixedsymmetrytestsE}
This is the most interesting case. The Type-B theory in $AdS_4$ does not differ much from the Type-A --- the spectrum consists of totally-symmetric HS fields. This is not the case in $d>3$ where the spectrum of Type-B contains mixed-symmetry fields with Young diagrams of hook shape \eqref{typebcurrent} in accordance with the singlet spectrum of free fermion $\Di$. Much less is known about these theories\footnote{Some cubic interaction vertices for mixed-symmetry fields in $AdS$ were constructed in \cite{Boulanger:2012dx,Boulanger:2011se,Boulanger:2011qt}. A part of the Type-B cubic action that contains $0-0-s$ vertices was found in \cite{Skvortsov:2015pea}.} except that they should exist in any dimension since $\Di$ and $\Rac$ do.

\paragraph{Zeta.} First of all we check that $\zeta(0)=0$ and thus the bulk contribution is well-defined. It is convenient to present a contribution of the $\bar{\psi}\psi$ operator and of the hooks for each height $p$ separately. Here $p$ can run over $0,...,d-2$ with $p=0$ corresponding to totally-symmetric HS fields. However, one can (and should) take into account only half of the hooks since the rest can be dualized back to $p+1\leq d/2$ and the zeta function is the same. The latter is in accordance with the generalized Flato-Fronsdal theorem, which we now write for $AdS_{2k+2}$:
\begin{align}
    \Di\otimes\Di &=\bullet \oplus \bigoplus_{n,i}\Yy{n,1^{k-i-1}}\,,
\end{align}
where there is one scalar and half of the hooks. For example, in $AdS_6$ we find
\begin{align}
    \zeta_{\bar{\psi}\psi}(0)&=-\frac{37}{7560}\,, & \zeta_p(0)=\left\{-\frac{1}{1512},\frac{1}{180}\right\}\,, &&\sum\zeta_p(0) =\frac{37}{7560}\,.
\end{align}
Here one can see the contribution of the Type-A fields with $s\geq1$, which is $-1/1512$. In Type-A this is canceled by the $\Delta=3$ scalar. Now, the contribution of $\bar{\psi}\psi$ is different, but there is the $p=1$ sector and $\zeta_{HS}(0)=0$. In $AdS_8$ we find
\begin{align}
    \zeta_{\bar{\psi}\psi}(0)&=-\frac{119}{32400}\,, & \zeta_p(0)=\left\{-\frac{127}{226800},\frac{1}{280},\frac{1}{1512}\right\}\,, &&\sum\zeta_p(0) =\frac{119}{32400}\,.
\end{align}
It can be checked for higher dimensions that the total $\zeta_{HS}(0)=0$. Now let us have a look at the minimal theories. The $O(N)$-singlet version of the Flato-Fronsdal theorem tells that
\begin{align}
    (\Di\otimes\Di)_{O(N)} =\bullet \oplus &\bigoplus_{n,i}\Yy{2n,1^{k-4i-1}}\oplus \Yy{2n,1^{k-4i-4}}\\
    &\bigoplus_{n,i}\Yy{2n+1,1^{k-4i-2}}\oplus \Yy{2n+1,1^{k-4i-3}}\,,
\end{align}
where the scalar is present whenever $(k-1)\mod4=0$ or $(k-2)\mod4=0$. Analogously to odd dimensions, simply taking anti-symmetric part of $\Di\otimes\Di$ can result in somewhat strange spectra, which may not contain graviton. Nevertheless, such spectra yield vanishing contribution to $\zeta_{HS}(0)$. For example, in $AdS_6$ we find
\begin{align}
    (\Di\otimes\Di)_{O(N)} =\bullet \oplus &\bigoplus_{n,i}\Yy{2n,1}\oplus\bigoplus_{n,i}\Yy{2n+1}\,, \label{eventypebmin}
\end{align}
and the contribution of all odd spin fields is zero, while hooks of even spins give exactly $\frac{37}{7560}$ to cancel that of the scalar. Similar pattern is true in higher dimensions and both minimal and non-minimal Type-B have $\zeta_{HS}(0)=0$.

\paragraph{Zeta Prime.}
The main computational problem is to find $\zeta'_{HS}(0)$. Below we give the summary of our results in several dimensions, with technicalities devoted to the Appendices. Let us note that despite some analytic regularization, which is needed to make sums over spins well-defined, there are non-trivial self-consistency checks for the computations: certain integrals cannot be evaluated but they cancel each other, also all complicated factors disappear from the final result. For non-minimal theories the total contribution to $-\tfrac12 \zeta'_{HS}(0)$ is:\footnote{We list here only those results that fit one line. See also a closely related paper \cite{Giombi:2016pvg}.}
\begin{align}
    AdS_4&: &&
    -\frac12 \zeta'_{HS}(0)=-\frac{\zeta (3)}{8 \pi ^2}\,,\\
    AdS_6&: &&
    -\frac12 \zeta'_{HS}(0)= -\frac{\zeta (3)}{96 \pi ^2}-\frac{\zeta (5)}{32 \pi ^4}\,, \\
    AdS_8&: &&
    -\frac12 \zeta'_{HS}(0)= -\frac{\zeta (3)}{720 \pi ^2}-\frac{\zeta (5)}{192 \pi ^4}-\frac{\zeta (7)}{128 \pi ^6}\,,\\
    AdS_{10}&: &&
    -\frac12 \zeta'_{HS}(0)=-\frac{\zeta (3)}{4480 \pi ^2}-\frac{7 \zeta (5)}{7680 \pi ^4}-\frac{\zeta (7)}{512 \pi ^6}-\frac{\zeta (9)}{512 \pi ^8}\,,     \\
    AdS_{12}&: &&
    -\frac12 \zeta'_{HS}(0)=-\frac{\zeta (3)}{25200 \pi ^2}-\frac{41 \zeta (5)}{241920 \pi ^4}-\frac{13 \zeta (7)}{30720 \pi ^6}-\frac{\zeta (9)}{1536 \pi ^8}-\frac{\zeta (11)}{2048 \pi ^{10}}\,.
\end{align}
The case of $AdS_4$ was studied in \cite{Giombi:2013fka}. The discrepancy with the sphere free energy of free fermion, $F^d_\psi$, is systematic, see Appendix \ref{app:MrCasimir} for some explicit values. However, these numbers are not random. They can be reproduced as a difference in the free energy via RG-flow induced by a double-trace operator $O^2_\Delta$. If the operator $O_\Delta$ is bosonic the general formula for $\delta \tilde F^\phi_\Delta=\tilde F_{IR}-\tilde F_{UV}$ can be found in \cite{Klebanov:2011gs}:\footnote{Here we pass to generalized sphere free energy $\tilde F$ that is defined as $-\sin(\tfrac{\pi d}{2})F$, see e.g. \cite{Giombi:2014xxa}. }
\begin{align}
   \delta \tilde F^\phi_{\Delta}&=\frac{1}{\Gamma (d+1)}\int_0^{\Delta-d/2} u \sin (\pi  u) \Gamma \left(\frac{d}{2}+u\right) \Gamma \left(\frac{d}{2}-u\right) \, du\,.
\end{align}
The values of the free scalar $F$-energy can also be computed as $F$-difference:
\begin{align}
    \tilde F^\phi_d&= -\delta \tilde F^\phi_{\Delta=\tfrac{d-2}2}= \delta F^\phi_{\Delta=\tfrac{d+2}2}\,.
\end{align}
The numbers that resulted from the tedious computations in $AdS_{2n+2}$ arrange themselves into the following sequence:
\begin{align}
    -\frac12 \zeta'_{HS}(0)&=\delta \tilde F^\phi_{\Delta=\tfrac{d-1}2}= -\delta \tilde F^\phi_{\Delta=\tfrac{d+1}2}\,. \label{FenergyA}
\end{align}
However, the dual of Type-B is supposed to be a fermionic theory, for which a generalization of \cite{Klebanov:2011gs} to fermionic $O_\Delta$ in any $d$ gives \cite{Giombi:2014xxa}:
\begin{equation}
     \delta \tilde F^\psi_{\Delta}= \frac{ 2}{\Gamma(d+1)}\int_0^{\Delta-d/2} \cos(\pi u) \Gamma\left(\frac{d+1}{2}+u\right)\Gamma\left(\frac{d+1}{2}-u\right)du\,.
\end{equation}
Again the free fermion $F$-energy can be computed as $F$-difference:
\begin{equation}
   \tilde F^\psi_d= \delta \tilde F^\psi_{\Delta=\frac{d+1}{2}}= -\delta \tilde F^\psi_{\Delta=\frac{d-1}{2}}\,.
\end{equation}
We observe that for $\Delta=\frac{d-2}{2}$ it will give $-\tfrac12 \zeta'_{HS}(0)$ up to a factor of $\pm1/4$:
\begin{equation}
   -\frac12 \zeta'_{HS}(0)=-\frac{1}{4} \delta \tilde F^\psi_{\Delta=\frac{d-2}{2}}=\frac{1}{4} \delta \tilde F^\psi_{\Delta=\frac{d+2}{2}}\,.\label{FenergyB}
\end{equation}
For the minimal theories the computations are even more involved, but the unwanted constants do cancel and we find\footnote{A word of warning is that the spectrum of the minimal Type-B is defined in \eqref{eventypebmin}. Other projections, e.g. the $usp$-constraint or various Majorana-Weyl projections, would result in a slightly different spectra, all of which yield similar numbers, i.e. the unwanted constants go away. } for the total contribution to $-\tfrac12 \zeta'_{HS}(0)$:
\begin{align*}
    AdS_4&: &&
    -\frac12 \zeta'_{HS}(0)=\frac{\log (2)}{8}-\frac{5 \zeta (3)}{16 \pi ^2}\,,\\
    AdS_6&: &&
    -\frac12 \zeta'_{HS}(0)=\frac{45 \zeta (5)}{128 \pi ^4}-\frac{3 \zeta (3)}{64 \pi ^2}-\frac{3 \log(2)}{64}\,,  \\
    AdS_8&: &&
    -\frac12 \zeta'_{HS}(0)=\frac{649 \zeta (3)}{23040 \pi ^2}-\frac{23 \zeta (5)}{1536 \pi ^4}-\frac{449 \zeta (7)}{1024 \pi ^6}+\frac{5 \log (2)}{256}\,,\\
    AdS_{10}&: &&
    -\frac12 \zeta'_{HS}(0)=\frac{315 \zeta (7)}{4096 \pi ^6}+\frac{3825 \zeta (9)}{8192 \pi ^8}-\frac{617 \zeta (3)}{43008 \pi ^2}-\frac{85 \zeta (5)}{4096 \pi ^4}-\frac{35 \log (2)}{4096}\,,\\
    AdS_{12}&: &&
    -\frac12 \zeta'_{HS}(0)=\frac{29 \zeta (7)}{49152 \pi ^6}+\frac{13579 \zeta (9)}{49152 \pi ^8}+\frac{31745 \zeta (11)}{32768 \pi ^{10}}-\frac{68843 \zeta (3)}{5160960 \pi ^2}-\frac{31033 \zeta (5)}{1105920 \pi ^4}-\frac{63 \log (2)}{8192}\,.
\end{align*}
Again, these numbers do not look random. Curiously enough the $AdS_6$ result equals $6F^{\phi}$.

\subsection{Tadpole}
\label{sec:tadpole}
In principle, higher-spin theories should be consistent as quantum theories to all loops as they are duals of well-defined CFT's that are, in general, either free fields or interacting vector-models. It is hard to say anything about higher loops or Feynman-Witten diagrams with legs due to the lack of the complete action. Also, any analog of the non-renormalization theorems for HS theories is not known at present. Moreover, it seems that vectorial super-symmetry cannot help too much and one should better stick to HS extensions of the usual SUSY. Still everything should boil down to the consistency of a simple bosonic HS theory, i.e. HS SUSY should improve the quantum properties, but the need for nontrivial summation over all spins appears unavoidable.

We can see that at least a part of the tadpole diagram vanishes for the reasons similar to the tests performed above. In \cite{Bekaert:2015tva} the quartic scalar vertex $0-0-0-0$ was reconstructed from the free scalar CFT at $d=3$. Though, the base of structures used there is over-determined and the coefficients are not known in explicit form it seems that the following should be true in any $d$. The quartic vertex is a double sum
\begin{align}
    V_4&= \sum a_{n_1,n_2} \square^{n_2}(\Phi\nabla^{n_1} \Phi) (\Phi\nabla^{n_1}\Phi)\,,
\end{align}
where we just meant to indicate that it is a doubly-infinite sum over all independent structures allowed by kinematics. The order of derivatives is unbounded, but the growth of the coefficients is suppressed by locality. The sum is doubly-infinite due to the four-point function it contributes to being the function of two conformally-invariant cross-ratios.

Let us consider the tadpole Feynman graph, $\bep(14,10)\put(0,0){\line(1,0){14}}\put(7,5){\circle{10}}\eep$. There is an infinite factor of various derivatives of the Green function at coincident points
\begin{align}
    \sum b_{n,m}(\Phi\nabla^{n} \Phi) (\nabla^{m} G(x,x))\,.
\end{align}
If we are in the simple $\Phi^4$ theory then the tadpole $\Phi^2 G(0)$ contributes to the mass of the field. Now, due to the fact that higher derivatives are present in $V_4$ we can have a contribution of the kinetic term $\Phi\nabla^2\Phi$, which would imply wave-function renormalization. Also, there are infinitely many of unwanted terms $\Phi\nabla^n\Phi$, $n>2$ with more than two derivatives, which are absent in the action.

The part of the tadpole that does not have derivatives on $G$, but can have arbitrarily many derivatives on $\Phi$'s, can be related to heat kernel $G(x,x)=\int K(x,x;t)$. Indeed,
\begin{align}
    \frac{\pl}{\pl M^2} \log \det [\square+M^2]=\frac{\pl}{\pl M^2}\mathrm{tr} \log [\square+M^2]&=\mathrm{tr}\, \frac{1}{[\square+M^2]}=\mathrm{vol}(AdS_{d+1}) G(x,x)\,.
\end{align}
Using the general relation between $M^2$ and conformal weight $\Delta$ we find
\begin{align}
    \mathrm{vol}(AdS_{d+1}) G_\Delta (x,x)&=\frac{1}{2\Delta-d}\frac{\pl}{\pl \Delta} \left[2\zeta(0) \log \Lambda l+ \zeta'(0)\right]\,.
\end{align}
$\zeta(0)$ was shown to vanish quite generally for $d$ even and any $\Delta$. Then, for $\zeta'(0)$ and for totally-symmetric fields we get, see Section \ref{sec:aanomaly},
\begin{align}
    \frac{1}{2\Delta-d}\frac{\pl}{\pl \Delta}\zeta'_s(0)&=g^B(s)\log R \frac{\Gamma (\Delta -1) (\Delta +s-1) (d-\Delta +s-1)}{\Gamma (d+1) \Gamma (-d+\Delta +2)}\,.
\end{align}
Formulae of this type have just been shown to facilitate the computation of $a$-anomaly as an integral of $\pl_\Delta \zeta'(0)$ over $\Delta$.

Assuming that all terms enter with the same coefficient and with the standard regularization we find that $G_{HS}(x,x)$ vanishes in all even dimensions $d$. Basically, we just computed $\zeta_{HS}(1)$. The contribution of $\Delta=d-2$ scalar is always zero, but the sum over HS fields is non-trivial (ghosts need to be subtracted as usual). For example in $AdS_5$, evaluation of $\zeta_{HS}(0)$ and $\zeta_{HS}(1)$ leads to
\begin{align}
   \zeta'_{HS}(0)&: &  {\log R}\sum_s \frac{1}{180} s^2 (s+1)^2 (14 s (s+1)+3)&=0\,,\\
   \zeta_{HS}(1)&: &{\log R}  \sum_s \frac{1}{24} s (s+1) (2 s+1)^2=0\,.
\end{align}
In can be easily checked for non-minimal Type-A theories in $AdS_{2n+1}$ that $\zeta_{HS}(1)=0$. Therefore, at least a part of the full HS tadpole should be zero. Also, we see that it is zero on its own, without any help from other diagrams. Of course, this fact does not ensure full one-loop consistency as similar divergences can come from other diagrams and will mix with the tadpole, so that only the total sum can vanish. It would be interesting to include tadpoles with derivatives, which requires the coincident point limit of  derivatives of the heat kernel. Also, one can investigate $\bep(20,10)\put(0,5){\line(1,0){9}}\put(14,5){\circle{10}}\eep$, which can lead to unwanted $\nabla^k\Phi$-terms. The identities observed above favour application of the heat kernel techniques to HS theories.

\section{F(4) Higher-Spin Theory and Romans Supergravity}
\label{sec:ffour}
Exceptional algebraic structures seldom occur in the higher-spin context, see \cite{Gaberdiel:2013vva} for the discussion of $D(2,1;\alpha)$ in application to HS AdS/CFT. Hence it is remarkable that there exists an exceptional  $AdS_6/CFT_5$  HS algebra \cite{Fernando:2014pya} that is based on the super-singleton of exceptional Lie superalgebra $F(4)$. More specifically it is realized as the enveloping algebra of the minimal unitary realization (super-singleton) of $F(4)$  obtained via the quasiconformal method \cite{Fernando:2014pya}. The super-singleton multiplet of $F(4)$ consists of an $SU(2)_R$ doublet of $\Rac$'s and a singlet $ \Di$.

As in other cases, one can take the $F(4)$ super-singleton as a free $5d$ CFT and consider the higher-spin theory dual to its singlet sector. The spectrum of fields can be computed as a tensor product of two $F(4)$ super-singletons. As we will show, such HS theory is closely related to the Romans $F(4)$ gauged supergravity in $AdS_6$ \cite{Romans:1985tw}.

The original motivation for this work stems from the goal to study
this exceptional $F(4)$ HS theory and the known one-loop tests were further developed so as to apply them to it. Below we review the construction of the $F(4)$ HS algebra and work out the full spectrum of HS fields. In particular, we shall prove that the Romans graviton supermultiplet  belongs to the spectrum of $F(4)$ HS theory. 

\subsection{Exceptional Lie Superalgebra \texorpdfstring{$F(4)$}{F(4)}}
The exceptional Lie superalgebra $F(4)$ has 24 even and  16 odd generators \cite{Kac:1977em}. The real form of $F(4)$ we are interested in has\footnote{We will always work with algebras and superalgebras while using the capitalized names for all of them for historical reasons. } $SO(5,2)\oplus SU(2)$ as its even subalgebra with the odd generators transforming in the $(8,2)$ representation. It is the unique simple superconformal  algebra in five dimensions. It can be realized  as a superconformal symmetry group of an exceptional superspace  coordinatized by the exceptional Jordan superalgebra which has no realization in terms of associative super-matrices \cite{Gunaydin:1989dq,Gunaydin:1989ur}.
The minimal unitary realization of $F(4)$ was obtained via quasiconformal methods relatively  recently \cite{Fernando:2014pya}, which we shall review below.

Following \cite{Fernando:2014pya} we shall denote the generators of $SO(5,2)$ as $M_{AB}$ where $A,B,...=0,1,\dots, 6$  which satisfy
\begin{equation}
\commute{M_{AB}}{M_{CD}}
= i \left(
    \eta_{BC} M_{AD} - \eta_{AC} M_{BD} - \eta_{BD} M_{AC} + \eta_{AD} M_{BC}
    \right)\,,
\end{equation}
where $\eta_{AB} = \mathrm{diag} \left( -,+,+,+,+,+,- \right)$. The generators $T_\la$ $(\la,\lb,...=1,2,3$) of the R-symmetry group $SU(2)_R$ satisfy:
\begin{equation}
[T_\la,T_\lb] =i \epsilon_{\la\lb\lc} T_\lc\,.
\end{equation}
The supersymmetry generators that transform in the $(8,2)$ representation of $SO(5,2)\times SU(2)$ are denoted as $\Xi_\alpha^r$ with $\alpha,\beta,\dots =1,2,\dots 8$ and $r,s,\dots =1,2$.
Their commutators with the generators $M_{AB}$ of $SO(5,2)$ can be written as follows:
\begin{equation}
\commute{M_{AB}}{\Xi_\alpha^r}
= - \left( \Sigma_{AB} \right)_{\alpha\beta} \Xi_\beta^r\,,
\end{equation}
where $\Sigma_{AB}$ are the matrices of the spinor representation of $SO(5,2)$.
Their anticommutators close into the generators of $SO(5,2)\times SU(2)$.
\begin{equation}
\anticommute{\Xi_\alpha^r}{\Xi_\beta^s}
=  i \epsilon^{rs} \, M_{AB} \left( \Sigma^{AB} \mathcal{C}_7 \right)_{\alpha\beta}
   + 3 i \, \left( \mathcal{C}_7 \right)_{\alpha\beta} ( i\sigma_2 \sigma^\la)^{rs} \, T_\la\,,
\end{equation}
where $\epsilon^{rs}$ is the two dimensional Levi-Civita tensor and $C_7$ is  the symmetric charge conjugation matrix $(C_7)_{\alpha\beta} =(C_7)_{\beta\alpha}$ in seven dimensions.

\subsection{Minimal Unitary Representation of \texorpdfstring{$SO(5,2)$ and its Unique Deformation}{SO(5,2)}}
The Hilbert space of the minimal unitary representation of $SO(d,2)$ obtained via the quasiconformal method is spanned by states in the tensor product  of the Fock space of $(d-2)$ bosonic oscillators  with the state space of the Calogero Hamiltonian or of conformal quantum mechanics \cite{Gunaydin:2001bt,Gunaydin:2006vz,Fernando:2015tiu}.
The  explicit expressions for  minimal unitary realization (minrep) of $SO(5,2)$ were given in \cite{Fernando:2014pya}. To show that the minrep of $SO(5,2)$ is a positive energy unitary representation that describes a  massless conformal scalar field in $5d$ one uses the compact 3-grading of $SO(5,2)$ with respect to the Lie algebra of its maximal compact subgroup
$SO(5)\times SO(2)$ whose  covering group is  $Spin(5)\times U(1)\equiv USp(4)\times U(1)$:
\begin{align}
    SO(5,2)&= \underbrace{B_{IJ}}_{-1}\oplus\underbrace{( U_{IJ} +  H )}_{0}\oplus\underbrace{\Bb_{IJ}}_{+1}\,,
\end{align}
where $U_{IJ}=U_{JI}$   are the $USp(4)$ generators with $I,J,... =1,2,3,4$ denoting the spinor indices of $USp(4)$,\footnote{We should note that the spinor indices  $I,J,...$ of $USp(4)$ in the compact three-grading go over to spinor indices of the Lorentz group $USp(2,2)$ in the noncompact three-grading determined by the generator of dilatations \cite{Fernando:2014pya}.} $H$ is the conformal Hamiltonian ($AdS$ energy)  and the grade $\pm1$ components transform as $5$ of $USp(4)$:
\begin{align} B_{IJ}&=-B_{JI}\,, & \Omega^{IJ} B_{IJ} &=0\,, \end{align}
where $\Omega_{IJ}$ is the symplectic invariant metric and $\bar{B}_{IJ} = B_{IJ}^\dagger $.
$USp(4)$ generators satisfy the Hermiticity property
\begin{equation}
U_{IJ} = \Omega_{IK} U_{KL}^\dag \Omega_{LJ} \,.
\end{equation}
The above generators of $SO(5,2)$ satisfy the commutation relations:
\begin{equation}
\begin{split}
\commute{U_{IJ}}{U_{KL}}
&= \Omega_{JK} \, U_{IL} + \Omega_{IK} \, U_{JL}
   + \Omega_{JL} \, U_{IK} + \Omega_{IL} \, U_{JK}\,,
\\
\commute{U_{IJ}}{\overline{B}_{KL}}
&= \Omega_{JK} \, \overline{B}_{IL}
   + \Omega_{IK} \, \overline{B}_{JL}
   - \Omega_{JL} \, \overline{B}_{IK}
   - \Omega_{IL} \, \overline{B}_{JK}\,,
\\
\commute{U_{IJ}}{B_{KL}}
&= \Omega_{JK} \, B_{IL} + \Omega_{IK} \, B_{JL}
   - \Omega_{JL} \, B_{IK} - \Omega_{IL} \, B_{JK}\,,
\\
\commute{H}{U_{IJ}}
&= \commute{\overline{B}_{IJ}}{\overline{B}_{KL}}
 = \commute{B_{IJ}}{B_{KL}}
 = 0\,,
\\
\commute{H}{\overline{B}_{IJ}}
&= + \overline{B}_{IJ}\,,
\qquad \qquad
\commute{H}{B_{IJ}}
 = - B_{IJ}\,.
\end{split}
\end{equation}
We should note that the Dynkin labels $(n_1,n_2)_D $ of $USp(4)$ are related to the Dynkin labels of   $Spin(5)$ by interchange of $n_1$ and $n_2$:
\begin{equation}
(n_1,n_2)_D \, \, \text{of}  \, \, \, USp(4) \Longleftrightarrow (n_2,n_1)_D \,\,\text{of} \, \,  Spin(5)\,.
\end{equation}
Furthermore we shall indicate the Gelfand-Zetlin labeling of the representations of $Spin(5)$ with the subscript $GZ$. They are related to Dynkin labeling as follows:
\begin{equation}
(j_1,j_2)_{GZ} = ( j_1-j_2,2j_2)_D\,.
\end{equation}

In the Hilbert space of the minrep there exists a unique state that is annihilated by all the grade -1 generators $B_{IJ}$ and is a singlet of $USp(4)$ with a definite eigenvalue $E=3/2$ (conformal weight) of the conformal Hamiltonian $H$. This shows that the minrep is a positive energy unitary representation of $SO(5,2)$.
The $5d$ Poincar\'e mass operator vanishes identically as an operator, $P_\mu P_\nu \eta^{\mu\nu} =0$, for the minrep. Hence it describes a massless conformal scalar field $\phi$ in $5d$, i.e. $\Rac$. 

A positive energy unitary irreducible representation of $SO(5,2)$ can be  uniquely labelled by its lowest energy irrep  $|E;(j_1,j_2)_{GZ}\rangle = |E; (j_1-j_2,2j_2 )_D\rangle $
where $(j_1,j_2)_{GZ}$ are the Gelfand-Zetlin labels of the lowest energy $Spin(5)$ irrep and $E$ is the eigenvalue of the conformal Hamiltonian (or AdS energy). The corresponding unitary representation of $SO(5,2)$ is then labelled as $D(E;(j_1,j_2)_{GZ})$ or as $D(E;(j_1-j_2,2j_2)_D)$. Thus, Rac  is simply the irrep $D(3/2;(0,0)_D )$. 
The decomposition of the minrep  $\Rac$ with respect to its maximal compact subgroup, which is called $K$-type decomposition, was given in \cite{Fernando:2015tiu}:
\begin{equation}
\Rac = D(3/2;(0,0)_D ) = \bigoplus_{s=0}^\infty | E=3/2+s; (s,0)_D \rangle \,.
\end{equation}

The minrep of $SO(5,2)$ admits a single deformation which is realized by adding a spin term to the generators of the little group $SO(3)$ of massless particles in $5d \,$ \cite{Fernando:2014pya}. The Hilbert space of the deformed minrep is spanned by the states which are in the tensor product of the Hilbert space of the minrep with the Fock space of two fermionic oscillators transforming as a spinor of the covering group $SU(2)$ of the little group $SO(3)$. There exist four states in the Hilbert space of the deformed minrep that are annihilated by the  generators $B_{IJ}$ and transform in the spinor representation of $USp(4)$ with definite eigenvalue ($E=2$)  of conformal Hamiltonian $H$. The deformed minrep describes a massless conformal symplectic Majorana spinor field $\psi_I$ in $5d$, which is the spinor singleton  $\Di= D(2;(0,1)_D)$. The K-type decomposition of the deformed minrep  $\Di$ is as follows \cite{Fernando:2014pya}
\begin{equation}
\Di =D(2;(0,1)_D)= \bigoplus_{s=0}^\infty | E=2+s; (s,1)_D \rangle \,.
\end{equation}

\subsection{Compact 3-grading of \texorpdfstring{$F(4)$ with respect to $OSp(2|4) \oplus SO(2)$}{(F(4)}}\label{sec:F(4)c3G}
The Lie superalgebra $F(4)$ admits  a 3-graded decomposition with respect to its compact subsuperalgebra $OSp(2|4) \oplus SO(2)_{\mathcal{H}} $ \cite{Fernando:2014pya}
\begin{equation}
F(4)
= \mathfrak{C}^- \oplus \mathfrak{C}^0 \oplus \mathfrak{C}^+\,,
\end{equation}
where
\begin{equation}
\begin{split}
\mathfrak{C}^-
&= B_{IJ} \oplus
   T_- \oplus
  \mathcal{Q}_I\,, \\
\mathfrak{C}^0
&= \mathcal{H} \oplus
   U_{IJ}  \oplus
  Z
 \oplus
  \mathcal{R}_I \oplus \overline{\mathcal{R}}_J\,, \\
\mathfrak{C}^+
&= \overline{B}_{IJ}  \oplus
   T_+ \oplus
   \overline{\mathcal{Q}}_I\,,\end{split}
\label{F(4)3grading}
\end{equation}
where  $\mathcal{H}= H + T_3 $ is the $SO(2)$ generator that determines the compact 3-grading of $F(4)$.  The generators of $SU(2)_R$ are denoted as $T_+,T_-$ and $T_3$ which satisfy
\begin{equation}
\commute{T_+}{T_-} = 2 T_3\,, \qquad \qquad\qquad\commute{T_3}{T_{\pm} }= \pm T_{\pm}\,.
\end{equation}
The subsuperalgebra $OSp(2|4)$  has the even subalgebra of $SO(2) \oplus USp(4)$ whose generators are $Z=H + T_3$ and $U_{IJ}$. The odd  generators of  $OSp(2|4)$ that transform as complex spinors of $USp(4)$ are  denoted as $\mathcal{R}_I$ and $\overline{\mathcal{R}}_I$ and satisfy
\begin{equation}
\overline{\mathcal{R}}_I
= \mathcal{R}_J^\dag \Omega_{JI}\,.
\end{equation}
The generators of $OSp(2|4)$ satisfy:
\begin{equation}
\begin{split}
\commute{U_{IJ}}{U_{KL}}
&= \Omega_{JK} \, U_{IL} + \Omega_{IK} \, U_{JL}
   + \Omega_{JL} \, U_{IK} + \Omega_{IL} \, U_{JK}\,,
\\
\anticommute{\mathcal{R}_I}{\mathcal{R}_J}
&= 0\,,
\qquad \qquad\qquad\qquad
\anticommute{\overline{\mathcal{R}}_I}{\overline{\mathcal{R}}_J}
= 0\,,
\\
\anticommute{\mathcal{R}_I}{\overline{\mathcal{R}}_J}
&= \Omega_{IJ} \, Z - U_{IJ}\,,
\\
\commute{\mathcal{Z}}{\mathcal{R}_I}
&= - \mathcal{R}_I\,,
\qquad\qquad\qquad\quad
\commute{\mathcal{Z}}{\overline{\mathcal{R}}_I}
= + \overline{\mathcal{R}}_I\,,
\\
\commute{U_{IJ}}{\mathcal{R}_K}
&= \Omega_{JK} \, \mathcal{R}_I + \Omega_{IK} \, \mathcal{R}_J\,,
\\
\commute{U_{IJ}}{\overline{\mathcal{R}}_K}
&= \Omega_{JK} \, \overline{\mathcal{R}}_I
   + \Omega_{IK} \, \overline{\mathcal{R}}_J\,.
\end{split}
\end{equation}
The odd generators that belong to grade $-1$ and  grade $+1$ subspaces are denoted as $\mathcal{Q}_I$ and $\overline{\mathcal{Q}}_I$, respectively, and are related by Hermitian conjugation
\begin{equation}
\overline{\mathcal{Q}}_I
= \mathcal{Q}_J^\dag \Omega_{JI}\,.
\end{equation}
The remaining commutation relations of the superalgebra $F(4)$ are given below:
\besubeqs
\begin{eqnarray}
\anticommute{\mathcal{Q}_I}{\overline{\mathcal{Q}}_J}
&=&\Omega_{IJ} \left( 3 \, \mathcal{H} - 2 \, \mathcal{Z} \right)
  + U_{IJ}\,, \\
\anticommute{\mathcal{Q}_I}{\mathcal{R}_J}
&=& + 3 \, \Omega_{IJ} \, T_-\,,
\\
\anticommute{\mathcal{Q}_I}{\overline{\mathcal{R}}_J}
&= &- B_{IJ}\,,
\\
\commute{\mathcal{Z}}{\mathcal{Q}_I}
&=& - 2 \, \mathcal{Q}_I\,,
\\
\commute{U_{IJ}}{\mathcal{Q}_K}
&= &\Omega_{JK} \, \mathcal{Q}_I + \Omega_{IK} \, \mathcal{Q}_J\,,
\\
\commute{U_{IJ}}{T_+}
&=& 0\, ,\\
\anticommute{\overline{\mathcal{R}}_I}{\overline{\mathcal{Q}}_J}
&=&- 3 \, \Omega_{IJ} \, T_+\,,
\\
\anticommute{\mathcal{R}_I}{\overline{\mathcal{Q}}_J}
&= & + \overline{B}_{IJ}\,,
\\
\commute{\mathcal{Z}}{\overline{\mathcal{Q}}_I}
&=& + 2 \, \overline{\mathcal{Q}}_I\,,
\\
\commute{U_{IJ}}{\overline{\mathcal{Q}}_K}
&= &\Omega_{JK} \, \overline{\mathcal{Q}}_I
   + \Omega_{IK} \, \overline{\mathcal{Q}}_J\,,
\\
\commute{U_{IJ}}{T_-}
&= &0\,, \\
\commute{\mathcal{Q}_I}{\bar{B}_{JK}} & = &-\Omega_{JK} \mathcal{R}_I - 2 \delta_{IJ} \Omega_{KL} \mathcal{R}_L  +2 \delta_{IK} \Omega_{JL} \mathcal{R}_L\,, \\
\commute{\overline{\mathcal{Q}}_I}{B_{JK}} & =& -\Omega_{JK} \overline{\mathcal{R}}_I - 2 \delta_{IJ} \Omega_{KL} \overline{\mathcal{R}}_L  +2 \delta_{IK} \Omega_{JL} \overline{\mathcal{R}}_L\,.
\end{eqnarray} 
\esubeqs

\subsection{Minimal Unitary Supermultiplet of \texorpdfstring{$F(4)$}{F(4)}}
In the minimal unitary realization of $F(4)$, as obtained via the quasiconformal method, the generators are expressed in terms of 3 bosonic oscillators, a singular oscillator and  two fermionic oscillators \cite{Fernando:2014pya}. One finds that  the supersymmetry generators of the minrep  satisfy certain special relations:
\besubeqs
 \begin{align}
\Omega_{IJ} \mathcal{Q}_I \mathcal{Q}_J &= 0\,,&
\Omega_{IJ} \overline{\mathcal{Q}}_I \overline{\mathcal{Q}}_J &= 0\,,
\label{SUSYconstraint}
\\
\Omega_{IJ} \mathcal{Q}_I \overline{\mathcal{R}}_J &= 0\,,
&
\Omega_{IJ} \mathcal{R}_I \overline{\mathcal{Q}}_J &= 0\,,
\\
\overline{\mathcal{Q}}_I \overline{\mathcal{Q}}_J &= - \Omega_{IK}\Omega_{JL} \bar{B}_{KL} T_+ \,, & \mathcal{Q}_I \mathcal{Q}_J &= - \Omega_{IK}\Omega_{JL} B_{KL} T_-\,,
\\
T_+ \overline{\mathcal{Q}}_I &=0\,, & T_- \mathcal{Q}_I &=0 \,,
\\
 \mathcal{Q}_I \mathcal{Q}_J \mathcal{Q}_K &= 0\,,
&
\overline{ \mathcal{Q}}_I \overline{\mathcal{Q}}_J \overline{\mathcal{Q}}_K &= 0\,.
\end{align}
\esubeqs
In the (super)Hilbert space of the minrep there exists a unique normalizable state $| E=\frac{3}{2};(0,0) \rangle^- $ that is annihilated by all the grade -1 generators $(T_- , B_{IJ} ,  \mathcal{Q}_I )$ of the compact 3-grading and is an eigenstate of $H $ and $T_3$. It is simply the tensor product of the lowest weight vector of the minrep of $SO(5,2)$ with the Fermionic Fock vacuum $|0\rangle_F $, which we shall denote as $|\Phi_{0}^-\rangle \equiv  | E=\frac{3}{2};(0,0) \rangle^- $
\begin{eqnarray}
H |\Phi_{0}^-\rangle = \frac32 |\Phi_{0}^-\rangle\,, \qquad\qquad\qquad
T_3 |\Phi_{0}^-\rangle = -\frac12 |\Phi_{0}^-\rangle \,.
\end{eqnarray}
This state is a singlet of the compact subsuperalgebra $OSp(2|4)$.
Acting on $|\Phi_{0}^-\rangle$ repeatedly  by the grade +1 generators $T_+, \bar{B}_{IJ} $ and $\overline{\mathcal{Q}}_J$ one generates the states that form the minimal unitary representation of $F(4)$. Acting on $ | E=\frac{3}{2};(0,0) \rangle^- $ with grade +1 generators $\bar{B}_{IJ}$ repeatedly one generates the minrep of $SO(5,2)$ corresponding to a conformal scalar with $t_3=-1/2$, i.e. $
\Rac$. Acting on  $ | E=\frac{3}{2};(0,0) \rangle^- $ with $T_+$ one generates the lowest weight vector of a second  copy of the minrep with $t_3=1/2$, the second $
\Rac$. Supersymmetry generator $\overline{\mathcal{Q}}_I$ on  $ | E=\frac{3}{2};(0,0) \rangle^- $  generates the lowest energy irrep of the deformed minrep that describes a massless conformal spinor field, i.e. $\Di$. No additional lowest energy irreps of $SO(5,2)$ are generated by further action of grade $+1$ generators. Therefore the minimal unitary supermultiplet consists of two complex scalar fields  in the doublet of R-symmetry group $SU(2)_R$ and a symplectic Majorana spinor.  The lowest weight vector corresponding to the second scalar field will be denoted as $|\Phi_{0}^+\rangle $ and the lowest vector of the deformed minrep that describes the massless conformal spinor field is denoted as $|\Psi^0_I\rangle $:
\begin{align}
|\Phi_{0}^+\rangle  &= T_+ |\Phi_{0}^-\rangle\,, &
|\Psi^0_I\rangle &= \overline{\mathcal{Q}}_J |\Phi_{0}^-\rangle\,.
\end{align}
Therefore the minimal unitary supermultiplet of $F(4)$ consists of two complex massless conformal scalar fields that transform in a doublet of $SU(2)_R$  and a massless symplectic Majorana spinor in five dimensions.
The corresponding unitary module consists of an infinite tower of supermultiplets of states that form representations of the subsuperalgebra $OSp(2|4)$ with definite eigenvalues of the $U(1)$ generator $\mathcal{H}$ that determines the compact 3-grading, namely
\begin{eqnarray*}
\text{Minrep of F(4)} &=& | E=\tfrac{3}{2};(0,0) \rangle^{-1/2}  \oplus  \\ &&   \sum_{s=0} \left( |E=s+\tfrac{3}{2};(s,0)\rangle^{+1/2}   \oplus |E=s+2; (s,1)\rangle^0 \oplus |E=s+\tfrac{5}{2} ; (s+1,0)\rangle^{-1/2} \right)\,, \nonumber
\end{eqnarray*}
where the state $| E=\frac{3}{2};(0,0) \rangle^{-1/2} $ is singlet of $OSp(2|4)$ and the states in the second line above form an irreducible supermultiplet of $OSp(2|4)$ for each value of $s$. The exponent on the kets indicates the eigenvalue of $T_3$.

\subsection{Romans \texorpdfstring{$F(4)$}{F(4)} Graviton Supermultiplet  in Compact 3-grading}
Since there is no invariant concept of mass in $AdS$ spacetimes  and hence no universal definition of masslesness it was proposed in \cite{Gunaydin:1984fk,Gunaydin:1984wc,Gunaydin:1985tc} that massless representations in $AdS_{d+1}$  should be defined as those representations that occur in the tensor product of singleton or doubleton representations of $SO(d,2)$. This definition agrees with the results of \cite{Flato:1978qz} which showed that tensor products of two singleton  representations of $SO(3,2)$ contain all the massless representations in $d=4$ by taking their Poincare limit. Dimension $4$ is special in that various definitions of masslessness in $AdS_4$ agree.
The graviton supermultiplet of gauged maximal supergravities in $d=4,5$ and $d=7$ dimensions were obtained from tensoring of singleton ($d=4$) or CPT self-conjugate doubleton supermultiplets  ($d=5,7$) of the corresponding $AdS$ superalgebras $OSp(8|4,\mathbb{R})$, $SU(2,2|4)$ and $OSp(8^*|4)$ \cite{Gunaydin:1984wc,Gunaydin:1984fk,Gunaydin:1985tc}. The singleton and doubleton supermultiplets of these $AdS$ superalgebras do not have a Poincare limit and, as pointed out in these references, their field theories live on the boundaries of $AdS$ spacetimes as superconformal field theories. Taking higher tensor products results in massive supermultiplets of the corresponding superalgebras. In the twistorial oscillator construction of the unitary representations of $AdS$ superalgebras of   \cite{Gunaydin:1984wc,Gunaydin:1984fk,Gunaydin:1985tc} tensoring is very straightforward and corresponds to simply increasing the number of colors of the (super)-oscillators since the generators  are realized as bilinears of these oscillators.  However tensoring of the supersingleton of  $F(4)$ is more subtle since its realization, as obtained via the  quasiconformal method \cite{Fernando:2014pya}, is nonlinear.

 We shall adopt the same definition of massless supermultiplets in $AdS_6$ and construct the massless graviton supermultiplet that underlies the $\mathcal{N}=2$ $AdS_6$ gauged supergravity of Romans \cite{Romans:1985tw} by tensoring two singleton supermultiplets of $F(4)$.
The fields of the Romans gauged $\mathcal{N}=2$ $AdS_6$ supergravity  are \cite{Romans:1985tw}: graviton $e^a_\mm$, four gravitini $\psi_{\mm;\alpha}$ satisfying the symplectic Majorana-Weyl condition, an anti-symmetric gauge two-form $B_{\mm\nn}$, an auxiliary abelian gauge vector $a_\mm$, three $SU(2)$ gauge vectors $A^\la_\mm$, four spin-half fields $\chi_{\alpha}$ and a scalar $\sigma$. Starting with the above fields Romans constructed the $AdS_6$ gauged supergravity with gauged $R$-symmetry group $SU(2)_R$. We should note that i)  the auxiliary vector $a_\mm$ can be combined with $B_{\mm\nn}$ as  $B_{\mm\nn}+ \pl_\mm a_\nn-\pl_\nn a_\mm$ and is a Stueckelberg field that can be gauged away and serves to make $B_{\mm\nn}$ into a massive two-form field a-la Higgs; (ii) the mass of the two-form is a free parameter in the Lagrangian.  However there is a unique vacuum that enjoys full $F(4)$ symmetry, which corresponds to setting $m=g/3$, where $g$ is the gauge coupling constant.

At the Lie superalgebra level tensoring is equivalent to taking a sum of two copies of the generators of $F(4)$. Let us denote the generators of $F(4)$ in compact three grading symbolically as follows:
\begin{equation}
F(4) = \mathcal{C}_A \oplus \mathcal{C}_A^B \oplus \bar{\mathcal{C}}^B \,,
\end{equation}
 where the upper (lower) index $A$ in $\bar{\mathcal{C}}^A $ (in $\mathcal{C}_A$)  runs over all the generators in grade +1 (-1) space. In taking direct sum of two copies of the generators we shall label the corresponding generators as follows:
 \begin{equation}
F(4) =( \mathcal{C}_A(1) + \mathcal{C}_A(2) ) \oplus (\mathcal{C}_A^B(1) + \mathcal{C}_A^B(2) ) \oplus (\bar{\mathcal{C}}^B(1) +\bar{\mathcal{C}}^B(2) )\,.
\end{equation}
In contrast to the maximal supergravity multiplets in $AdS_{4,5,7}$ the tensor product of the lowest weight vectors  $ |\Phi_{0}^-(1)\rangle $ and $ |\Phi_{0}^-(2) \rangle $ of two singleton supermultiplets does not lead to a supermultiplet that includes the graviton. One finds that  that the following set of tensor product states
  \begin{equation}
  \{|\Omega^A(1,2)\rangle \} = \{ \bar{\mathcal{C}}^A(1)|\Phi_{0}^-(1)\rangle  |\Phi_{0}^-(2) \rangle -  |\Phi_{0}^-(1) \rangle \bar{\mathcal{C}}^A(2)|\Phi_{0}^-(2)\rangle \} \label{groundlevel} \,.
  \end{equation}
are annihilated by all the grade -1 generators $ \mathcal{C}_A(1) + \mathcal{C}_A(2)$ and transform irreducibly under the subsuperalgebra $OSp(2|4)$. Acting on these states repeatedly by the grade +1 generators $  (\bar{\mathcal{C}}^B(1) +\bar{\mathcal{C}}^B(2) )$ one obtains an infinite set of states that form a unitary irreducible supermultiplet of $F(4)$ that includes the graviton and is precisely the  Romans supermultiplet.
We shall call the states in equation \eqref{groundlevel} ground level of the unitary irrep in compact three grading. They decompose into three irreps of $USp(4)\times U(1)_E$ corresponding to the lowest energy irreps of $SO(5,2)$. By acting with grade $+1$ generators $\bar{B}_{IJ} =\bar{B}_{IJ}(1)+\bar{B}_{IJ}(2) $ on these irreps one generates the infinite tower of states that form the bases of the corresponding  unitary representations of $SO(5,2)$. By acting with the grade +1 generator $T_+=T_+(1)+T_+(2)$  one generates their irrep with respect to the R-symmetry group $SU(2)_R$.  The resulting irreps of $SO(5,2)\times SU(2)_R$ are
  \begin{equation}
  [ D(4;(1,0)_D) , 3]  \oplus [D(3;(0,0)_D , 1] \oplus D[7/2;(0,1)_D),2]
  \end{equation}
corresponding to 3 massless vector fields, one  massless scalar and two massless symplectic Majorana Weyl spinor fields in $AdS_6$. By acting with the supersymmetry generators $\overline{\mathcal{Q}}_I=\overline{\mathcal{Q}_I}(1)+\overline{\mathcal{Q}}_I(2)$ one generates new lowest energy irreps of $SO(5,2)\times SU(2)$ corresponding to the irreps:
  \begin{equation}
  [ D(4;(0,2)_D) , 1]   \oplus D[(9/2;(1,1)_D),2]
  \end{equation}
which describe an anti-symmetric tensor field and two gravitini in $AdS_6$. Finally  the action of the commutator of two supersymmetry generators $\overline{\mathcal{Q}}_I $ leads to the lowest energy irrep of $SO(5,2) $ corresponding to the representation:
  \begin{equation}
  [D(5;(2,0)_D), 1]
  \end{equation}
that describes the graviton in $AdS_6$. The resulting supermuliplet is simply the Romans graviton supermultiplet without the auxiliary vector field. This is a general feature of the manifestly unitary oscillator or quasiconformal construction which involve only the physical fields.

This unitary graviton supermultiplet of $F(4)$ can be decomposed into an infinite set of supermultiplets of $OSp(2|4)$ labelled by the eigenvalues of the compact generator $\mathcal{H}$ that determines the compact 3-grading
    \begin{equation}
  \mathcal{H}= H + T_3 \,,
  \end{equation}
  where $H$ is the conformal Hamiltonian. The $U(1)$ generator $Z$ inside $OSp(2|4)$ is given by
  \begin{equation}
  Z= H+ 3 T_3=\mathcal{H}+ 2T_3
  \end{equation}
  and its supertrace in a given irrep of $OSp(2|4)$ vanishes.

\subsection{Compact 5-grading of \texorpdfstring{$F(4)$ with respect to $USp(4) \oplus SU(2) \oplus SO(2)$}{F(4)}}
In going from  $SO(5,2)$ to the Lie superalgebra $F(4)$ the compact 3-grading of $SO(5,2)$ admits an extension to  compact five-graded decomposition of $F(4)$  with respect to its maximal compact Lie subalgebra $USp(4)\oplus SO(2)\oplus SU(2)_R$. This  5-grading is the compact analog of the natural noncompact five grading with respect to the subgroup $SO(4,1)\times SO(1,1) \times SU(2)$ with grade $\pm1/2$ subspaces corresponding to Poincare and special conformal supersymmetry generators \cite{Fernando:2014pya}. The compact decomposition is as follows:
\begin{align}
    F(4)&= \underbrace{B_{IJ}}_{-1}\oplus\underbrace{ Q^r_I}_{-1/2}\oplus \left(\underbrace{U_{IJ}\,, T_{\la}\,, H}_{0} \right) \oplus\underbrace{\Qb^r_I}_{+1/2} \oplus\underbrace{\Bb_{IJ}}_{+1}\,,
\end{align}
where $T_{\la} \, (\la=1,2,3) $ are the generators of $SU(2)_R$,  $Q^r_I$ and $\Qb^r_I$  ($ r,s,... =1,2$) are the supersymmetry generators that transform in the  $(2, 4)$ representation of $SU(2)_R\oplus USp(4)$. The supersymmetry generators $Q^r_I$ and $\bar{Q}^r_I$ in the 5-grading are related to the supersymmetry generators in 3-grading as follows:
\begin{eqnarray}
\bar{Q}^1_I &=  \mathcal{\bar{Q}}_I\,, \qquad \qquad\qquad  \bar{Q}^2_I =  -\mathcal{R}_I\,, \\
Q^1_I & =  \mathcal{\bar{R}}_I\,, \qquad  \qquad\qquad  Q^2_I = -\mathcal{Q}_I \,.
\end{eqnarray}
The generators in the above 5-grading satisfy the super-commutation relations
\besubeqs
\begin{align}
    \commute{U_{IJ}}{Q_K^r}&=\Omega_{JK} Q_I^r + \Omega_{IK} Q^r_J\,, \\
    \commute{U_{IJ}}{\bar{Q}^r_K} &=\Omega_{JK} \bar{Q}_I^r + \Omega_{IK} \bar{Q}^r_J\,, \\
    \commute{T_\la}{Q^r_I} &=\frac{1}{2} (\sigma_\la)^{sr} Q^s_I\,,\\
    \commute{T_\la}{\bar{Q}^r_I}&=\frac{1}{2} (\sigma_\la)^{sr}{\bar{Q}^s_I}\,, \\
    \commute{\bar{Q}_I^r}{B_{JK}}&=\,-2 \delta_{IJ} \Omega_{KL} Q^r_L + 2 \delta_{IK} \Omega_{JL} Q^r_L - \Omega_{JK} Q^r_I , \\
    \commute{Q^r_I}{\bar{B}_{JK}}&=\,-2 \delta_{IJ} \Omega_{KL} \bar{Q}^r_L + 2 \delta_{IK} \Omega_{JL} \bar{Q}^r_L - \Omega_{JK} \bar{Q}^r_I , \\
    \anticommute{Q^r_I}{Q^s_J}&=\,-\epsilon^{rs} B_{IJ}, \\
    \anticommute{\bar{Q}^r_I}{\bar{Q}^s_J}&=\,\epsilon^{rs} \bar{B}_{IJ}, \\
    \anticommute{Q^r_I}{\bar{Q}^s_J}&=\,\epsilon^{rs} \Omega_{IJ} H -  3 \Omega_{IJ} (i \sigma_2 \sigma_\la)^{rs} T^\la  + 2 \epsilon^{rs} \Omega_{IK}\Omega_{JL} U_{KL} .
\end{align}
\esubeqs
For the minimal unitary realization of $F(4)$ one finds that the $SU(2)_R$ covariant supersymmetry generators satisfy
\begin{eqnarray}
\Omega_{IJ} Q^r_I Q^s_J  =0\,, \qquad \qquad \qquad\Omega_{IJ} \bar{Q}^r_I \bar{Q}^s_J =0\,. \end{eqnarray}
In this  Hilbert space there exist only two states $|\phi^r \rangle $, two $\Rac$'s, that are annihilated by all the negative grade generators $B_{IJ}$ and $Q_I^r$:
\begin{equation}
    Q_I^r |\phi_0^s \rangle =0 \,, \qquad\qquad\qquad B_{IJ} |\phi_0^s \rangle =0\,.
\end{equation}
They correspond to the lowest weight vectors of an $SU(2)_R$ doublet of scalar singletons that describe massless conformal scalars in five dimensions.  Acting on these two states with the supersymmetry generators $\bar{Q}^r_I$ one finds
\begin{equation}
\bar{Q}^r_I |\phi_0^s \rangle = \epsilon^{rs} |\psi^0_I \rangle\,,
\end{equation}
where $|\psi^0_I \rangle $ is the lowest energy irrep of the spinor singleton,  $\Di$,
\begin{equation}
B_{IJ} |\psi^0_K \rangle =0\,.
\end{equation}
 There are no other states transforming irreducibly under $USp(4)$ and are annihilated by $B_{IJ}$. Hence the minimal unitary supermultiplet of $F(4)$ describes a supermultiplet of two complex massless conformal scalars transforming as a doublet of $SU(2)_R$ and a massless spinor field as we found using the compact three grading of $F(4)$. Since the lowest energy irrep $|\psi^0_I \rangle $ transforms in the spinor representation of $USp(4)$ the spinor singleton describes a symplectic Majorana spinor field in $d=5$. 

\subsection{Romans \texorpdfstring{$F(4)$}{F(4)} Graviton Supermultiplet in Compact 5-grading}
\label{sec:}
The minrep, i.e. $\Rac$, of $SO(5,2)$ inside the minimal unitary supermultiplet of $F(4)$  occurs with multiplicity two and transforms as a doublet of $SU(2)_R$. Tensoring two copies of the supersingletons corresponds to taking direct sum of two copies of the generators of $F(4)$ and tensoring  of the corresponding Hilbert spaces.
In the tensor product space there is a unique lowest weight vector that is a singlet of $SU(2)_R$, is annihilated by all the negative grade generators of $F(4)$ in the compact 5-grading and leads to the graviton supermultiplet of Romans' theory. This singlet state is
\begin{equation}
  |\Omega_0 \rangle = \epsilon_{rs} |\phi_0^r(1)\rangle |\phi^s_0(2)\rangle\,.
\end{equation}
The additional lowest weight vectors of $SO(5,2)$ inside the resulting unitary representation of $F(4)$ are \cite{Klebanov:2002ja,Sezgin:2002rt,Sezgin:2003pt} obtained by acting with antisymmetrized products of the supersymmetry generators $\bar{Q}^r_I$ which transform in the (2,4) representation of $SU(2)_R\times USp(4)$. We list below the lowest weight irreps of $SO(5,2)$  that make up the graviton supermultiplet, their transformation under $USp(4)\times SU(2)_R$ and the corresponding $6d$ fields:
\begin{equation}
\begin{tabular}{ccc}
    Lowest Energy Irreps &       $SO(2)\oplus USp(4)\oplus SU(2)$ & Romans Field \\ \hline
    \rule{0pt}{18pt} $\Romans$ &  $\left(3,\bullet,\bullet\right)$ & scalar \\
    \rule{0pt}{18pt} $\Qb^i_I\Romans$ & $\left(\tfrac72,\YoungpA,\YoungpA\right)$ & complex spinor in a doublet of $SU(2)_R$\\
    \rule{0pt}{28pt} $(\Qb^r_I\Qb^s_J)_A \Romans$ & $\left(4,\YoungpB,\YoungpAA\right)\oplus\left(4,\YoungpAA,\YoungpB\right)$ & \parbox{6cm}{two-form field $B_{\mm\nn}$ and $SU(2)_R$ triplet of vector fields  $A^\la_\mm$}\\
     \rule{0pt}{18pt}$(\Qb^r_I\Qb^s_J\Qb^t_K)_A\Romans$ & $\left(\tfrac92,\YoungpBA,\YoungpBA\right)$ & complex gravitinos in a doublet of $SU(2)_R$\\
    \rule{0pt}{18pt} $(\Qb^r_I\Qb^s_J\Qb^t_K\Qb^u_L)_A \Romans$ & $\left(5,\YoungpBB,\YoungpBB\right)$ &  graviton
\end{tabular}\notag
\end{equation}
The first column in the above table list the lowest energy irreps of $SO(5,2)$ generated by the action of anti-symmetrized products of  the supersymmetry generators
$\Qb^r_I$. The second column lists their transformation properties under  $USp(4)\oplus SU(2)$.

Using the $Spin(5)\times U(1)$  labelling of the unitary representations corresponding to the $AdS_6$ fields the Romans supermultiplet decomposes  as follows:
\begin{subequations}\label{RomansMult}
\begin{eqnarray}
\text{Scalar:} & \quad \quad D(3;(0,0)_D) \\
\text{Spinors:} &  \quad \quad D^r(7/2;(0,1)_D) \\
\text{Tensor field:} & \quad \quad D(4;(0,2)_D) \\
\text{Vector fields:} & \quad \quad D^\la(4;(1,0)_D) \\
\text{Gravitinos:} & \quad \quad D^r ( 9/2;(1,1)_D) \\
\text{Graviton:} & \quad \quad D(5;(2,0)_D)
\end{eqnarray}
\end{subequations}
where $r=1,2$ and $\la=1,2,3$ are the spinor and adjoint $SU(2)_R$ symmetry indices.

On expanding the Romans supergravity Lagrangian \cite{Romans:1985tw} around the unique $F(4)$ supersymmetric vacuum it turns out that the mass of the scalar field is $(-6)$, which is that of a conformally-coupled scalar and the $AdS$ energy is $3$; the $AdS$ energy of $B_{\mm\nn}$ is $4$; the $AdS$ energy of $\chi$ is $7/2$; the $AdS$ energies of graviton, gravitini and $SU(2)$ gauge field are fixed by gauge symmetry to be $5$, $9/2$ and $4$, respectively. Therefore, the supermultiplet \eqref{RomansMult} thus obtained by tensoring of two $F(4)$ supersingletons with the lowest weight vector $|\Omega_0\rangle$ in the compact 5-grading  is precisely the Romans supermultiplet, agreeing with the result obtained in compact $3$ grading.

\subsection{F(4) HS Theory Spectrum and One-loop tests}
In addition to the Romans supermultiplet, the tensor product of two $F(4)$ super-singletons contains an infinitely many massless  $F(4)$ supermultiplets that have higher-spin fields. In fact they include an infinite tower of massless higher spin supermultiplets that extend the graviton supermultiplet which we list below:
\begin{align}
\text{Scalar tower: } & \quad \quad D(3+s;(s,0)_D) && \parbox{40pt}{\RectARow{4}{$s$}}\\
\text{Spinor tower: } &  \quad \quad D^r(7/2 +s ;(s,1)_D) &&\parbox{40pt}{\RectARow{4}{$s$}}_{\tfrac12}\\
\text{Tensor field tower: } & \quad \quad D(4+s;(s,2)_D) && \parbox{50pt}{\bep(40,20)\put(0,0){\YoungA}\put(0,10){\RectARow{5}{$s+1$}}\eep}\\
\text{Vector field tower: } & \quad \quad D^\la(4+s;(s+1,0)_D) && \parbox{50pt}{\RectARow{5}{$s+1$}}\\
\text{ Gravitino tower: } & \quad \quad D^r ( 9/2+s;(s+1,1)_D)&&\parbox{50pt}{\RectARow{5}{$s+1$}}_{\tfrac12} \\
\text{Graviton tower: } & \quad \quad D(5+s;(s+2,0)_D) && \parbox{60pt}{\RectARow{6}{$s+2$}}
\end{align}
where $r=1,2$ and $\la=1,2,3$ are the spinor and adjoint indices of $SU(2)_R$ symmetry and $s=0,1,2,...$. In the rightmost column we displayed the Young symmetries of the corresponding HS fields, spin-tensors having $\tfrac12$ subscript. For each $s$ they describe an irreducible unitary supermultiplet of $F(4)$. We shall refer to this infinite tower of massless supermultiplets labelled by $s$ as Roman's tower.

The lowest energy irreps of the infinite towers of irreducible representations of $SO(5,2)$ labelled by $s$ given above  form unitary supermultiplets of the compact subsuperalgebra $OSp(2|4)$ of $F(4)$  for each value of $s$ which we give below:
\begin{align}
\begin{aligned}
 &| 3+s;(s,0)_D\rangle^0   \oplus  \\  &| 7/2 +s ;(s,1)_D\rangle^{1/2} \oplus  | 7/2 +s ;(s,1)_D\rangle^{-1/2}  \oplus \\ &|4+s;(s,2)_D\rangle^0  \oplus \\     &|4+s;(s+1,0)_D\rangle^{-1}    \oplus  |4+s;(s+1,0)_D\rangle^{0} \oplus |4+s;(s+1,0)_D\rangle^{+1}  \oplus \\  &|9/2+s;(s+1,1)_D \rangle^{-1/2}     \oplus  |9/2+s;(s+1,1)_D \rangle^{+1/2}   \oplus \\ &|5+s;(s+2,0)_D\rangle  \,,
\end{aligned}
\end{align}
where the superscript over a ket indicates the eigenvalue of $T_3$. For generic $s$ the above supermultiplet is a long supermultiplet of $OSp(2|4)$. For $s=0$ it decomposes as the sum  of three  short supermultiplets namely the supermultiplet with the eigenvalues of $\mathcal{H}=H+T_3= 3,4,5$:
\besubeqs
\begin{eqnarray}
\mathcal{H}= 3 &\Longrightarrow  &| 3;(0,0)_D\rangle^0   \oplus   | 7/2  ;(0,1)_D\rangle^{-1/2} \oplus |4;(1,0)_D\rangle^{-1} \,, \\
\mathcal{H}= 4 &\Longrightarrow &
   |7/2;(0,1)_D\rangle^{1/2} \oplus   |4;(0,2)_D\rangle^0       \oplus  |4;(1,0)_D\rangle^{0}  \oplus |9/2;(1,1)_D \rangle^{-1/2} \,,  \\
\mathcal{H} = 5 &    \Longrightarrow &   |9/2;(1,1)_D \rangle^{+1/2}   \oplus |5;(2,0)_D\rangle \oplus |4;(1,0)_D\rangle^{+1}  \,.
\end{eqnarray}
\esubeqs
We should perhaps note that despite the fact that the fields of various supergravity multiplets can occur in the spectrum of HS theories, their appearance is somewhat different. For example the massive two-form that shows up in the product of two $F(4)$ singletons is represented as a matter-like anti-symmetric rank-two tensor in the higher-spin theory. In the Romans $F(4)$ gravity it is realized as a gauge two-form field $B_{\mm\nn}$ that is Higgsed via an additional $SO(2)$ gauge field $a_{\mm}$.

Interestingly we find that the full spectrum obtained by tensoring two $F(4)$ super-singleton multiplets contains  the entire Romans tower of  massless $F(4)$ supermultiplets plus a single short supermultiplet which we denote as $L(8|8)$:
\begin{equation}
\left[ F(4) \, \, \text{Super Singleton} \right]^2 = \text{Romans Tower} \, \oplus \,  L(8|8)\,,
\end{equation}
where the supermultiplet $L(8|8)$ decomposes as follows:
\begin{equation}
    L(8|8) = D(4;(1,0)_D) \oplus D^r( 7/2;(0,1)_D) \oplus D^\la(3;(0,0)) \oplus D(4;(0,0))\,,
\end{equation}
where $\la=1,2,3$ and $r=1,2$. $L(8|8)$ consists of a vector field, two spinor fields and four scalars which decompose as a triplet plus a singlet of $SU(2)_R$. We should note that the conformal weight of the singlet scalar is 4 while the triplet of scalars have conformal weight 3. Appearance of this short supermultiplet may look surprising at first sight
since the infinite scalar tower corresponds to the gauge fields of the standard bosonic higher spin theory in $AdS_6$  and the Romans tower corresponds to a supersymmetric extension of this standard bosonic HS theory that includes the fields of Romans gauged supergravity  at the lowest level.
However, when we apply
the one-loop tests of the previous Sections  to the $F(4)$ theory  we find that the Romans tower by itself does not make $\zeta_{HS}(0)$ vanish, i.e. the one-loop contribution for such a spectrum is ill-defined since the $\log \Lambda$-terms do not vanish. This computation requires the following information: $\zeta(0)$ is zero for the sum of all totally-symmetric HS fields and for the sum of all HS fermionic fields; it equals $1/1512$ for weight-$3$ scalar, $-(1271/3780)$ for massless vector field, $271/15120$ for spin-half and the net contribution of height-one hook fields gives $1/180$. One finds that the total contribution of the Romans tower to $\zeta(0)$ is $3/8$.

On the other hand, the full spectrum of the tensor square of the $F(4)$ super-singleton ($2\Rac\oplus \Di$) passes the one-loop tests. Since the full spectrum contains the $L(8|8)$ supermultiplet we conclude that $AdS/CFT$ requires that HS theory of Romans tower must be coupled to the fields of the  $L(8|8)$ supermultiplet. Indeed, given that $\zeta(0)$ for the weight-four scalar is $-(37/7560)$, the contribution of the $L(8|8)$ multiplet is exactly $-3/8$ and cancels that of the Romans tower.

The same computation can be presented in a way that makes the power of the $F(4)$ symmetry more manifest. While the zeta-functions for each individual field of the $F(4)$ HS theory are quite complicated, see Appendices, the value of $\zeta(0)$ for the spin-$s$ supermultiplet of the Romans tower, where $s$ is the highest spin in the multiplet, is remarkably simple:
\begin{align}
\zeta_{\text{Romans},s}(0)&=-\frac{3}{8} s^4
\end{align}
In particular, $\zeta(0)$ for the Romans supergravity multiplet is $-6$. Summing over the whole Romans tower with $e^{-\epsilon s}$ regulator we get $3/8$, which is then canceled by the contribution from the $L(8|8)$ supermultiplet.

Remarkably the supermultiplet $L(8|8)$ corresponds simply to the linear multiplet which plays a crucial role in the off-shell formulations of $5d$ conformal supergravity  and their matter couplings \cite{Bergshoeff:2002qk,Bergshoeff:2001hc,Ozkan:2016csy}.  It is also related  to the off-shell (improved)  vector multiplet in 5d. Therefore we conclude that the consistent formulation of $F(4)$ HS theory must be based on the reducible multiplet extending the Romans supergravity multiplet by the supermultiplet  $L(8|8)$, which plays the role of compensating supermultiplet in $5d$ conformal supergravity, coupled to the infinite set of higher-spin fields belonging to the Romans tower.
The resulting  $F(4)$ HS theory passes the one-loop tests by Casimir Energy and its Type-A and the fermionic parts are in agreement with the free energy on five-sphere. The Type-B part reveals a puzzle, which is a general feature of type-B theories that we discuss in the Conclusions.

\section{Discussion and Conclusions}
\label{sec:conclusions}
Our results are as follows:
\begin{itemize}
    \item the spectral zeta-function is derived for arbitrary  mixed-symmetry fields;

    \item to the list of known one-loop tests we added those that are based on zeta-function for fermions and specific mixed-symmetry fields that arise in Type-B theories;

    \item fermionic HS fields were shown to pass both the Casimir Energy and the zeta-function tests quite easily since they are not expected to generate any one-loop corrections at all, which is what we observed. However, vanishing of the fermions contribution is still nontrivial and involves the summation over all spins;

    \item knowing the zeta-function for a generic mixed-symmetry field allowed us to derive a very simple formula for the derivative $\pl_\Delta a(\Delta)$ of the $a$-anomaly 
    that allows one to integrate it to full  $ a(\Delta)$. A similar feature was observed for the second derivative of the Casimir Energy $\pl_\Delta^2 E_c$;

    \item we showed that $\zeta_{HS}(1)=0$ at least in some of the cases, which is a different type of equality relying on the spectrum of HS theories. This fact should be related to vanishing of the tadpole diagram, which can be problematic in HS theory;

    \item the spectrum of the Type-B theories, which should be generically dual to a free fermion and involve mixed-symmetry fields  in the bulk, passes the zeta-function tests for $AdS_{2n+1}/CFT^{2n}$, where for the minimal Type-B theories one finds the $a$-anomaly of free fermion. But they fail naively for $AdS_{2n+2}/CFT^{2n+1}$, which was first observed for $AdS_4$ in \cite{Giombi:2013fka}. Nonetheless we show that the bulk one-loop results can be computed as a change in $F$-energy, \eqref{FenergyA} and \eqref{FenergyB};

    \item the tensor product of two $F(4)$ super-singletons, which consist of  a doublet of  $\Rac$s and a singlet $\Di$, was evaluated and decomposed into irreducible unitary supermultiplets of $F(4)$. The resulting spectrum contains the multiplet of Romans gauged supergravity in $AdS_6$ as well as an infinite series of HS $F(4)$ supermultiplets that contain fermionic HS fields, totally-symmetric HS fields and height-one hook fields of Type-B. The spectrum of the $F(4)$ HS theory consists of the infinite Romans tower plus  a single additional short supermultiplet  $L(8|8)$. The multiplet $L(8|8)$ corresponds to the linear multiplet of $5d$ conformal supergravity and its contribution to $F(4)$ HS theory is critical to pass the one loop tests. 

    \item partially-massless fields arising in the duals of the non-unitary higher-order singletons $\square^k\phi=0$, both minimal and non-minimal, were shown to pass the Casimir Energy tests, see also \cite{Basile:2014wua}. They also pass the zeta-function tests in $AdS_{2n+1}$, where for the minimal models the result equals the $a$-anomaly of higher-order singletons. Such theories provide examples of HS theories with massive HS fields. In addition this series of theories has relation to the $A_k$ series of Lie algebra, see Appendix \ref{app:PMfields};

    \item higher-spin doubletons with $j>1$, which are unitary as representations of conformal algebra but pathological from the CFT point of view in not having a local stress tensor, were shown not to pass the Casimir Energy test in $AdS_5/CFT^4$, see Appendix \ref{app:HSgletons}.
\end{itemize}

While the tests successfully passed require no further comments, let us discuss the cases where we discovered a mismatch between AdS and CFT sides.

As it was already mentioned, for $AdS_{2n+1}/CFT^{2n}$ the list of unitary conformal fields includes higher-spin doubletons, in addition to the omnipresent $\Rac$ and $\Di$. It was shown in \cite{Beccaria:2014zma,Beccaria:2014qea} that the spin-one, $j=1$, doubleton in $AdS_5$, i.e. the dual of the Maxwell field, and in $AdS_7$, i.e. the dual of the self-dual tensor, are consistent with the duality. We observed that for $j>1$ the excess of the Casimir energy in the bulk cannot be compensated by a simple modification of $G^{-1}\sim N$ relation. However, higher-spin doubletons are pathological as CFT's so we should not worry that they do pass the test.

We observed that there is a general puzzle about Type-B HS theories that are dual to free fermion. It has been already noted in \cite{Giombi:2013fka} that there is a discrepancy in $AdS_4/CFT^3$ Type-B duality. At least in $AdS_4/CFT^3$ it can be explained almost without computations. The free spectrum of single-trace operators built out of free fermion is identical to that of the $3d$ critical boson at $N=\infty$, which was noted in \cite{Leigh:2003gk}. Therefore, unless a miracle happens the two theories --- Type-A with $\Delta=2$ boundary condition for the scalar field and Type-B --- cannot pass the one-loop test simultaneously.

Our computations extend this puzzle to any $AdS_{2n+2}/CFT^{2n+1}$. The fact that the discrepancy is for fermions and it is in odd dimensions makes one think that the problem is due to parity anomaly \cite{Giombi:2013fka}. The issue could have been easily resolved by allowing fractional coupling constant in the bulk HS theory, i.e. by having a more complicated $G^{-1}(N)$-relation. Indeed, the bulk constant has to be quantized \cite{Maldacena:2011jn}, but the precise mechanism of how this happens in the bulk is unclear. In particular, it is uncertain if  $G^{-1}$ has to be of the form $a(N+\text{integer})$ or not. However, the need for fractional shift of $N$ would spoil the whole logic of one-loop tests. Moreover, it would render SUSY HS theories inconsistent since the $G^{-1}\sim N$ relation for the Type-A subsector of any SUSY HS theory is canonical.

Also, it is not obvious what is the field-realization of the singlet constraint in higher dimensions. At least in $d=3$ there is a natural candidate --- Chern-Simons matter theories  --- that imposes the singlet constraint when coupling is small and provides a family of models that interpolate between free/critical boson/fermion \cite{Giombi:2011kc}. Therefore, there is no 'sharp difference' between bosons and fermions in $3d$. However, the spectrum of single-trace operators of Type-A and Type-B is cleary different in $d>3$. In addition, there does not seem to be any natural candidate to impose the singlet constraint.

Lastly, as the sum over spins requires regularization one cannot exclude the possibility that a different kind of regularization is needed for Type-B theories. The latter is unlikely since the same regularization works for Type-B in odd dimensions and all Type-A theories and fermionic HS fields. Therefore, it seems to be crucial to understand the nature of the singlet constraint and explain the discrepancy for the Type-B.

The one-loop tests performed in this paper show that the heat kernel and zeta-functions techniques provide us with powerful  tools to investigate quantum properties of higher-spin theories. While most of the one-loop tests produced the results to be expected, there is an interesting puzzle about Type-B theories in $AdS_{2n}$. We have also shown  that the graviton supermultiplet of the Romans gauged supergravity in $AdS_6$ belongs to the spectrum of  the unique supersymmetric HS theory based on the exceptional Lie superalgebra $F(4)$ studied in this paper. This remarkable supersymmetric HS theory passed all the one-loop tests we performed modulo the puzzle with Type-B theories in even dimensional $AdS$ spacetimes. Resolution of this puzzle as well as the introduction of  interactions in $F(4)$ HS theory and its dual CFT will be left to future studies. 

\section*{Acknowledgments}
\label{sec:Aknowledgements}
We would like to thank Nicolas Boulanger, Maxim Grigoriev, Shailesh Lal, Ruslan Metsaev and Arkady Tseytlin for the very useful discussions, comments and explanations. T.T. is grateful to Igor Klebanov for a helpful correspondence.  We are grateful to Stuart Dowker for the very useful correspondence. The work of E.S. was supported by the Russian Science Foundation grant 14-42-00047 in association with Lebedev Physical Institute and by the DFG Transregional Collaborative Research Centre TRR 33 and the DFG cluster of excellence ”Origin and Structure of the Universe”. E.S. also would like to thank the organizers of the QFTG'2016 conference, Tomsk, Russia where some of the results were reported. One of us (M.G.) would like to thank the hospitality of Ludwig-Maximilians-Universit\"at in M\"unchen and NORDITA in Stockholm for their hospitality where part of this work
was performed.

{\it Note added: after the completion of our work we learned that Giombi, Klebanov and Tan have independently obtained some of the results on the one-loop tests of  higher-spin theories presented in this paper, see \cite{Giombi:2016pvg}.}

\begin{appendix}
\renewcommand{\thesection}{\Alph{section}}
\renewcommand{\theequation}{\Alph{section}.\arabic{equation}}
\setcounter{equation}{0}\setcounter{section}{0}

\section{Characters, Dimensions and all that}
\label{app:dimschars}
We collect below some useful formulas for the dimensions of various irreducible representations. The classical general formulae for the dimensions of irreducible representations were found by Weyl and for the case of $so(2k)$ and $so(2k+1)$ read:
{\allowdisplaybreaks\besubeqs\label{dimweyl}
\begin{align}
    \mathbb{Y}^{so(2k)}(s_1,...,s_k)&: && \prod_{1\leq i<j\leq k} \frac{(s_i-s_j-i+j)(s_i+s_j-i-j+2k)}{(j-i)(2k-i-j)}\,,\\
    \mathbb{Y}^{so(2k+1)}(s_1,...,s_k)&: && \prod_{1\leq i<j\leq k} \frac{(s_i-s_j-i+j)}{(j-i)}  \prod_{1\leq i\leq j\leq k} \frac{(s_i+s_j-i-j+2k+1)}{(2k+1-i-j)}\,,
\end{align}
\esubeqs}\noindent
where the representation is defined by Young diagram $\mathbb{Y}(s_1,...,s_k)$ with the $i$-th row having length $s_i$ or $s_i-\tfrac12$ if all $s_i$ are half-integer. For some of the particular cases of use we find for $so(d)$:
{\allowdisplaybreaks\besubeqs\label{soddims}
\begin{align}
    \mathbb{Y}(s)&: && \frac{(d+2 s-2) \Gamma (d+s-2)}{\Gamma (d-1) \Gamma (s+1)}\,,\\
    \mathbb{Y}_{\tfrac12}(s)&: && \frac{\Gamma (d+s-1) 2^{\left[\frac{d}{2}\right]}}{\Gamma (d-1) \Gamma (s+1)}\,,\\
    \mathbb{Y}(a,b)&: && \frac{(a-b+1) (2 a+d-2) (2 b+d-4) (a+b+d-3) \Gamma (a+d-3) \Gamma (b+d-4)}{\Gamma (a+2) \Gamma (b+1) \Gamma (d-3) \Gamma (d-1)}\,,\\
    \mathbb{Y}_{\tfrac12}(a,b)&: && \frac{(a-b+1) (a+b+d-2) \Gamma (a+d-2) \Gamma (b+d-3) 2^{\left[\frac{d}{2}\right]}}{ (a+1)!  b! \Gamma (d-3) \Gamma (d-1)}\,,\\
    \mathbb{Y}(s,1^p)&: && \frac{(N+2 s-2) \Gamma (N+s-1)}{(p+s) \Gamma (p+1) \Gamma (s) (N-p+s-2) \Gamma (N-p-1)}\,,\\
    \mathbb{Y}(a,b,1^h)&: &&\frac{(a-b+1) (2 a+d-2) (2 b+d-4) (a+b+d-3) \Gamma (a+d-2) \Gamma (b+d-3)}{(a+h+1) a! (b+h) \Gamma (b) \Gamma (d-1) h! (a+d-h-3) (b+d-h-4) \Gamma (d-h-3)}\,,
\end{align}
\esubeqs}\noindent
where we use $\mathbb{Y}_{\tfrac12}(m_1,...)$ to denote spinorial representations. For example, $\mathbb{Y}_{\tfrac12}(m)$ is a symmetric rank-$m$ spin-tensor $T^{a(s);\alpha}$, i.e. it has spin $s=m+\tfrac12$. Similar formula for symplectic algebra $sp(N)$ yields:
\begin{align}
    \mathbb{Y}(a,b)&: && \frac{(a-b+1) (a+b+N-1) \Gamma (a+N-1) \Gamma (b+N-2)}{\Gamma (a+2) \Gamma (b+1) \Gamma (N-2) \Gamma (N)}\,,
\end{align}
which allows to compute the dimension of any representation of $so(5)\sim sp(4)$:
\begin{align}
    \mathbb{Y}(a,b)&: &&
\frac{1}{6} (3 + 2 a) (1 + a - b) (2 + a + b) (1 + 2 b)\,,\\
    \mathbb{Y}_{\tfrac12}(s)&: &&\frac{2}{3} (s+1) (s+2) (s+3)\,,
\end{align}
where $a$, $b$ can be half-integers. Analogously, for special linear algebra $sl(d)$:
\begin{align}
    \mathbb{Y}(a,b,c)&: && \frac{(b+c) \Gamma (b) c! (a+b-c-2) \Gamma (a-c-1) \Gamma (a+d) \Gamma (b+d-1) \Gamma (c+d-2)}{(a+2 b-2) \Gamma (d-2) \Gamma (d-1) \Gamma (d) \Gamma (a+b-1)}\,.
\end{align}
The isomorphism $su(4)\sim so(6)$ gives for $so(6)$:
\begin{align}\notag
    \mathbb{Y}(a,b,c)&: &&\frac{(2 a-2)! (a+b+3)! (a-c-1)! (a-c+2)! (a+c-2)! (b-c)! (b-c+1)! (a+b-2 c)}{12 (2 a-3)! (3 a+b-2 (c+1)) (2 a+b-c-2)!}\,.
\end{align}
Note that the dimension \eqref{dimweyl} in the even case $so(2k)$ is the dimension of irreducible representation, while \eqref{soddims} formulas pack (anti)-selfdual representations together, so that \eqref{soddims} sometimes gives twice that of \eqref{dimweyl}.

\paragraph{Characters.} We will discuss only one-particle partition-functions without extra chemical potentials. Character of a generic representation with spin $\mathbb{S}$ is obtained by counting $\pl^k$-descendants assuming there are no relations among them:
\begin{align}
    \chi_{\Delta,\mathbb{S}}&=\mathrm{dim}\, \mathbb{S} \times \frac{q^\Delta}{(1-q)^d}\,.
\end{align}
The characters of more complicated representations are obtained from the resolvent thereof. The simplest representations given by a short exact sequence correspond to partially-massless HS fields:
\begin{align}
    0\longrightarrow V(\Delta,\mathbb{S}') \longrightarrow V(\Delta-t,\mathbb{S}) \longrightarrow D(\Delta-t,\mathbb{S}) \longrightarrow0\,,
\end{align}
where $V(...)$ denotes generalized Verma module, which can be reducible, and $D$ is the irreducible module. Here, $\Delta=d+s_i-1-i$ and $\mathbb{S}'$ is the spin of the gauge parameter in $AdS_{d+1}$ or, equivalently, the symmetry type of the conservation law for a higher-spin current.\footnote{In the case of massless totally-symmetric fields we have
\begin{align}
    0\longrightarrow V(d+s-2,s-1) \longrightarrow V(d+s-2,s) \longrightarrow D(d+s-2,s) \longrightarrow0\,.
\end{align}}
An additional parameter $t$ is the depth of partially-masslessness \cite{Deser:2001us} and $t=1$ for massless fields. The Casimir Energy of a massless field is simply the difference between that of the two Verma modules --- field and its gauge symmetries. Generalization for long exact sequences is straightforward.

In the case of free scalar, $\Rac$, and free fermion, $\Di$, the sequence is short but different. The singular vectors are  associated with $\square\phi$ and $\slashed{\pl}\psi$:
\begin{align}
    \Rac&: &&     0\longrightarrow V(\tfrac{d+2}{2},0) \longrightarrow V(\tfrac{d-2}{2},0) \longrightarrow D(\tfrac{d-2}{2},0) \longrightarrow0\,,\\
    \Di&: &&    0\longrightarrow V(\tfrac{d+1}{2},\tfrac{1}{2}) \longrightarrow V(\tfrac{d-1}{2},\tfrac{1}{2}) \longrightarrow D(\tfrac{d-1}{2},\tfrac{1}{2}) \longrightarrow0\,.
\end{align}
Below we collect some of the blind characters of $so(d,2)$. The dimensions of irreducible $so(d)$ representations can be found above
\begin{align*}
    \chi(\phi_\Delta)&=(1-q)^{-d} q^{\Delta }\,, && \text{scalar of dimension } \Delta\,,\\
    \chi(Rac)&=\chi(\phi_\Delta)-\chi(\phi_{\Delta+2})\Big|_{\Delta=\frac{d-2}2}=\left(1-q^2\right) (1-q)^{-d} q^{\frac{d}{2}-1}\,,\\
    \chi(O_{\Delta,s})&=\frac{(1-q)^{-d} (d+2 s-2) q^{\Delta } \Gamma (d+s-2)}{\Gamma (d-1) \Gamma (s+1)}\,,&& \text{symmetric tensor operator}\,,\\
    \chi(J_s)&=\chi(O_{\Delta,s})-\chi(O_{\Delta+1,s-1})\Big|_{\Delta=d+s-2}\,, && \text{conserved tensor}\,,\\
    \chi(\psi_\Delta)&=(1-q)^{-d} q^{\Delta }2^{[\tfrac{d}2]}\,,&& \text{fermion of dimension } \Delta\,,\\
    \chi(Di)&=\chi(\psi_\Delta)-\chi(\psi_{\Delta+1})\Big|_{\Delta=\frac{(d-1)}{2}}\,.
\end{align*}
The simplest instance of the Flato-Fronsdal theorem then follows from
\begin{align}
    \chi^2(Rac)=\sum_s \chi(J_s)\,.
\end{align}
Given a character $Z(q=e^{-\beta})$, the (anti)-symmetric parts of the tensor product can be extracted in a standard way:
{\allowdisplaybreaks\begin{align}
    \text{symmetric}&: && \frac12 Z^2(\beta)+\frac12 Z(2\beta)\,,\\
    \text{anti-symmetric}&: && \frac12 Z^2(\beta)-\frac12 Z(2\beta)\,.
\end{align}}\noindent
The character of the weight-$\Delta$ spin-$(s,1^h)$ operator and the associated conserved current are:
\begin{align}
   \chi(O_{s,1^h})&=\frac{(1-q)^{-d} (d+2 s-2) q^{\Delta } \Gamma (d+s-1)}{(h+s) \Gamma (h+1) \Gamma (s) (d-h+s-2) \Gamma (d-h-1)}\,,\\
   \chi(J_{\Delta,s,1^h})&=\chi(O_{\Delta,s,1^h})-\chi(O_{\Delta+1,s-1,1^h})\Big|_{\Delta=d+s-2}\,.
\end{align}
Fermionic spin-tensor conformal quasi-primary operator $O_{\alpha;a(s)}$ obeys $\gamma^m{}\fud{\beta}{\alpha}O_{\beta;ma(s-1)}=0$, which allows to compute its character and the character of the conserved higher-spin super-current:
\begin{align*}
    \chi(O)&=\frac{(1-q)^{-d} q^{\Delta } \Gamma (d+s-1) 2^{\left[\frac{d}{2}\right]}}{\Gamma (d-1) \Gamma (s+1)}\,,\\
    \chi(J_s^F)&=\chi(O_{\Delta,s})-\chi(O_{\Delta+1,s-1})\Big|_{\Delta=d+s-3/2}=\frac{(1-q)^{-d} q^{d+s-\frac{3}{2}} (d-q s+s-2) \Gamma (d+s-2) 2^{\left[\frac{d}{2}\right]}}{\Gamma (d-1) \Gamma (s+1)}\,.
\end{align*}

\paragraph{Tensor Products of Spinors.} To derive the decomposition of $\Di\otimes \Di$ together with its (anti)-symmetric projections we need to know how to take tensor product of two $so(d)$ spinors. For $d$ odd we have Dirac spinors, which we denote $\sD$. For $d$ even there are two Weyl spinors, which we denote $\sW$ and $\sWb$.\footnote{Various other possibilities like  symplectic Majorana-Weyl spinors in some dimensions will be ignored.} There are three distinct cases: $so(2k+1)$, $so(4k)$ and $so(4k+2)$. Consulting math literature we can find out that:
\begin{align}
    so(2k+1)&: &&
    \left\{\begin{aligned}
    (\sD\otimes \sD)_S&=\bigoplus \Yy{1^{k-4i}}\oplus\Yy{1^{k-4i-3}}\\
    (\sD\otimes \sD)_A&=\bigoplus \Yy{1^{k-4i-1}}\oplus\Yy{1^{k-4i-2}}
    \end{aligned}\right.\\
    so(4k)&: &&
    \left\{\begin{aligned}
    (\sW\otimes \sW)_S&=\Yy{1^{2k}}_+\oplus\bigoplus \Yy{1^{2k-4i}}\\
    (\sW\otimes \sW)_A&=\bigoplus \Yy{1^{2k-4i-2}}\\
    (\sW\otimes \sWb)_{\phantom{A}}&=\bigoplus \Yy{1^{2k-2i-1}}\\
    \end{aligned}\right.  \\
    so(4k+2)&: &&
    \left\{\begin{aligned}
    (\sW\otimes \sW)_S&=\Yy{1^{2k+1}}_+\oplus\bigoplus \Yy{1^{2k+1-4i}}\\
    (\sW\otimes \sW)_A&=\bigoplus \Yy{1^{2k-4i-1}}\\
    (\sW\otimes \sWb)_{\phantom{A}}&=\bigoplus \Yy{1^{2k-2i}}\\
    \end{aligned} \right.
\end{align}
where the sums are from $i=0$ to the maximal value it can take in each of the cases.  
Defining in even dimensions $\sD=\sW\oplus\sWb$ we observe:
\begin{align}
   so(2k+1)&: & \sD\otimes\sD&=\bigoplus_{i=0} \Yy{1^{k-i}}\,,\\
   so(2k)&: &   \sD\otimes\sD&=\Yy{1^{k}}_+\oplus\Yy{1^{k}}_-\oplus2\bigoplus_{i=1} \Yy{1^{k-i}}\,.
\end{align}
The decomposition of $\Di\otimes \Di$ is known and is quoted in the main text. Let us work out the spectrum of the $O(N)$-singlet free fermion. In the case of even $d$ we introduce $\Wi$ as free Weyl fermion. It should be taken into account that higher-spin currents dress the tensor product $\bar{\psi}(x_1)\psi(x_2)$ with a Gegenbauer polynomial in derivatives that is (anti)-symmetric for (odd)even number of derivatives in the current. Combing the symmetry of the product of two spinorial representation with the symmetry of the derivative-dressing we find\footnote{Fermionic fields anti-commute, so $O(N)$-singlets belong to the anti-symmetric part of the tensor product $\Di\otimes\Di$. When dealing with SUSY HS theories one can refer to the anti-symmetric product as symmetric in the superalgebra sense.}
\begin{align}
    so(2k+1)&: &&
    \left\{\begin{aligned}
    (\Di\otimes \Di)_A=&\bigoplus \Yy{2n+1,1^{k-4i-1}}\oplus\Yy{2n+1,1^{k-4i-4}}\oplus\\ &\bigoplus \Yy{2n,1^{k-4i-2}}\oplus\Yy{2n,1^{k-4i-3}}
    \end{aligned}\right. \\
    so(4k)&: &&
    \left\{\begin{aligned}
    (\Wi\otimes \Wi)_A=&\Yy{2n+1,1^{2k-1}}_+\oplus\bigoplus \Yy{2n+1,1^{2k-4i-1}}\oplus\\
    &\bigoplus \Yy{2n,1^{2k-4i-3}}\oplus \begin{cases}
    \bullet\,, & k=2m+1\\ \emptyset\,,& k=2m \end{cases}
    \end{aligned}\right.  \\
    so(4k+2)&: &&
    \left\{\begin{aligned}
    (\Wi\otimes \Wi)_A=&\Yy{2n+1,1^{2k}}_+\oplus\bigoplus \Yy{2n+1,1^{2k-4i}}\oplus\\
                &\bigoplus \Yy{2n,1^{2k-4i-2}}
    \end{aligned} \right.
\end{align}
where we indicated the $so(d)$-spin of the singlet quasi-primary operators, the conformal weight being obvious from $\Di\otimes \Di$. The above formulae generalize the Flato-Fronsdal theorem to the $O(N)$-singlet sector of free fermion theory in any dimension. Other versions of the singlet constraint follow from the above results.

\section{Amusing Numbers}
\label{app:MrCasimir}
We collect below various numbers associated to the fields discussed in the main text: Casimir Energy, sphere free energy, Weyl $a$-anomaly coefficients.
\paragraph{Casimir Energy.}
Casimir Energy, $E_c$, is given by a formally divergent sum
\begin{align}
    E_c&=(-)^F \frac12\sum_n d_n\omega_n\,,
\end{align}
for which the standard regularization is to use the $\exp[ -\epsilon\omega_n]$ as a cut-off and then remove all poles in $\epsilon$. All the data can be extracted from the characters. We see that the spin degrees of freedom factor out for massive fields and the Casimir energy is given by
\begin{align*}
   (-)^F E_c(\chi_{\Delta,\mathtt{S}})&= \frac12 \mathrm{dim}\, \mathtt{S} \left.\sum \frac{\Gamma[d+n]}{n!\Gamma[d]} (\Delta+n) e^{-\epsilon(\Delta+n)} \right|_{\text{finite}}=\mathrm{dim}\, \mathtt{S} \left.\frac{ e^{-(\Delta +1) \epsilon } \left(d+\Delta(  e^{\epsilon }-1)\right)}{\left(1-e^{-\epsilon }\right)^{d+1} }\right|_{\text{finite}}\,.
\end{align*}
Casimir Energy for a massive scalar field of weight $\Delta$:
\begin{center}
\begin{tabular}{c|c}
     $d$ & $E_c$ \\ \hline
     $2$ & $\frac{1}{24} (\Delta -1) \left(2 \Delta ^2-4 \Delta +1\right) $ \\
     $3$ & $\frac{1}{480} \left(-10 \Delta ^4+60 \Delta ^3-120 \Delta ^2+90 \Delta -19\right) $ \\
     $4$ & $\frac{1}{1440} (\Delta -2) \left(6 \Delta ^4-48 \Delta ^3+124 \Delta ^2-112 \Delta +27\right) $  \\
     $5$ & $\frac{-84 \Delta ^6+1260 \Delta ^5-7350 \Delta ^4+21000 \Delta ^3-30240 \Delta ^2+19950 \Delta -4315}{120960} $  \\
     $6$ & $\frac{(\Delta -3) \left(12 \Delta ^6-216 \Delta ^5+1494 \Delta ^4-4968 \Delta ^3+8112 \Delta ^2-5904 \Delta +1375\right)}{120960} $
\end{tabular}
\end{center}
allows one to get the Casimir Energy for any massive representation by multiplying it by $\mathrm{dim}\,\mathbb{S}$. 
Formulas for massless representations are obtained as differences of the massive ones according to exact sequences. Some of the formulae below can be found in \cite{Gibbons:2006ij,Ozcan:2006jn}.\footnote{There is a typo in one of the expressions in the latter paper.} The Casimir Energies for higher-spin bosonic fields in lower dimensions are:
\begin{center}
\begin{tabular}{c|c}
     $d$ & $E_c$ \\ \hline
     $3$ & $\frac{1}{240} \left(30 s^4-20 s^2+1\right)$ \\
     $4$ & $-\frac{1}{1440} s (s+1) \left(18 s^4+36 s^3+4 s^2-14 s-11\right)$  \\
     $5$ & $\frac{(s+1)^2 \left(84 s^6+504 s^5+994 s^4+616 s^3-308 s^2-504 s-31\right)}{120960}$  \\
     $6$ & $-\frac{(s+1)^2 (s+2)^2 \left(12 s^6+108 s^5+338 s^4+408 s^3+32 s^2-282 s-31\right)}{483840}$  \\
\end{tabular}
\end{center}
Note that $d=3$ and $s=0$ case is special in that the fake ghost contribution does not vanish automatically and the right value is $E_c=\tfrac{1}{480}$. Casimir Energies for higher-spin fermionic fields in lower dimensions are:
\begin{center}
\begin{tabular}{c|c}
     $d$ & $E_c$ \\ \hline
     $3$ & $\frac{1}{240} \left(-30 s^4+20 s^2-1\right)$ \\
     $4$ & $\frac{(2 s+1)^2 \left(18 s^4+36 s^3-8 s^2-26 s+3\right)}{2880}$  \\
     $5$ & $-\frac{(2 s+1) (2 s+3) \left(84 s^6+504 s^5+910 s^4+280 s^3-532 s^2-280 s+11\right)}{241920}$  \\
     $6$ & $\frac{(2 s+1) (2 s+3)^2 (2 s+5) \left(12 s^6+108 s^5+314 s^4+264 s^3-144 s^2-162 s-3\right)}{1935360}$  \\
\end{tabular}
\end{center}
Note that $d=3$ and $s=\tfrac12$ the general formula does not oversubtract the fake descendants and the right value is still $E_c=\tfrac{17}{1920}$.
Casimir Energies for $\Rac$'s and $\Di$'s in lower dimensions $d=2,3,...$ are:\footnote{The fermion is always a Dirac one. $E_c$ for the Weyl fermion is half of the value in the table.}
\begin{align}
    E_c(\Rac)&=\left\{-\frac{1}{12},0,\frac{1}{240},0,-\frac{31}{60480},0,\frac{289}{3628800},0,-\frac{317}{22809600},0,\frac{6803477}{2615348736000}\right\}\,,\\
    E_c(\Di)&=\left\{-\frac{1}{24},0,\frac{17}{960},0,-\frac{367}{48384},0,\frac{27859}{8294400},0,-\frac{1295803}{851558400},0,\frac{5329242827}{7608287232000}\right\}\,.
\end{align}
Casimir Energies for massive $\Delta=d-1$ anti-symmetric tensors $\mathbb{Y}(1^h)$, $h=2,3,...$:\footnote{When self-duality applies it is the Casimir energy of the two fields.}
\begin{center}
\begin{tabular}{c|c|c|c}
     $d$ & $E_c$ \\ \hline
     $4$ & $-\frac{1}{20 h! \Gamma (5-h)} $  \\
     $5$ & $\frac{221}{1008 h! \Gamma (6-h)} $  \\
     $6$ & $-\frac{95}{84 h! \Gamma (7-h)} $
\end{tabular}
\end{center}
The Casimir Energies for massless hooks $\mathbb{Y}(s,1^p)$:
\begin{center}
\begin{tabular}{c|c}
     $d$ & $E_c,p=1$ \\ \hline
     $4$ & $\frac{1}{720} (-s (s+1) (2 s (s+1) (9 s (s+1)-22)+19)-3) $  \\
     $5$ & $\frac{3 s (s+2) (42 (s-1) s (s+2) (s+3) (2 s (s+2)+1)+221)+221}{120960} $  \\
     $6$ & $-\frac{(s+1) (s+2) (s (s+3) (2 s (s+3) (s (s+3) (6 s (s+3)-11)-54)+111)+95)}{120960} $
\end{tabular}
\end{center}

\paragraph{Sphere Free Energy.} Also, we will need the free energy on a sphere for free scalar and fermion, see e.g. \cite{Klebanov:2011gs},
\begin{align}
    F^3_\phi&=\frac{1}{16}(2\log 2-\frac{3\zeta(3)}{\pi^2})\,,
    && F^5_\phi=\frac{-1}{2^8}(2\log 2+\frac{2\zeta(3)}{\pi^2}-\frac{15\zeta(5)}{\pi^4})\,,\\
    F^3_\psi&=\frac{1}{16}(2\log 2+\frac{3\zeta(3)}{\pi^2})\,,
    && F^5_\psi=\frac{-1}{2^8}(6\log 2+\frac{10\zeta(3)}{\pi^2}+\frac{15\zeta(5)}{\pi^4})\,.
\end{align}

\paragraph{Weyl Anomaly.} The general formula for Weyl anomaly $a$ for real conformal scalar \cite{Casini:2010kt} and fermion \cite{Aros:2011iz} gives for $d=4,6,8,...$:\footnote{We changed normalization as compared to \cite{Aros:2011iz}.}
\begin{align}
    a_\phi&=\left\{\frac{1}{90},-\frac{1}{756},\frac{23}{113400},-\frac{263}{7484400},\frac{133787}{20432412000}\right\}\,,\\
    a_\psi&=\left\{\frac{11}{180},-\frac{191}{7560},\frac{2497}{226800},-\frac{14797}{2993760},\frac{92427157}{40864824000}\right\}\,.
\end{align}
\paragraph{Volumes.} The volume of $d$-sphere and the regularized volume of the hyperbolic space, which is Euclidean anti-de Sitter space, are \cite{Diaz:2007an}:
\begin{align}
    \mathrm{vol}\, S^d&=\frac{2 \pi ^{(d+1)/2}}{\Gamma \left(\frac{d+1}{2}\right)}\,, &&
    \mathrm{vol}\, \mathbb{H}^{d+1}=
        \begin{cases}
        \frac{2 (-\pi )^{d/2} }{\Gamma \left(\frac{d}{2}+1\right)}\log R\,, & d=2k\,,\\
        \pi ^{d/2} \Gamma \left(-\frac{d}{2}\right)\,, & d=2k+1\,.
        \end{cases}
\end{align}

\section{More HS Theories}
\label{app:strangeHS}
In this Section we discuss higher-spin doubletons that result in more general mixed-symmetry fields and higher-order singletons that lead to partially-massless fields and mixed-symmetry fields.

\subsection{Higher-Spin Doubletons}
\label{app:HSgletons}
An interesting possibility that $AdS_5$ offers (and more generally any $AdS_{2n+1}$, $n>1$) are higher-spin doubletons \cite{Gunaydin:1984fk,Gunaydin:1984wc,Metsaev:1995jp,Bekaert:2009fg,Fernando:2015tiu} as conformal fields in $CFT^{2n}$. These are parametrized by (half)-integer spin $J$, with $J=0,\tfrac12$ being the usual \Rac{} and \Di.\footnote{The Young diagram of $so(2n)$ that determines the spin of the field has a form of a rectangular block of length $J$ and height $n$, i.e. the labels are $(J,...,J)$. One can also consider higher-spin representations of more complicated symmetry type, however they may be  non-unitary.} The $J=1$ is free massless spin-one field, i.e. Maxwell. For $J>1$ the HS doubletons are unusual CFT's in not having a local stress-tensor, while they still are unitary representations of the conformal algebra.

In \cite{Beccaria:2014zma,Beccaria:2014xda} it was conjectured that there should exist an AdS HS theory that is dual to $N$ free Maxwell fields, called Type-C in analogy with Type-A, $J=0$, and Type-B, $J=\tfrac12$. It was found that one-loop tests are successfully passed, but already the non-minimal theory requires the bulk coupling to be $G^{-1}=2N-2$, i.e. modified. Similar conclusions were arrived at in \cite{Beccaria:2014qea} for the $J=1$ doubleton in $AdS_7/CFT^6$ \cite{Gunaydin:1984wc}.

Let us show that all Type-D,E,... theories, i.e. those with $J>1$, do not pass the one-loop test. The Casimir Energy of the spin-$J$ doubleton is easy to find:\footnote{For $J=0$ it gives the Casimir Energy of two real scalars. For lower spins $J=0,\tfrac12,1$ we therefore find $E_c=\tfrac{1}{240}, \tfrac{17}{960},\tfrac{11}{120}$. }
\begin{align}
    E_{c,J}&=\frac{1}{120} (-1)^{2 J} \left(30 J^4-20 J^2+1\right)\,.
\end{align}
The spectrum of Type-X theory can be found by evaluating the tensor product of two spin-$J$ doubletons \cite{Dolan:2005wy,Boulanger:2011se,Beccaria:2014zma}:
\begin{align}
    (J,0)\otimes(J,0)&=\sum_{k=0}^{2J}\Verma{2+2J}{k}{0}\oplus \sum_{k=1}\Verma{2+2J+k}{2J+\tfrac{k}2}{\tfrac{k}2}\,,\\
    (J,0)\otimes(0,J)&=\sum_{k=0}\Verma{2+2J+k}{J+\tfrac{k}2}{J+\tfrac{k}2}\,,
\end{align}
where in the first line we see massive and massless mixed-symmetry tensors and massless symmetric HS fields in the second line. The absence of the stress-tensor reveals itself in that the spectrum of massless HS fields is bounded from below by $2J$. In particular, there is no dynamical graviton for $J>1$.

The Casimir Energies for the three parts of the spectrum: massive, mixed-symmetry massless, and symmetric massless, can be computed with the net result:
\begin{align}
    E_c^J&=-\frac{1}{630} J (2 J-1) (2 J+1) \left(288 J^4-208 J^2-3\right)\,.
\end{align}
We see that the total Casimir energy vanishes for $J=0,\tfrac12$ in accordance with \cite{Giombi:2014yra}. It does not vanish for $J=1$ \cite{Beccaria:2014zma,Beccaria:2014xda}, rather it equals that of the two Maxwell fields, which still can be compensated by shifting the bulk coupling. However, for $J>1$ there does not seem to be any natural way of compensating the excess of the Casimir energy.

The same problem can be understood at the level of characters, which is a simpler approach. The blind character of the spin-$j$ doubleton is, see e.g. \cite{Beccaria:2014zma}:
\begin{align}
    Z_j&=\sum_k (2j+k+1)(k+1)q^{j+1+k}=\frac{(2 j (q-1)-q-1) q^{j+1}}{(q-1)^3}\,.
\end{align}
The singlet partition function is $[Z_j]^2$. It is symmetric in $\beta$, $q=e^\beta$, for $j=0,\tfrac12$. For $j=1$ it is not symmetric but the anti-symmetric part can be expressed as a multiple of $Z_1$, which can be compensated by modifying $G^{-1}=N$ \cite{Beccaria:2014zma}. However, for $j>1$ the anti-symmetric part cannot be compensated this way, but can be expanded in terms of $Z_{i\leq j}$.

Therefore, we see that the duals of HS doubletons $J>1$ should have pathologies as quantum theories. Classically, it should be possible to manufacture some interaction vertices in $AdS$ such that they reproduce the correlation functions of conserved HS currents
\begin{align}
    \langle j_{s_1}...j_{s_k}\rangle&=\text{Witten diagrams}\,, && s_i\geq 2J\,.
\end{align}
The generating function of three-point correlators was constructed in \cite{Zhiboedov:2012bm}. That such reconstruction is possible for three-point function follows from counting the number of independent structures that can contribute to $\langle j_{s_1}j_{s_2}j_{s_3}\rangle$ \cite{Costa:2011mg} and to the cubic vertex $V_{s_1,s_2,s_3}$ of three massless HS fields \cite{Metsaev:1993ap,Metsaev:2005ar}. This number is the same $n=\min(s_1,s_2,s_3)+1$ and is given by the minimal spin, which is related to the fact that the currents that one can construct from a spin-$J$ doubleton must have $s\geq 2J$, see \cite{Gelfond:2006be} for the explicit form in $4d$. Indeed, only those doubletons can give a contribution to $\langle j_{s_1}j_{s_2}j_{s_3}\rangle$ that have $2J\leq \min (s_1,s_2,s_3)$.

The above considerations pose a puzzle: we see that most of the cubic vertices that exist in principle cannot be a part of any consistent unitary HS theory.\footnote{HS doubletons exist for even boundary dimension only. However, the number of independent correlators $\langle j_{s_1}j_{s_2}j_{s_3}\rangle$ seems to be indifferent to this fact, as if one could formally define HS doubletons in odd dimensions as well.} In \cite{Boulanger:2011se} it was shown that deformations of HS spin algebras in any $d$  that are consistent with unitarity in the sense that gauging of such algebras leads to unitary (mixed-symmetry) fields can depend on at most one continuous parameter. In references \cite{Govil:2014uwa,Govil:2013uta,Fernando:2015tiu} a one-to-one correspondence between $AdS_{d+1}/CFT_d$ HS algebras and their  deformations and supersymmetric extensions and the  massless unitary representations of conformal algebras and superalgebras in  $d$ dimensional Minkowskian space-times was established. Only in $d=4$ is the deformation  parameter  continuous \cite{Fradkin:1989md,Fernando:2009fq,Govil:2013uta,Manvelyan:2013oua} corresponding to helicity \cite{Govil:2013uta}, while in $d>4$ deformations are discrete. The HS algebras resulting from HS doubletons belong to this family as well. We see that restriction to HS doubletons with spin $0,\tfrac12,1$ eliminates a considerable part of the mixed-symmetry fields. Therefore, only very restricted Young shapes can arise in HS theories with massless mixed-symmetry fields --- no more than two columns of height greater than one. Still a large fraction of massless mixed-symmetry fields is not embedded in any kind of AdS/CFT duality. Perhaps, they can be brought to existence as duals of non-unitary spinning conformal fields $\phi^{\mathbb{S}}$ that obey $\square \phi^{\mathbb{S}}+...=0$. Massive mixed-symmetry fields of any admissible Young shape are present in string theory, so it should be important to be able to incorporate massless limits thereof into HS theories.

\subsection{Partially-Massless Fields}
\label{app:PMfields}
As it was already noted, the list of free CFT's becomes  infinitely richer if the unitarity is abandoned. The simplest one-parameter family corresponds to higher-order singletons:
\begin{align}
    \Rac_k&: & \square^k \phi&=0\,, &&\Delta=\frac{d}2-k\,.
\end{align}
The spectrum of single-trace operators contains partially-conserved currents \cite{Dolan:2001ih}
\begin{align}
    J_s&=\phi\square^i \pl^s\phi+...\,, && \pl^{k-i}\cdot J_s=0\,.
\end{align}
The spectrum is encoded in the tensor product of two $\Rac_k$ \cite{Bekaert:2013zya}:
\begin{align}
    \Rac_k\otimes \Rac_k &= \sum_{s=0}^{\infty}\sum_{i=1}^{i=k} D(d+s-2i,s)\,.
\end{align}
The fields that are dual to partially-conserved currents are partially-massless fields \cite{Deser:2001us,Skvortsov:2006at}:
\begin{align}
    \pl^m...\pl^m J_{m(t)a(s-t)}&=0 &&\Longleftrightarrow && \delta \Phi^{\ua(s)}=\nabla^\ua...\nabla^{\ua}\xi^{\ua(s-t)}+...\,,
\end{align}
where $t$ is the depth of partially-masslessness. Massless fields occur at $t=1$. Therefore, the spectrum of a theory that is dual to $\Rac_k$ is a nested tower of (partially)-massless fields with the $\Rac_{k-1}$ tower contained in the $\Rac_k$ one. In particular, usual massless HS fields are present. Note that only odd depths $t$ are found in $\Rac_k\otimes \Rac_k$.

We can call the dual of $\Rac_k$ as Type-$A_k$, which is not meaningless for the following reason \cite{Alkalaev:2014nsa}.
One can define HS algebra for the generalized free field of weight-$\Delta$. This algebra is naturally described as a centralizer of $hs(\lambda)$,\footnote{$hs(\lambda)$ is defined as a quotient of $U(sl(2))$ by the two-sided ideal generated by $C_2-\lambda$, \cite{Vasiliev:1989re,Feigin}. It 'interpolates' between matrix algebras.} where $\Delta$ is related to $\lambda$. The HS algebras defined by $\Rac_k$ can be understood as quotients of this algebra that arise at exactly the same values where the dual algebra $hs(\lambda)$ acquires and ideal and reduces to $sl(k)$. Therefore, the duals of $\Rac_k$ are related to the $A$-series of Lie algebras. The (anti)-symmetric parts of $\Rac_k\otimes \Rac_k$ should then be related to the $B,C,D$ series  of algebras.

It is important that the operators with $s<i$ are not conserved tensors and are dual to massive fields, which for $k>2$ also contain massive HS fields. Therefore, duals of $\Rac_k$ provide an example of HS theories that contain HS gauges fields and HS massive fields with a spin bounded from above.

As a simple test of the AdS/CFT duality we can check the vanishing of Casimir Energy in the non-minimal Type-$A_k$ theory, see also \cite{Basile:2014wua}. On general grounds the Casimir Energy of $\Rac_k$ vanishes in odd dimensions. For example, for the simplest case of $\Rac_t$ we find in $d=3,4,...$:{\footnotesize
\begin{align*}
    E_c&=\{0,-\frac{1}{720} t \left(6 t^4-20 t^2+11\right),0,-\frac{t \left(12 t^6-126 t^4+336 t^2-191\right)}{60480},0,-\frac{t \left(10 t^8-240 t^6+1764 t^4-4320 t^2+2497\right)}{3628800}\}
\end{align*}}\noindent
The Casimir Energy of a depth-$t$ partially-massless spin-$s$ field can be computed in a standard way. For example, in the  $d=3$ case we find ($g=2s+1$):
\begin{align}
    E_c&=\frac{t \left(5 g (g-2 t) \left(3 g^2-6 g t+4 t^2-6\right)-17\right)}{1920}\,.
\end{align}
Consider the simplest case of $\Rac_2$. The spectrum contains that of Type-A and massive fields $\Phi$, $\Phi_\ua$, $\Phi_{\ua\ua}$ plus depth-$3$ partially-massless fields $s=3,4,...$. The sum over the Type-A spectrum was already found to vanish \cite{Basile:2014wua}. At least for odd $d$ we have to ensure that the sum over the rest vanishes as well. Using the standard exponential cut-off $\exp[-\epsilon(s+x)]$ we find that this is the case for $x=(d-5)/2$. Therefore, different parts of the spectrum should be summed with different regulators.

The dual of $\Rac_3$ contains the spectrum of Type-$A$=Type-$A_1$, the fields we have just studied plus massive fields $\Phi_{\ua(k)}$, $k=0,1,2,3,4$ and depth-$5$ partially-massless fields. The sum of the Casimir Energies of this last part gives zero for $x=(d-7)/2$.

Let us turn to the minimal Type-$A_k$ theory. It is useful to recall that the Casimir Energy can also be computed as
\begin{align}
    E_c&=(-)^F\frac12 \zeta(-1)\,, &\zeta(z)&=\frac{1}{\Gamma(z)}\int \beta^{z-1}d\beta\, Z(q=e^{-\beta})\,.
\end{align}
As it was already noted \cite{Giombi:2014yra}, the non-zero contribution to $E_c$ comes from the $\beta^{-1}$ pole, which is absent if $Z(\beta)$ is an even function of $\beta$. This is typically the case for the tensor product of two singletons, but is not for the (anti)-symmetric projections, which results in
\begin{align}
    Z_{\text{sing}}&=\frac12 Z^2(\beta)\pm \frac12 Z(2\beta)\,,
\end{align}
where the first term is an even function of $\beta$ in most cases. Then the contribution to the Casimir Energy is equal to that of the free field due to the last term. A slight  generalization of  \cite{Basile:2014wua,Bekaert:2013zya} implies that the minimal type-$A_2$ contains fields of even spins only. The excess of the Casimir Energy can be reduced to a linear combination of $\Rac_k$ by expressing the $\beta$-odd part of $(\Rac_k\otimes\Rac_k)_S$:
\begin{align}
    \beta-\text{odd part}\left[(\Rac_k\otimes \Rac_k)_S- \frac12 Z_k(2\beta)\right]&=0\,,
\end{align}
where $Z_k$ is the character of $\Rac_k$:
\begin{align}
    Z_k(q)&=(1-q)^{-d} \left(1-q^{2 k}\right) q^{\frac{1}{2} (d-2 k)}\,.
\end{align}
This identity directly implies that the Casimir energy of the minimal type-$A_k$ theory is equal to that of one $\Rac_k$, $E_c^k$. If instead we sum over spins with $\exp[-\epsilon(s+x)]$ cut-off we will have to use $x=(d-3)/2$ for depth-$1$ fields, $x=(d-5)/2$ for depth-$2$ fields etc. In particular, for type-$A_2$ the sum over its type-$A$ sub-sector gives $E_c$ of $\Rac_1$, while the sum over the depth-$2$ fields gives $E^2_c-E^1_c$ with the total result $E^2_c$, as before.

Also, it can be checked that the tensor product $\Rac_n\otimes \Rac_m$ with $m\neq n$ gives zero contribution to the Casimir Energy. Such products should arise in a theory built of several different higher-order singletons.

With the help of the zeta-function we can also check that $-2^{-1}\zeta'(0)$ matches the $a$-anomaly of $\square^k \phi=0$ free field. The latter can be extracted from the same zeta-function according to $a_{CHS}=-2a_{HS}$ where the conformal field dual to the order-$k$ singleton has weight $(d+2k)/2$. The summation over spins can be done as before and we should not forget that the depth-$t$ partially-massless field of spin-$s$ has $AdS$ energy $\Delta=d+s-t-1$ and the ghost has spin $(s-t)$ and weight $d+s-1$. Lastly, the contribution of the massive (possibly HS fields) that appear in the tensor product of two higher-order singletons need to be separated. For example, let us consider $AdS_5$ and set $k=2$ as above. We find:
\begin{align}
    \zeta'_{Type-A}(0)&=0\,, &\zeta'_{PM}(0)&=\frac{{\log R}}{15}\,, &\zeta'_{\text{massive}}(0)&=-\frac{{\log R}}{15}\,,
\end{align}
so that the total contribution is zero. For the minimal Type-$A_2$ model, i.e. the one above truncated to even spins only, we have:
\begin{align}
    \zeta'_{min, Type-A}(0)&=-\frac{{\log R}}{45}\,, &\zeta'_{PM, even}(0)&=\frac{{\log R}}{3}\,, &\zeta'_{\text{massive, even}}(0)&=\frac{14\, {\log R}}{45}\,,
\end{align}
the total contribution being $-2^{-1}\zeta'(0)=-\tfrac{1}{45} (14\, {\log R})$, which is exactly the value of the zeta-function
\begin{align}
     \frac{1}{180} (\Delta -2)^3 {\log R} (s+1)^2 \left(5 (s+1)^2-3 (\Delta -2)^2\right)
\end{align}
at $s=0$ and $\Delta=(d+4)/2$. Using the explicit form of $\zeta'(0)$ for $d=2k$ it is easy to extract the $a$-anomaly of higher-order singletons.

Therefore, despite non-unitarity, higher-order singletons that lead to partially-massless fields seem to be consistent at one-loop.

\section{On the Computations in Even Dimensions}
\label{app:generalzetaHS}
\setcounter{equation}{0}
In this Section we discuss the computations of $\zeta$ and $\zeta'$ in even dimensions. We presume that the full zeta-function is given in the form
\begin{align}
    \zeta(z)&=\int_0^\infty du\, \frac{\tilde{\mu}(u)}{[u^2+\nu^2]^z} h(u)\,, && \tilde{\mu}(u)=\sum_k \mu_k u^k\,,
\end{align}
where $\nu=\Delta-d/2$ and $h(u)$ is either $\tanh \pi u$ or $\coth{\pi u}$. The computation of $\zeta(0)$ can be done by using
\begin{align}
    \tanh x&=1+\frac{-2}{1+e^x}\,, &  \coth x&=1+\frac{2}{-1+e^x}\,,
\end{align}
which leads to
\begin{align}
    \zeta(z)&=\int_0^\infty du\, \frac{\tilde{\mu}(u)}{[u^2+\nu^2]^z}\mp2\int_0^\infty du\, \frac{\tilde{\mu}(u)}{[u^2+\nu^2]^z(e^{2\pi u}\pm1)}=I+II\,.
\end{align}
The first integral can be done for large enough $z$ and then continued to $z=0$. The second one is perfectly convergent and we can set $z=0$ and use
\begin{align}
  \int  \frac{-2u^k}{e^{2\pi u}+1}&= -4^{-k} \left(2^k-1\right) \pi ^{-k-1} \zeta (k+1) \Gamma (k+1) \label{D.4}\,,\\
  \int  \frac{2u^k}{e^{2\pi u}-1}&=2^{-k} \pi ^{-k-1} \text{Li}_{k+1}(1) \Gamma (k+1) \label{D.5}\,.
\end{align}
To compute $\zeta'(0)$ we first differentiate $\zeta(z)$ with respect to $z$. This can be directly done for the first part $I$, with two contributions produced:
\begin{align}
  \frac{\pl}{\pl z} I\Big|_{z=0}&= p_1(\nu)+ \log \nu \times p_2(\nu)\,,
\end{align}
where $p_{1,2}$ are polynomials. In the second part $II$ we find no problem with convergence, but a quite complicated integral
\begin{align}
  \frac{\pl}{\pl z} II\Big|_{z=0}&=\pm 2\int_0^\infty du\, \frac{\tilde{\mu}(u)\log[u^2+\nu^2]}{(e^{2\pi u}\pm1)}\,.
\end{align}
Using $\log[u^2+\nu^2]=\log u^2+ \int_0^{\nu} dx\, 2x(x^2+u^2)^{-1}$ we can split it into two parts:
\begin{align}
  II.1&=\pm2\int_0^\infty du\, \frac{\tilde{\mu}(u)\log[u^2]}{(e^{2\pi u}\pm1)}=\pm 2\sum_k \mu_k c^{\pm}_k\,,\\
    II.2&=\pm 2\int_0^\infty du\, \frac{\tilde{\mu}(u)}{(e^{2\pi u}\pm1)}\int_0^{\nu} dx\, \frac{2x}{(x^2+u^2)}\,.
\end{align}
Now we introduce two types of auxiliary integrals
\begin{align}
  c^{\pm}_n&=\int_0^\infty du\, \frac{u^n\log[u^2]}{(e^{2\pi u}\pm1)}\,, &
  J^{\pm}_n&=\int_0^\infty du\, \frac{u^n}{(x^2+u^2)(e^{2\pi u}\pm1)}\,.
\end{align}
The first one we will not attempt to evaluate since all $c_n$ will cancel in the final expressions. The second one can be done iteratively by first finding
\begin{equation}
    J_1^{\pm}=\int_0^{\infty}du\frac{du}{(x^2+u^2)(e^{2 \pi u}\pm 1)}\,,
\end{equation}
where in [3.415, Table of integral],
\begin{equation}
    J_1^-=\int_0^{\infty}\frac{udu}{(u^2+x^2)(e^{2\pi u}-1)}=\frac{1}{2}\left(\log(x)-\frac{1}{2x}-\psi(x)\right)\,.
\end{equation}
Together with a useful formula in \cite{Camporesi:1991nw}, $J_n^+(2\pi)=J_n^-(2\pi)-2J_n^-(4\pi)$, one can get
\begin{equation}
    J_1^+ = \frac{1}{2}\psi(x+1/2)-\frac{1}{2}\log x\,.
\end{equation}
Consider the following equation
\begin{equation}
    \int_0^{\infty} \frac{u^ndu}{e^{2 \pi u}\pm 1}\log(a u^2 + x^2) = \log a \int_0^{\infty}\frac{u^ndu}{e^{2\pi u}\pm 1}+ \int_0^{\infty} \frac{u^ndu}{e^{2\pi u}\pm 1}\log(u^2+x^2/a)\,.
\end{equation}
Taking the derivative at $a=1$ on both sides, we obtain
\begin{equation}
    J_{n+2}^{\pm}=\int_0^{\infty}\frac{u^ndu}{e^{2\pi u}\pm 1}-x^2 J_n^{\pm}\,.
\end{equation}
Therefore, $J_n^{\pm}$ will contain two types of contributions:
\begin{align}
  J_n^+&= q_n^+(x) \psi(x+1/2)+[\tilde{p}^+_2(x)\log x+\tilde{p}_3^+(x)]\,,\\
  J_n^-&=q_n^-(x)\psi(x)+[\tilde{p}^-_2(x)\log x+\tilde{p}_3^-(x)]\,.
\end{align}
The second terms in each equation can be easily integrated over $x$:
\begin{align}
\pm 2\int_0^{\nu} dx\,2x [\tilde{p}^{\pm}_2(x)\log x+\tilde{p}^{\pm}_3(x)]=p_3(\nu)-p_2(\nu)\log \nu\,.
\end{align}
Importantly, all $\log \nu$ now cancel because $p_2(\nu)$ is the same as the one at $\partial_z I\big|_{z=0}$. The purely polynomial leftovers $p_1$ and $p_3$ from $J_n^{\pm}$ and $\partial_z I\big|_{z=0}$ can be added up. We also need to add $II.1$ to them. Then $\nu$ is replaced with $\Delta-d/2$ and we can sum over all spins as usual. This contribution we call $P=\sum P_{\nu,s}-P_{\nu+1,s-1}$. Importantly, all coefficients $c_n$ will be gone and we do not need to deal with their real form, both for Type-A and Type-B.

Now we are left with the contribution that we call $Q=\sum Q_{\nu,s}-Q_{\nu+1,s-1}$, which consists of either $\psi(x+1/2)$ or $\psi(x)$ times a polynomial in $x$, where
\begin{align}
    Q_{\nu,s}&=\ \ \ 4 \ \ \sum_{s,k}\int_0^{\Delta -d/2} dx\,\mu_k q_k(x)\psi(x+1/2)\,,\quad (\text{for bosons})\,,\\
    Q_{\nu,m}&=-4 \sum_{s=m+\frac{1}{2},k}\int_0^{\Delta -d/2} dx\,\mu_k q_k(x)\psi(x)\,,\qquad \quad (\text{for fermions})\,.
\end{align}
It can be simplified by using the integral representation for $\psi(x)$:
\begin{align}
\psi(x)&=\int_0^{\infty} dt\,\left[\frac{e^{-t}}{t}-\frac{e^{-t x}}{1-e^{-t}}\right] \label{D.21}\,.
\end{align}
Next, the integral over $x$ can be done and the sum over the spectrum is taken. As a result we are left with
\begin{align}
Q&=\sum f^{n,m}_{a,b,c}\int dt\, \frac{e^{bt} t^a}{(1-e^{-t})^{n+1}(1+e^{-t})^{m+1}}\,.
\end{align}
The summands can be expressed as derivatives at $z=1$ and $z=-1$ of Hurwitz-Lerch function \cite{Giombi:2013fka,Giombi:2014iua}
\begin{equation}
    \Phi(z,s,\nu)=\frac{1}{\Gamma(s)}\int_0^{\infty}dt \frac{t^{s-1}e^{\nu t}}{1-ze^{-t}}\,,
\end{equation}
which in return, can be analytically continued into Hurwitz zeta function $\zeta(s,\nu)$. It is worth noting that only in the minimal higher-spin theories there will be $(1+e^{-t})^m$ in the denominator. Using this zeta regularization scheme, we will display the results of for HS theories in different even dimensions, which are subdivided into four categories in the following appendices: Type-A (non-minimal and minimal), HS fermions, Hook fields and the result for Hooks and Type-A can be added up to get Type-B theories (non-minimal and minimal). The case of $AdS_6$ is presented in more detail while for other dimensions we only show the main intermediate steps.

\section{Zeta Function in \texorpdfstring{$AdS_6$}{AdS(6)}}
\setcounter{equation}{0}
First of all, let us show explicitly how to calculate the zeta function in $AdS_6$ for Type-A, fermionic HS theory, hook fields and Type-B.
\subsection{Type-A}
\paragraph{Zeta.} Starting with Vasiliev type A theory, we recall the zeta-function in the main text
\begin{equation}
     \tilde{\mu}(u)=-\frac{u  \left(u ^2+\frac{1}{4}\right) (s+1) (s+2) (2 s+3) \tanh (\pi  u ) \left(u ^2+\left(s+\frac{3}{2}\right)^2\right)}{720 }\,.
\end{equation}
With $\tanh x= 1 - \frac{2}{e^{2\pi x}+1}$, we can write the spectral zeta function as
\begin{equation}
\begin{split}
\zeta^{\mathbb{H}}(z) &= -\frac{1}{720} (s+1)(2s+3)(s+2) \Bigg[\lim_{z\rightarrow 0} \int_0^{\infty} du \frac{u(u^2+1/4)\left(u^2+(s+3/2)^2\right)}{(u^2+\nu^2)^z}  \\
&- 2\int_0^{\infty} du \frac{u(u^2+1/4)\left(u^2+(s+3/2)^2 \right)}{(1+e^{2\pi u})} \Bigg]\,.
\end{split}
\end{equation}
Using \eqref{D.4}, one can obtain easily the zeta function for the Type-A HS theory \cite{Giombi:2014iua}
\begin{equation} \label{E.3}
\begin{split}
\zeta_{(\Delta,s)}(0) &=- \frac{(s+1)\left(2s+3\right)(s+2)}{29030400}\Big[-1835-714s(s+3) \\&-420 \nu^2(27-60\nu^2+16\nu^4+s(36-72\nu^2)+s^2(12-24\nu^2)) \Big]\,.
\end{split}
\end{equation}
The total contribution from HS fields and ghosts is
{\allowdisplaybreaks
\begin{align}
    \zeta^{Type-A}(0) &= \sum_{s=0}^{\infty}\zeta_{(\Delta,s)}(0) - \zeta_{(\Delta+1,s-1)}(0)\notag\\  \label{E.4} &=\zeta_{(3,0)}+\sum_{s=1}^{\infty} \zeta_{(\Delta,s)} - \zeta_{(\Delta+1,s-1)}\\
    &=\frac{1}{1512} -\sum_{s=1}^{\infty} \frac{(1+s)^2(-20+28s+378s^2+868s^3+847s^4+378s^5+63s^6)}{30240}\,,\notag
    \end{align}}\noindent
where $\Delta=s+3$ and $\nu=s+\frac{1}{2}$. We use the exponential cut-off $\text{exp}[-\epsilon(s+\frac{d-3}{2})]$ to take the summation with $d=5$. A straightforward calculation shows that
\begin{equation}
    \zeta^{Type-A}=\zeta^A=0\,.
\end{equation}
The vanishing of zeta function is also true for the minimal Type-A theory, where $s=0,2,...$.
\begin{eqnarray}
    \zeta^{Type-A}_{min}=\zeta^A_{min}=\zeta_{(3,0)}+\sum_{s=2,4,...}^{\infty}  \zeta_{(\Delta,s)} - \zeta_{(\Delta+1,s-1)} = 0\,.
\end{eqnarray}

\paragraph{Zeta-prime.} After making sure that the  conformal anomaly does not contribute to the free energy, we now can take the $z$-derivative of $\zeta$ at $z=0$ to calculate $\zeta'(0)$. One can easily obtain
\begin{equation}\notag
\begin{split}
    \zeta'(0)&=-\frac{(s+1)(s+2)(2s+3)}{720}\Bigg[\frac{1}{288}\nu^2\Big(-81+270\nu^2-88\nu^4+108s(-1+3\nu^2)+36s^2(-1+3\nu^2)\\
    &+3\big(27-60\nu^2+16\nu^4+s(36-72\nu^2)+s^2(12-24\nu^2)\big)\log(\nu^2)\Big)\\
    &+2 \int_0^{\infty}du \frac{u(u^2+\frac{1}{4})(u^2+(s+\frac{3}{2})^2)\log(u^2)}{e^{2\pi u}+1}
    +4\int_0^{\infty} du \int_{0}^{\nu}dx x\frac{u(u^2+\frac{1}{4})(u^2+(s+\frac{3}{2})^2)}{(e^{2\pi u}+1)(u^2+x^2)} \Bigg]\,.
    \end{split}
\end{equation}
Following Appendix \ref{app:generalzetaHS}, the first integral is therefore
\begin{equation}
    II.1=-\frac{(s+1)(s+2)(2s+3)}{360}\left[c_5^+ +c_3^+ \left(\frac{1}{4}+\left(s+\frac{3}{2}\right)^2\right)+\frac{c_1^+}{4}\left(s+\frac{3}{2}\right)^2\right]\,.
\end{equation}
The second integral is just
{\allowdisplaybreaks
\begin{align*}
    II.2&=-\frac{(s+1)(s+2)(2s+3)}{180}\int_0^v dx x \left(J_5^+ +\left(\frac{1}{4}+\left(s+\frac{3}{2}\right)^2\right)J_3^+ +\frac{1}{4}\left(s+\frac{3}{2}\right)^2 J_1^+ \right) \\
    &=-\frac{(s+1)(s+2)(2s+3)}{720}\Bigg[\frac{1}{2880} \nu^2 \Big(3 (377 + 160 s (3 + s)) - 120 (8 + 3 s (3 + s)) \nu^2 +
    160 \nu^4\\
    &+
    60 (-3 (3 + 2 s)^2 + 12 (5 + 2 s (3 + s)) \nu^2 - 16 \nu^4) \log(\nu)\Big)\\ &-\frac{1}{8}\int_0^{\nu} x (9 + 12 s + 4 s^2 - 4 x^2) (-1 + 4 x^2) \psi(1/2 + x)\Bigg]\,.
    \end{align*}}\noindent
It is easy to see that the $\log$ constribution in (E.7) and (E.9) cancel each other. In the end, we are left with
\begin{equation}
    \zeta'^A(0)= P_{\nu,s}+Q_{\nu,s}\,,
\end{equation}
where,
\begin{equation}
\begin{split}
    P_{\nu,s}&=-\frac{(s + 1) (s + 2) (2 s + 3)}{720} \Bigg[\frac{
   \nu^2 (107+580\nu^2-240\nu^4+120s(1+6\nu^2)+40s^2(1+6\nu^2))}{960}\\
      &+ \frac{ c_1^+}{2} \big(s + \frac{3}{2}\big)^2 +
   2 c_3^+ \big(\big(s + \frac{3}{2}\big)^2 + 1/4\big) + 2 c_5^+\Bigg]\,,
   \end{split}
   \end{equation}
   \begin{equation}
       Q_{\nu,s}=\frac{(s+1)(s+2)(2s+3)}{5760}\int_0^{\nu} x (9 + 12 s + 4 s^2 - 4 x^2) (-1 + 4 x^2) \psi(1/2 + x)\,.
   \end{equation}
Using the cut-off method, the evaluation of $P=\sum_s P_{\nu,s} - P_{\nu+1,s-1}$ in the case of all spins and in the case of even spins only leads to the same result of zero, i.e the contribution of $P_{\nu,s}$ to $\zeta'(0)$ vanishes for both cases. The evaluation of $Q_{\Delta,s}$ is a little bit harder if one wishes to obtain an analytical result. We write the di-gamma function in its integral representation \eqref{D.21} and obtain
\begin{equation}
     Q = \sum_{s=0}^{\infty}Q_{\nu,s}-Q_{\nu+1,s-1} =0\,.
\end{equation}
Hence,
\begin{equation}
    \sum_{s=1}^{\infty}Q_{\nu,s}-Q_{\nu+1,s-1}=-Q_{\frac{1}{2},0}\,,
\end{equation}
where,
\begin{equation}
    Q_{\frac{1}{2},0}=-\frac{1}{120}\left(\frac{1181}{11520} - \frac{211 \log(2)}{4032} - \frac{23 \log A}{16} + \frac{5 \zeta(3)}{
 4 \pi^2} + \frac{15 \zeta(5)}{4 \pi^4} -
 \frac{63}{16} \zeta'(-5) + \frac{35}{8} \zeta'(-3)\right)
\end{equation}
here, $A=e^{\frac{1}{12}-\zeta'(-1)}$ is the Glaisher-Kinkelin constant. Above, we used the exponential cut-off $\text{exp}[-\epsilon \nu]$ to evaluate the sum over all spins.
For minimal Type-A theory, a straightforward calculation shows that the $\zeta'(0)_{min}$ is just
\begin{equation}
   \zeta'(0)_{min}= Q_{\frac{1}{2},0}+\sum_{s=2,4...} Q_{s+\frac{1}{2},s}-Q_{s+\frac{3}{2},s-1} = \frac{1}{2^7}\left(2\log 2 + \frac{2\zeta(3)}{\pi^2}-\frac{15\zeta(5)}{\pi^4}\right)=-2F^\phi_5\,,
\end{equation}
where,
\begin{equation}
\begin{split}
    \sum_{s=2,4...}Q_{s+\frac{1}{2},s}-Q_{s+\frac{3}{2},s-1} &=-\frac{1}{180}\Bigg[-\frac{1181}{7680} - \frac{7349 \log(2)}{2688} + \frac{69 \log A}{32} - \frac{
 75 \zeta(3)}{16 \pi^2} \\
 &+ \frac{495 \zeta(5)}{32 \pi^4} +
 \frac{189}{32}\zeta'(-5) -\frac{105}{16} \zeta'(-3)\Bigg]\,.
 \end{split}
\end{equation}

\subsection{Fermionic HS fields}

\paragraph{Zeta.} Above, we showed explicitly how to evaluate the zeta-function for the Type-A case. For fermionic HS fields, the computation is similar with the change of variable $s=m+1/2$. We recall the spectral function for fermions from the main text
\begin{eqnarray}
   \tilde{\mu}(u)= -\frac{u  \left(u ^2+1\right) \left(s+\frac{1}{2}\right) \left(s+\frac{3}{2}\right) \left(s+\frac{5}{2}\right) \coth (\pi  u ) \left( u^2+\left(s+\frac{3}{2}\right)^2\right)}{180 }\,.
\end{eqnarray}
We write  $s= m+1/2$, so that we can take the sum from $m=0$ to $\infty$. The degeneracy becomes
\begin{equation}
    g(m)\sim (m+1)(m+2)(m+3)\,.
\end{equation}
As we shall see the overall normalization factor does not affect the final result for fermions. Using \eqref{D.5}, we get
{\footnotesize    \begin{align*}
    \zeta_{\frac{1}{2}} \sim \sum_{m=0}^{\infty}\frac{1}{168} (-542 - 99 m + 8094 m^2 + 22806 m^3 + 28497 m^4 + 19404 m^5 +
   7448 m^6 + 1512 m^7 + 126 m^8) = 0\,.
\end{align*}}\noindent
\paragraph{Zeta-prime.} To find $\zeta'_{\frac{1}{2}}$, the integral that one needs to evaluate is
\begin{equation}
\begin{split}
        &\quad \partial_{z}\Big|_{z=0}g(m) \int_0^{\infty} \frac{u(u^2+1)(u^2+(m+2)^2)}{(\nu^2 + u^2)^z}\left(1+\frac{2}{e^{2 \pi u} -1}\right)\\
        &\sim \partial_{z}\Big|_{z=0} \left(\int_0^{\infty} \frac{u(u^2+1)(u^2+(m+2)^2)}{(\nu^2+u^2)^z} + \int_0^{\infty} \frac{2u(u^2+1)(u^2+(m+2)^2)}{(e^{2\pi u}-1)(\nu^2+u^2)^z}\right)\,.
        \end{split}
\end{equation}
We ignore $g(m)$ at the moment for simplicity. The first integral equals with
\begin{equation}
\begin{split}
    I&=\frac{1}{72} \nu^2 \Bigg[-144 + 135 \nu^2 - 22 \nu^4 + 36 m (-4 + 3 \nu^2) +
   9 m^2 (-4 + 3 \nu^2) \\
   &- 6 (-24 + 15 \nu^2 - 2 \nu^4 + 12 m (-2 + \nu^2) + 3 m^2 (-2 + \nu^2)) \log
     \nu^2\Bigg]\,.
    \end{split}
\end{equation}
The second integral is just $II=II.1+II.2$, where
\begin{equation}
 II.1= 2\left(2c_1^-(m+2)^2+2c_3^-((m+2)^2+1)+2c_5^-\right)\,,
\end{equation}
\begin{equation}
\begin{split}
    II.2 &= -4\int_0^{\nu}xdx \int_0^{\infty} du \frac{ u (1 + u^2) ((2 + m)^2 + u^2)}{( -1 + e^{2 \pi u}) (u^2 + x^2)}\\
    &=-4 \int_0^{\nu} xdx \left[(2+m)^2 J_1^-  + ((2+m)^2+1)J_3^- + J_5^-\right]\,.
    \end{split}
\end{equation}
Repeating the same algorithm as in the case of bosonic theory, we get
\begin{equation}
\begin{split}
   P_{\nu,m}&= -g(m) \Bigg[-\frac{1}{120}\nu(-480+51\nu+200\nu^2-155\nu^3-24\nu^4+30\nu^5-40m(12-\nu-4\nu^2+3\nu^3)\,,\\
   &-10m^2(12-\nu-4\nu^2+3\nu^3))-2c_1^-(m+2)^2-2c_3^-((m+2)^2+1)-2c_5^-\Bigg]\,,
   \end{split}
\end{equation}
\begin{equation}
   Q_{\nu,m}= -2g(m) \int_0^{\nu} dxx(x^2-1)(x^2-(m+2)^2)\psi(x)\,,
\end{equation}
where, we have returned the degeneracy into the calculation.
\begin{equation}
     P= \sum_{m=0}^{\infty} \left(e^{-\epsilon(m+1)}P_{m+1,m}-e^{-\epsilon(m+2)}P_{m+2,m-1}\right)=-\frac{1787}{3402000}\,,
\end{equation}
and $Q$ is just
\begin{equation}
\begin{split}
     Q &=  \sum_{m=0}^{\infty}\left( e^{-\epsilon(m+1)}Q_{m+1,m}-e^{-\epsilon(m+2)}Q_{m+2,m-1}\right)\\
    &=-\frac{1}{180}\int_0^{\infty} dt \Bigg[\frac{1440(e^{3t}-7e^{4t}-12e^{5t}-7e^{6t}-e^{7t})}{(-1+e^{t})^9t}-\frac{72e^{2t}(3+47e^t+47e^{2t}+3e^{3t})}{3(1+e^t)^6t^2}\\
    &\qquad \qquad +\frac{120e^{2t}(1+24e^t+33e^{2t})}{(-1+e^t)^9t^3}+\frac{360e^{2t}(1+19e^t+19e^{2t}+e^{3t})}{(-1+e^t)^6t^4}\\
    &\qquad \qquad +\frac{1440e^{2t}(1+4e^t+e^{2t})}{(-1+e^t)^5t^5}
    +\frac{1440e^{2t}(1+e^t)}{(-1+e^t)^4t^6}\Bigg]\\
    &=\frac{1787}{3402000}\,.
\end{split}
\end{equation}
Hence, $\zeta'(0)_{\frac{1}{2}}=0$, which guarantees that the consistency of SUSY HS theories relies on the bosonic part thereof.

\subsection{Height-one Hook HS fields}
\paragraph{Zeta.} To get to the Type-B theory we need to calculate the contribution of hook fields in $AdS_6$. The zeta-function is
\begin{equation}
   \tilde{\mu}(u)=-\frac{u  \left(u ^2+\frac{9}{4}\right) s (s+3) (2 s+3) \tanh (\pi u ) \left(u ^2+\left(s+\frac{3}{2}\right)^2\right)}{ 240}\,.
\end{equation}
Since $\Delta=s+3$ with $s=1,2,...$ and $\nu=s+1/2$, we can repeat the same calculation as for bosonic HS fields. The zeta function is therefore
\begin{equation}
    \zeta^{Hook} =-\frac{1}{240}\sum_{s=1}^{\infty} \frac{74}{63} - \frac{58 s}{7} - \frac{1109 s^2}{21} - 94 s^3 - \frac{337 s^4}{6} + 14 s^5 + \frac{
 91 s^6}{3} + 12 s^7 + \frac{3 s^8}{2}= \frac{1}{180}\,.
\end{equation}
While the result of zeta-function for even spin case is
 \begin{equation}
    \zeta^{Hook}_{min} = -\frac{1}{240}\sum_{s=2,4,...}^{\infty} \frac{74}{63} - \frac{58 s}{7} - \frac{1109 s^2}{21} - 94 s^3 - \frac{337 s^4}{6} + 14 s^5 + \frac{
 91 s^6}{3} + 12 s^7 + \frac{3 s^8}{2}= \frac{37}{7560}\,.
\end{equation}
It is easy to see that the zeta function for hook fields is not zero, which is not a problem since they make only a part of the Type-B spectrum.

\paragraph{Zeta-prime.} The $\zeta'=P_{\nu,s}+Q_{\nu,s}$ can be obtained by using the same treatment for bosonic theory, where we find that
\begin{equation}
\begin{split}
    P_{\nu,s}&= -\frac{s (3 + s) (3 + 2 s)}{240} \Bigg[2 c_5^+ + \frac{9}{8} c_1^+ (3 + 2 s)^2 +
   c_3^+ (9 + 6 s + 2 s^2)\\
   &+
   \frac{\nu^2}{960} \Big(187 + 1060 \nu^2 - 240 \nu^4 + 120 s (1 + 6 \nu^2) +
      40 s^2 (1 + 6 \nu^2)\Big)\Bigg]\,,
      \end{split}
\end{equation}
and
\begin{equation}
    Q_{\nu,s}= -\frac{s(s+3)(2s+3)}{1920}\int_0^{\nu} dx\ x(-9+4x^2)(-9-12s-4s^2+4x^2)\psi(x+1/2)\,.
\end{equation}
Summing over all spins, the result of $P$ is
\begin{equation}
    P^{Hook}= \sum_{s=1}^{\infty} P_{s+1/2,s}-P_{s+3/2,s-1} = \frac{1}{300}\,,
\end{equation}
while for the minimal case of Type-B, one needs to have
\begin{equation}
    P^{Hook}_{min}=\sum_{s=2,4,...}^{\infty}P_{s+\frac{1}{2},s}-P_{s+\frac{3}{2},s-1}=\frac{197}{51200} + \frac{3 c_1^+}{320} + \frac{c_3^+}{24} + \frac{c_5^+}{60}\,.
\end{equation}
Next, we evaluate the $Q^{Hook}$ for the non-minimal and minimal Type-B. We find for all spins:
\begin{equation}
    \begin{split}
         Q^{Hook}&= -\frac{623}{21600} + \frac{\log A}{6} + \frac{1}{6} \zeta'(-4) -
 \frac{1}{3} \zeta'(-3) + \frac{1}{3} \zeta'(-2)\\
    &=-\frac{623}{21600} +\frac{\log A}{6} + \frac{\zeta(5)}{8\pi^4}-\frac{\zeta(3)}{12 \pi^2}-\frac{1}{3}\zeta'(-3)\,,
    \end{split}
    \end{equation}
and for even spins only:
    \begin{equation}
         Q^{Hook}_{min}= -\frac{1433}{51200} + \frac{52709 \log(2)}{483840} + \frac{99 \log A}{640} +
\frac{\zeta(3)}{64 \pi^2} - \frac{93 \zeta(5)}{128 \pi^4} -
 \frac{21}{640} \zeta'(-5) - \frac{19}{64} \zeta'(-3)\,,
    \end{equation}
where we utilized,
\begin{equation}
\zeta'(-2n)=\frac{(-1)^n\zeta(2n+1)(2n)!}{2^{2n+1}\pi^{2n}}\,.
\end{equation}
Having these results at hand, we are now able to compute the $\zeta'_B$ for the non-minimal and minimal Type-B theories.

\subsection{Non-minimal Type-B }
In order to calculate the zeta function for Type-B, we need to collect all the information from Type-A, scalar field with $\Delta=4$ and the above hook fields. From (E.4), one can easily obtain the $\zeta^A_{s >0}$ for non-minimal which is $-\frac{1}{1512}$. For the scalar with $\Delta=4$, we simply get from \eqref{E.3} that
\begin{equation}
    \zeta_{4,0} = -\frac{37}{7560}\,.
\end{equation}
The spectrum of non-minimal Type-B involves the spectrum of Type-A theory with $s\geq1$, a scalar with $\Delta=4$ and the hook fields with $s\geq1$.
\begin{equation}
    \zeta^B=\zeta^A+\zeta_{4,0}+\zeta^{Hook}=-\frac{1}{1512}-\frac{37}{7560}+\frac{1}{180}=0\,.
\end{equation}
Below, we will list all the components in terms of their $P$ and $Q$ to calculate the $\zeta'^B$
\begin{equation}
\begin{tabular}{c|c}
     $Type$ & $P$  \\ \hline
     $P^A$ & $\frac{79}{153600}+\frac{3c_1^+}{320}+\frac{c_3^+}{24}+\frac{c_5^+}{60}$ \\
     $P^A_{\frac{3}{2},0}$ & $-\frac{197}{51200} - \frac{3 c_1^+}{320} -\frac{c_3^+}{24} - \frac{c_5^+}{60}$ \\
     $P^{Hook}$ & $\frac{1}{300}$\\
\end{tabular}
\end{equation}
It is easy to recognize that $P^B=P^A+P^A_{\frac{3}{2},0}+P^{Hook}=0$, i.e there is no contribution from $P$ in the Type-B theory. The relevant $Q$-terms are
\begin{equation}
\begin{tabular}{c|c}
     $Type$ & $Q$  \\ \hline
     $Q^A$ & $\frac{1}{120}\left(\frac{1181}{11520} - \frac{211 \log(2)}{4032} - \frac{23 \log A}{16} + \frac{5 \zeta(3)}{
 4 \pi^2} + \frac{15 \zeta(5)}{4 \pi^4} -
 \frac{63}{16} \zeta'(-5) + \frac{35}{8} \zeta'(-3)\right)$ \\
     $Q^A_{\frac{3}{2},0}$ & $\frac{1433}{51200} + \frac{211 \log(2)}{483840} - \frac{99 \log A}{640} + \frac{
 3 \zeta(3)}{32\pi^2} - \frac{3 \zeta(5)}{32 \pi^4}+
 \frac{21}{640} \zeta'(-5)+\frac{19}{64} \zeta'(-3)$ \\
     $Q^{Hook}$ & $-\frac{623}{21600} + \frac{\log A}{6} + \frac{\zeta(5)}{8\pi^4}-\frac{\zeta(3)}{12 \pi^2}-\frac{1}{3}\zeta'(-3)$
\end{tabular}
\end{equation}
Bringing everything together, we obtain
\begin{equation}
    \zeta'_B=\zeta'_{A,s\geq 1} + \zeta'_{Hook, s\geq 1} + \zeta'_{4,0}
    =\frac{\zeta(3)}{48 \pi^2}+\frac{\zeta(5)}{16\pi^4}\,.
\end{equation}
As explaining in the main text, this number is not random.

\subsection{Minimal Type-B}
From \eqref{E.4}, the zeta-function of Type-A with odd spins only is 0. One can read off the minimal Type-B $\zeta^B_{min}$ by considering the symmetric traceless fields with odd spins only, the hook fields with even spin and a scalar with $\Delta=4$.
\begin{equation}
    \zeta^B_{min}=\zeta^A_{odd}+\zeta_{4,0}+\zeta^{Hook}_{even}=0-\frac{37}{7560}+\frac{37}{7560}= 0\,.
\end{equation}
Therefore, the zeta function for Type-B is vanishing in both non-minimal and minimal cases. Next, we list the result for the minimal Type-B in terms of $P$ and $Q$
\begin{equation}
\begin{tabular}{c|c}
     Type & $P$  \\ \hline
     $P^A$ & $0$ \\
     $P^A_{\frac{3}{2},0}$ & $-\frac{197}{51200} - \frac{3 c_1^+}{320} - \frac{c_3^+}{24} - \frac{c_5^+}{60}$ \\
     $P^{Hook}$ & $\frac{197}{51200} + \frac{3 c_1^+}{320} + \frac{c_3^+}{24} + \frac{c_5^+}{60}$
\end{tabular}
\end{equation}
\begin{equation}
\begin{tabular}{c|c}
     Type & $Q$  \\ \hline
     $Q^A$ & $-\frac{\log(2)}{64} - \frac{\zeta(3)}{64 \pi^2} + \frac{15\zeta(5)}{128 \pi^4}$ \\
     $Q^A_{\frac{3}{2},0}$ & $\frac{1433}{51200} + \frac{211 \log(2)}{483840} - \frac{99 \log A}{640} + \frac{
 3 \zeta(3)}{32\pi^2} - \frac{3 \zeta(5)}{32 \pi^4} +
 \frac{21}{640} \zeta'(-5) +\frac{19}{64} \zeta'(-3)$ \\
     $Q^{Hook}$ & $ -\frac{1433}{51200} + \frac{52709 \log(2)}{483840} + \frac{99 \log A}{640} +
\frac{\zeta(3)}{64 \pi^2} - \frac{93 \zeta(5)}{128 \pi^4} -
 \frac{21}{640} \zeta'(-5) - \frac{19}{64} \zeta'(-3)$
\end{tabular}
\end{equation}
The $\zeta'^B_{min}$ for the minimal Type-B theory is just that:
\begin{equation}
    \zeta'^B_{min} =\zeta'_{A,odd} + \zeta'_{Hook, even} + \zeta'_{4,0}
    =\frac{3}{32}\log 2 +\frac{3\zeta(3)}{32\pi^2}-\frac{45\zeta(5)}{64\pi^4}\,.
\end{equation}
In the following appendices, we list the result of zeta function of Type-A, fermions, hook fields and Type-B in various dimensions, which can be used for later work.

\section{Summary of the Results in Other Even Dimensions}

\subsection{Type-A}
We first evaluate the zeta function in term of spin-$s$. Following the algorithm in the Appendix D, the results are listed below
\begin{equation}
\begin{tabular}{c|c}
     $d$ & $\zeta_{\Delta,s}-\zeta_{\Delta+1,s-1}$  \\ \hline
     $3$ & $\frac{1}{180} (-2 + 15 s^2 - 75 s^4)$ \\
     $5$ & $\frac{(1+s)^2(-20+28s+378s^2+868s^3+847s^4+378s^5+63s^6)}{30240}$\\
     $7$ & $\frac{(2+s)^2(-3048+1024s+55568s^2+162632s^2+228337s^4+188892s^5+98397s^6+32688s^7+6723s^8+780s^9+39s^{10})}{21772800}$ 
\end{tabular}
\end{equation}
The sum over spins will make $\zeta(0)$ vanish in both non-minimal and minimal cases.\footnote{We used the cut-off exponential $\exp[-\epsilon(s+\frac{d-3}{2})]$. The case with $d=3$ is special since one should start the sum from $s \geq 1$ and then add the scalar to have vanishing zeta function.} Next, we compute $P_{\nu,s}$ and $Q_{\nu,s}$ \\ \\
\textbf{Table for $P_{\nu,s}$:\footnote{From here, it is very easy to evaluate $P=\sum_s P_{\nu,s}-P_{\nu+1,s-1}$ by the exponential cut-off.}} \\
{\small$
\begin{aligned}
     d=3: \frac{(2s+1)(12 c_1^+ + 48 c_3^+ + 48 c_1^+ s + 48 c_1^+ s^2 + \nu^2 + 6 \nu^4)}{144}
\end{aligned}$}\\ \\
{\small $\begin{aligned}
     d=5: -\frac{(s + 1) (s + 2) (2 s + 3)}{691200} &\Big[1080 c_1^+ + 4800 c_3^+ + 1920 c_5^+ + 1440 c_1^+ s + 5760 c_3^+ s + 480 c_1^+ s^2 +
  1920 c_3^+ s^2 \\
  &+ 107 \nu^2 + 120 s \nu^2 + 40 s^2 \nu^2 + 580 \nu^4 +
  720 s \nu^4 + 240 s^2 \nu^4 - 240 \nu^6\Big]
\end{aligned}$}\\ \\
{\small$\begin{aligned} d=7:&\frac{(1 + s) (2 + s) (3 + s) (4 + s) (5 + 2 s) }{48771072000}\Big[567000 c_1^+ + 2610720 c_3^+ + 1411200 c_5^+ + 161280 c_7^+ + 453600 c_1^+ s \\
&+ 2016000 c_3^+ s + 806400 c_5^+ s + 90720 c_1^+ s^2 + 403200 c_3^+ s^2 +
 161280 c_5^+ s^2 + 343345 \nu^2 + 271740 s \nu^2 \\&+ 54348 s^2 \nu^2 -
 667674 \nu^4 - 512400 s \nu^4 - 102480 s^2 \nu^4 + 255920 \nu^6 +
 145600 s \nu^6 + 29120 s^2 \nu^6 - 23520 \nu^8\Big]
    \end{aligned}$}\\ \\
\textbf{Table of $Q_{\nu,s}$:}
\begin{equation}
\begin{tabular}{c|c}
     $d$ & $Q_{\nu,s}$  \\ \hline
     $3$ &  $\frac{1}{3}(2s+1)\int_0^{\nu}dx \left[(s+\frac{1}{2})^2x-x^3\right]\psi(x+\frac{1}{2})$\\
     $5$ & $\frac{(s+1)(s+2)(2s+3)}{5760}\int_0^{\nu} x (9 + 12 s + 4 s^2 - 4 x^2) (-1 + 4 x^2) \psi(1/2 + x)$\\
     $7$ & $\frac{(s+1)(s+2)(s+3)(s+4)(2s+5)}{604800}\int_0^{\nu}dx \frac{x}{32} (25 + 20 s + 4 s^2 - 4 x^2) (9 - 40 x^2 + 16 x^4) \psi(
   x+\frac{1}{2})$
\end{tabular}
\end{equation}
\paragraph{Non-minimal Type-A.}
The result for $P$ in both non-minimal and minimal theory are zero, i.e $P$ vanishes. Hence, one only needs to deal with $Q=\sum_s Q_{\nu,s}-Q_{\nu+1,s-1}$. The sum is evaluated with $\text{exp}[-\epsilon \nu]$ for $Q_{\nu,s}$ and with $\text{exp}[-\epsilon (\nu+1)]$ for $Q_{\nu+1,s-1}$. Analytical computation in the non-minimal Type-A shows that $Q$ also vanishes.

\paragraph{Minimal Type-A.}
In minimal theory, the story is a little bit different. Using the method of analytical continuation of Appendix D, we get
\begin{equation}
\begin{tabular}{c|c}
     $d$ & $Q$  \\ \hline
     $3$ &  $-\frac{1}{2^3}\left(2\log 2 - \frac{3\zeta(3)}{\pi^2}\right)$\\
     $5$ & $\frac{1}{2^7}\left(2\log 2 + \frac{2\zeta(3)}{\pi^2}-\frac{15\zeta(5)}{\pi^4}\right)$\\
     $7$ & $-\frac{1}{2^{11}}\left(4 \log 2 + \frac{82\zeta(3)}{15\pi^2}   -\frac{10\zeta(5)}{\pi^4} - \frac{63\zeta(7)}{\pi^6}\right)$
\end{tabular}
\end{equation}
These results can also be found in \cite{Giombi:2014iua,Klebanov:2011gs}.

\subsection{HS Fermions}
Above, we showed that $\zeta_{\frac{1}{2}}$ and $\zeta'_{\frac{1}{2}}$ is zero for $AdS_6$. In this Appendix, let us rewrite the result in $d=3,5$ and then make a general statement about higher dimensional cases. First of all, one needs to make the change of variable $s=m+\frac{1}{2}$. The zeta-functions with the ghost subtracted are
\begin{equation}
\begin{tabular}{c|c}
     $d$ & $\zeta_{\Delta,s}-\zeta_{\Delta+1,s-1}$  \\ \hline
     $3$ & $\frac{-47 - 360 m - 1560 m^2 - 2400 m^3 - 1200 m^4}{2880}$ \\
     $5$ & $\frac{542 + 99 m - 8094 m^2 - 22806 m^3 - 28497 m^4 - 19404 m^5 -
   7448 m^6 - 1512 m^7 - 126 m^8}{30240}$
\end{tabular}
\end{equation}
Summing over all spin starting from $m=0$ with the cut-off $\exp[-\epsilon(m+\frac{d-2}{2})]$, we see that the total zeta-functions in $d=3,5$ vanished. As a simple check, one can confirm that for higher dimensions this statement is also true.\\
Next, to calculate the $\zeta'$-function, we again split it into $P_{\nu,m}$ and $Q_{\nu,m}$.
\paragraph{Table for $P_{\nu,m}$:}
{\footnotesize
\begin{align}
d=3&:&&\begin{aligned}
      -\frac{(1 + m) (24 c_1^- + 24 c_3^- + 48 c_1^- m + 24 c_1^- m^2 - 12 \nu -
   24 m \nu - 12 m^2 \nu + \nu^2 + 4 \nu^3 - 3 \nu^4)}{36}
\end{aligned}\,,\\
d=5&:&&\begin{aligned}
     &-\frac{(1 + m) (2 + m) (3 + m)}{21600} \Big[960 c_1^- + 1200 c_3^- + 240 c_5^- +
    960 c_1^- m + 960 c_3^- m + 240 c_1^- m^2 + 240 c_3^- m^2 \\&- 480 \nu - 480 m \nu -
    120 m^2 \nu + 51 \nu^2 + 40 m \nu^2 + 10 m^2 \nu^2 + 200 \nu^3 +
    160 m \nu^3 + 40 m^2 \nu^3 - 155 \nu^4 \\&- 120 m \nu^4 - 30 m^2 \nu^4 -
    24 \nu^5 + 30 \nu^6\Big]\,.
\end{aligned}\end{align}}\noindent
Summing over all spins leads to
\begin{center}
\begin{tabular}{c|c}
     $d$ & $P$  \\ \hline
     $3$ & $-\frac{11}{270}$ \\
     $5$ & $\frac{1787}{3402000}$\\
\end{tabular}
\end{center}
One can see that for fermions $P$ is non-zero which is different from Type-A theories. For $Q_{\nu,m}$ we get
\begin{center}
\begin{tabular}{c|c|c}
     $d$ & $Q_{\nu,m}$&$Q$  \\ \hline
     $3$ &$-\frac{2(m+1)}{3}\int_0^{\nu} dx (x^3-(m+1)^2x)$ &$\frac{11}{270}$\\
     $5$ &$\frac{(m+1)(m+2)(m+3)}{90}\int_0^{\nu}dx(x^3-x)(x^2-(m+2)^2)\psi(x)$ & $-\frac{1787}{3402000}$\\
\end{tabular}
\end{center}
It is easy to see that $P$ and $Q$ always cancel each other. A further check confirms that $\zeta'(0)$ is zero in higher  dimensions.
\subsection{Hook fields}
The hook fields only appear in dimensions higher than four. For the computation of the spectral density function $\mu(u)$ of hooks with different $p$, the reader can refer to Section 3.2.2.
\subsubsection{Zeta}
In $d=5$, we only have $p=1$, while in $d=7$, $p$ can be one or two.\footnote{Due to the length of the final results, we only list the zeta function for $d=5,7$ here.}
{\footnotesize\begin{align*}
d=5\,,p=1&:&&\begin{aligned}
      \frac{148 - 1044 s - 6654 s^2 - 11844 s^3 - 7077 s^4 + 1764 s^5 +
 3822 s^6 + 1512 s^7 + 189 s^8}{30240}
\end{aligned}\,, \\
d=7\,,p=1&:&&\begin{aligned}
      &-\frac{(2 + s)}{5573836800} \Big[-81336637326 - 260554380359 s -
    287920256390 s^2 - 124396596105 s^3 \\
    &+ 7147903040 s^4 +
    30702694976 s^5 + 14557085760 s^6 + 3622437600 s^7 +
    540003840 s^8 \\&+ 48318720 s^9 + 2388480 s^{10} + 49920 s^{11}\Big]
\end{aligned}\,,\\
d=7\,,p=2&:&&\begin{aligned}
      &-\frac{
 s (4 + s)}{2786918400} \Big[-79449809509 - 151977792308 s - 101475411753 s^2 -
    17276191808 s^3 \\&+ 13378662464 s^4 + 9277153920 s^5 +
    2721896160 s^6 + 451660800 s^7 + 43687680 s^8 \\&+ 2288640 s^9 +
    49920 s^{10}\Big]\,.
\end{aligned}\end{align*}}\noindent
We will list the result of $\zeta$-function in both the non-minimal and minimal theory for hook fields below since it is important for our computation of Type-B theory\footnote{The hook fields of minimal theory in $d=5$ come with even spins while the hook fields with $p=1$ in $d=7$ come with odd spins and $p=2$ come with even spins.}
\begin{equation}
\begin{tabular}{c|c|c}
     $d$ & $p$ &$(\zeta, \zeta_{min})$  \\ \hline
     $5$ &$1$ &$\left(\frac{1}{180},-\frac{37}{7560}\right)$ \\
     $7$ & $1$ &$\left(\frac{1}{280}, -\frac{23}{226800} \right)$\\
     & $2$ &$\left(\frac{1}{1512},\frac{23}{226800} \right)$
\end{tabular}
\end{equation}
It is interesting that the zeta function for hook fields alone is not zero as in bosonic and fermionic theory. However, when one considers the whole spectrum of Type-B theory, the zeta function will again vanish.
\subsubsection{Zeta-prime}
Below are the tables for $P_{\nu,s}$ and $Q_{\nu,s}$ of hook fields.
\paragraph{Table for $P_{\nu,s}$:}
{\footnotesize\begin{align*}
d=5\,,p=1&:&&\begin{aligned}
      &-\frac{
 s (3 + s) (3 + 2 s)}{230400} \Big[9720 c_1^+ + 8640 c_3^+ + 1920 c_5^+ + 12960 c_1^+ s +
    5760 c_3^+ s + 4320 c_1^+ s^2 \\&+ 1920 c_3^+ s^2 + 187 \nu^2 + 120 s \nu^2 +
    40 s^2 \nu^2 + 1060 \nu^4 + 720 s \nu^4 + 240 s^2 \nu^4 - 240 \nu^6\Big]
\end{aligned}\,,\\
d=7\,,p=1&:&&\begin{aligned}
      &\ \ \frac{s (2 + s) (3 + s) (5 + s) (5 + 2 s)}{9754214400} \Big[1575000 c_1^+ +
   6804000 c_3^+ + 2056320 c5 + 161280 c_7^+ + 1260000 c_1^+ s \\&+
   5241600 c_3^+ s + 806400 c_5^+ s + 252000 c_1^+ s^2 + 1048320 c_3^+ s^2 +
   161280 c_5^+ s^2 + 149557 \nu^2 + 112140 s \nu^2 \\&+ 22428 s^2 \nu^2 +
   828786 \nu^4 + 646800 s \nu^4 + 129360 s^2 \nu^4 - 255920 \nu^6 -
   100800 s \nu^6 - 20160 s^2 \nu^6 + 18480 \nu^8\Big] \end{aligned}\,,\\
d=7\,,p=2&:&&\begin{aligned}
      &\ \ \frac{s (1 + s) (4 + s) (5 + s) (5 + 2 s)}{4877107200} \Big[14175000 c_1^+ +
   10836000 c_3^+ + 2378880 c_5^+ + 161280 c_7^+ + 11340000 c_1^+ s \\&+
   6854400 c_3^+ s + 806400 c_5^+ s + 2268000 c_1^+ s^2 + 1370880 c_3^+ s^2 +
   161280 c_5^+ s^2 + 234733 \nu^2 + 145740 s \nu^2 \\&+ 29148 s^2 \nu^2 +
   1329426 \nu^4 + 848400 s \nu^4 + 169680 s^2 \nu^4 - 296240 \nu^6 -
   100800 s \nu^6 - 20160 s^2 \nu^6 + 18480 \nu^8\Big]\,.
\end{aligned}\end{align*}}\noindent
Summing over spins leads to
\begin{equation}
\begin{tabular}{c|c|c}
     $d$ & $p$ &$(P,P_{min})$  \\ \hline
     $5$ &$1$ &$\left(\frac{1}{300},\frac{197}{51200} + \frac{3 c_1^+}{320} + \frac{c_3^+}{24} + \frac{c_5^+}{60}\right)$ \\
     $7$ & $1$ &$\left(\frac{1361}{264600}, \frac{508061}{6502809600} + \frac{5 c_1^+}{3584} + \frac{37 c_3^+}{5760} + \frac{c_5^+}{288} + \frac{c_7^+}{2520}\right)$\\
     & $2$ &$\left(\frac{61}{158760},-\frac{508061}{6502809600} - \frac{5 c_1^+}{3584} - \frac{37 c_3^+}{5760} - \frac{c_5^+}{288} - \frac{c_7^+}{2520} \right)$
\end{tabular}
\end{equation}
\paragraph{Table for $Q_{\nu,s}$:}
{\footnotesize \begin{align*}
d=5\,,p=1&:&&\begin{aligned}
      &-\frac{s(s+3)(2s+3)}{1920}\int_0^{\nu} dx\ x(-9+4x^2)(-9-12s-4s^2+4x^2)\psi(x+1/2)
\end{aligned}\,, \\
d=7\,,p=1&:&&\begin{aligned}
      &\ \ \frac{s(s+2)(s+3)(s+5)(2s+5)}{120960}\int_0^{\nu} dx \frac{x}{32}(25+20s+4s^2-4x^2)(25-104x^2+16x^4)\psi(x+\frac{1}{2}) \end{aligned}\,,\\
d=7\,,p=2&:&&\begin{aligned}
      &\ \ \frac{s (s + 1) (s + 4) (s + 5) (2 s +
   5)}{60480}\int_0^{\nu}dx\frac{x}{32} (25 + 20 s + 4 s^2 - 4 x^2) (225 - 136 x^2 + 16 x^4)\psi(x+\frac{1}{2})\,.
\end{aligned}\end{align*}}
\paragraph{Non-minimal Type-B.}
Following the method in appendix D, we list the results of $Q$ in $d=5,7$.
\begin{equation}
\begin{tabular}{c|c|c}
     $d$ & $p$ &$Q$  \\ \hline
     $5$ & $1$& $-\frac{623}{21600} +\frac{\log A}{6} + \frac{\zeta(5)}{8\pi^4}-\frac{\zeta(3)}{12 \pi^2}-\frac{\zeta'(-3)}{3}$\\
     $7$ &$1$ &$-\frac{26777}{1058400} + \frac{7 \log A}{60} -\frac{113\zeta(3)}{1440\pi^2}+\frac{13\zeta(5)}{96\pi^4} -\frac{\zeta(7)}{32\pi^6}-\frac{\zeta'(-3)}{3}-\frac{\zeta'(-5)}{20} $ \\
     & $2$ & $-\frac{991}{317520} + \frac{\log A}{60}-\frac{7\zeta(3)}{1440\pi^2}-\frac{\zeta(5)}{96\pi^4}+\frac{\zeta(7)}{32\pi^6}+\frac{\zeta'(-5)}{60}  $
\end{tabular}
\end{equation}

\paragraph{Minimal Type-B.} In the minimal theory, the computations are much longer since there are more derivatives involved when one calculates the Hurwitz-Lersch functions.
\begin{equation}\notag
\begin{tabular}{c|c|c}
     $d$ & $p$ &$Q$  \\ \hline
     $5$ & $1$&  {\footnotesize${-\frac{1433}{51200} + \frac{52709 \log(2)}{483840} + \frac{99 \log A}{640} +
\frac{\zeta(3)}{64 \pi^2} - \frac{93 \zeta(5)}{128 \pi^4} -
 \frac{21\zeta'(-5)}{640}  - \frac{19\zeta'(-3)}{64} }$}\\
     $7$ &$1$ &{\footnotesize$\frac{2545 \log (A)}{21504}+\frac{535 \zeta '(-4)}{2304}+\frac{4787 \zeta '(-2)}{11520}-\frac{139 \zeta '(-5)}{3072}-\frac{1037 \zeta '(-3)}{3072}-\frac{487 \zeta '(-6)}{11520}-\frac{17 \zeta '(-7)}{21504}-\frac{6610955}{260112384}-\frac{4067243 \log (2)}{232243200} $} \\
     & $2$ & {\footnotesize$\frac{181 \log (A)}{107520}+\frac{73 \zeta '(-5)}{15360}+\frac{113 \zeta '(-4)}{1152}+\frac{389 \zeta '(-2)}{1152}-\frac{13 \zeta '(-3)}{3072}-\frac{17 \zeta '(-7)}{21504}+\frac{1205 \zeta (7)}{1024 \pi ^6}-\frac{755987}{6502809600}-\frac{13592843 \log (2)}{232243200} $}
\end{tabular}
\end{equation}

\subsection{Type-B}
We can now combine the results above to get the results for Type-B models. The spectrum of such models is given in Section \ref{sec:mixedsymmetrytestsE}.
\subsubsection{Non-minimal}
\paragraph{Scalar Field.}
The scalar in Type-B has $\Delta_B^{\phi}=\Delta_A^{\phi}+1$, where $\Delta_A^{\phi}$ is the conformal weight of the scalar in Type-A theory. One can use this to compute $\zeta,P,Q$ using all the formulas in Type-A:
\begin{equation}\notag
\begin{tabular}{c|c}
     $d$ & $\zeta_{\Delta_B,0}$  \\ \hline
     $5$ &$ -\frac{37}{7560}$\\
     $7$ & $-\frac{119}{32400}$
\end{tabular}
\qquad \qquad \qquad
\begin{tabular}{c|c}
     $d$ & $P^{\phi}$  \\ \hline
     $5$ & $-\frac{197}{51200} - \frac{3 c_1^+}{320} -\frac{c_2^+}{24} - \frac{c_5^+}{60}$\\
     $7$ &  $-\frac{1317595}{260112384}+\frac{5c_1^+}{3584}+\frac{37c_3^+}{5760}+\frac{c_5^+}{288}+\frac{c_7^+}{2520}$
\end{tabular}
\end{equation}
\begin{equation}\notag
\begin{tabular}{c|c}
     $d$ & $Q^{\phi}$  \\ \hline
     $5$ & $\frac{1433}{51200} + \frac{211 \log(2)}{483840} - \frac{99 \log A}{640} + \frac{
 3 \zeta(3)}{32\pi^2} - \frac{3 \zeta(5)}{32 \pi^4}+
 \frac{21\zeta'(-5)}{640} +\frac{19 \zeta'(-3)}{64}$\\
     $7$ &  $\frac{6610955}{260112384}-\frac{15157\log(2)}{232243200}-\frac{2545\log A}{21504}+\frac{23\zeta(3)}{288\pi^2}-\frac{25\zeta(5)}{192\pi^4}+\frac{5\zeta(7)}{128\pi^6}+\frac{1037\zeta'(-3)}{3072}+\frac{139\zeta'(-5)}{3072}+\frac{17\zeta'(-7)}{21504}$
\end{tabular}
\end{equation}

\paragraph{Summary.}
    In non-minimal Type-B theory, we have one scalar with $\Delta_B=\Delta_A+1$, Type-A with $s\geq 1$, and the hook fields with $s\geq 1$. The total contribution to the zeta-function gives zero
\begin{center}
\begin{tabular}{c|c|c}
     $d$ & $\zeta_A + \zeta_{Hook} + \zeta^{\phi}_{\Delta,s}$ & $\zeta_B$ \\ \hline
     $5$ & $-\frac{1}{1512}+\frac{1}{180}-\frac{37}{7560}$ & $0$\\
     $7$ & $ -\frac{127}{226800}+\frac{1}{280}+ \frac{1}{1512}-\frac{119}{32400}$ & $0$ \\
\end{tabular}
\end{center}
For higher dimensions, this is also true and we can confirm that the zeta-function for non-minimal Type-B is always zero by combining all the component fields. Next, we need $\zeta'_B=\zeta'_{\Delta_B,0}+\zeta'_{A,s\geq1}+\zeta'_{Hook}$:
\begin{equation}
\begin{tabular}{c|c}
     $d$ & $\zeta'_B$  \\ \hline
     $5$ & $\frac{\zeta(3)}{48 \pi^2}+\frac{\zeta(5)}{16\pi^4}$\\
     $7$ &  $\frac{\zeta(3)}{360 \pi^2} + \frac{\zeta(5)}{96 \pi^4} +
 \frac{\zeta(7)}{64 \pi^6}$
\end{tabular}
\end{equation}
In the main text, our results were generated up to $AdS_{12}$ or $d=11$, but we checked up to $AdS_{18}$ that they agree with the change of $F$-energy.

\subsubsection{Minimal}
We need to combine the scalar field from the previous sub-section with the results for odd/even spins that can be found above. The final results can be found in the main text.

\end{appendix}

\setstretch{1.0}
\providecommand{\href}[2]{#2}\begingroup\raggedright\endgroup

\end{document}